\documentclass[11pt]{article}
\usepackage[margin=1in]{geometry}
\usepackage{amsmath, amssymb, braket, dsfont,authblk}
\usepackage{graphicx}
\usepackage{rotating, mathtools}
\usepackage[matrix,arrow,cmtip]{xy}
\usepackage[colorlinks]{hyperref}
\usepackage[numbers,sort&compress]{natbib}

\hypersetup{linkcolor={blue}}

\newcommand{\Tr}{\operatorname{Tr}}
\newcommand{\SU}[1]{{\hat{\mathfrak{su}}(#1)}}

\newcommand{\D}{{\mathcal{D}}}
\newcommand{\T}{{{\cal T}}} 
\newcommand{\modT}{{\mathbf{T}}}
\newcommand{\modS}{{\mathbf{S}}}

\usepackage{xifthen}

\newcommand{\Z}{\mathord{\vcenter{\hbox{\includegraphics[scale=1]{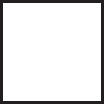}}}}}

\newcommand{\ZRxRw}{\mathord{\vcenter{\hbox{\includegraphics[scale=1]{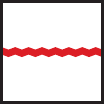}}}}}
\newcommand{\ZGxG}{\mathord{\vcenter{\hbox{\includegraphics[scale=1]{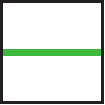}}}}}

\newcommand{\ZxRRw}{\mathord{\vcenter{\hbox{\includegraphics[scale=1]{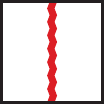}}}}}
\newcommand{\ZxGG}{\mathord{\vcenter{\hbox{\includegraphics[scale=1]{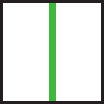}}}}}

\newcommand{\ZRRxw}{\mathord{\vcenter{\hbox{\includegraphics[scale=1]{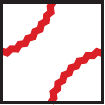}}}}}
\newcommand{\ZGGx}{\mathord{\vcenter{\hbox{\includegraphics[scale=1]{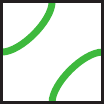}}}}}

\newcommand{\ZGRGw}{\mathord{\vcenter{\hbox{\includegraphics[scale=1]{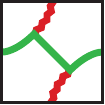}}}}}

\newcommand{\ZPants}{\mathord{\vcenter{\hbox{\includegraphics[scale=.25]{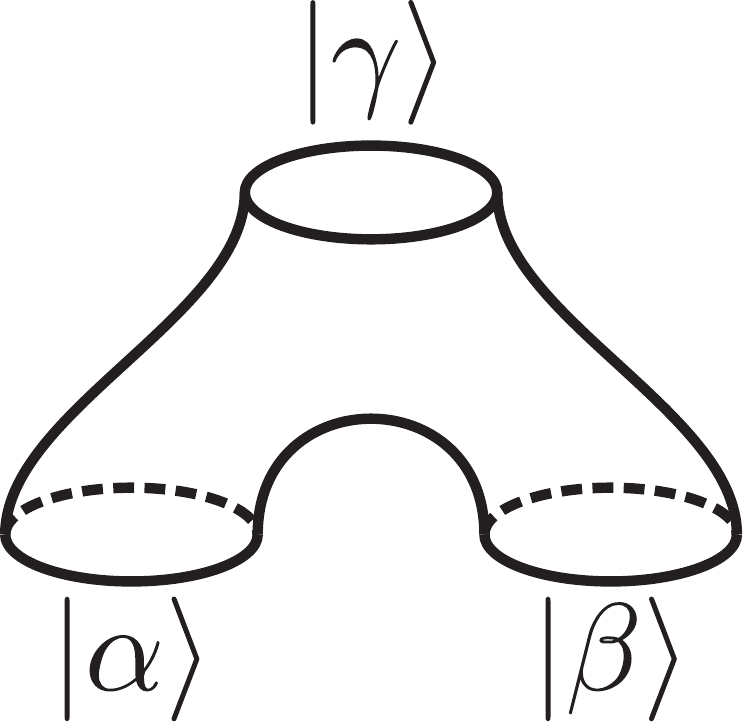}}}}}
\newcommand{\ZPantsa}{\mathord{\vcenter{\hbox{\includegraphics[scale=.25]{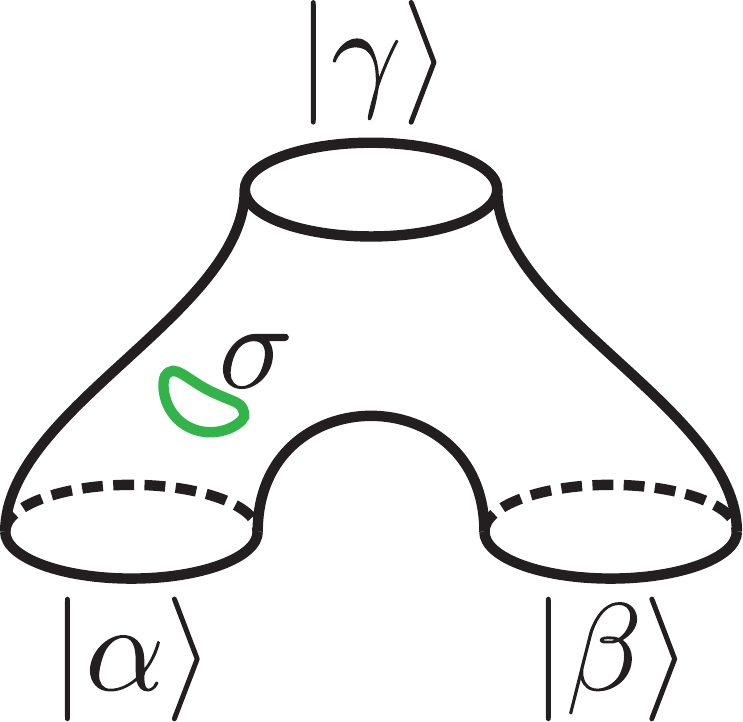}}}}}
\newcommand{\ZPantsb}{\mathord{\vcenter{\hbox{\includegraphics[scale=.25]{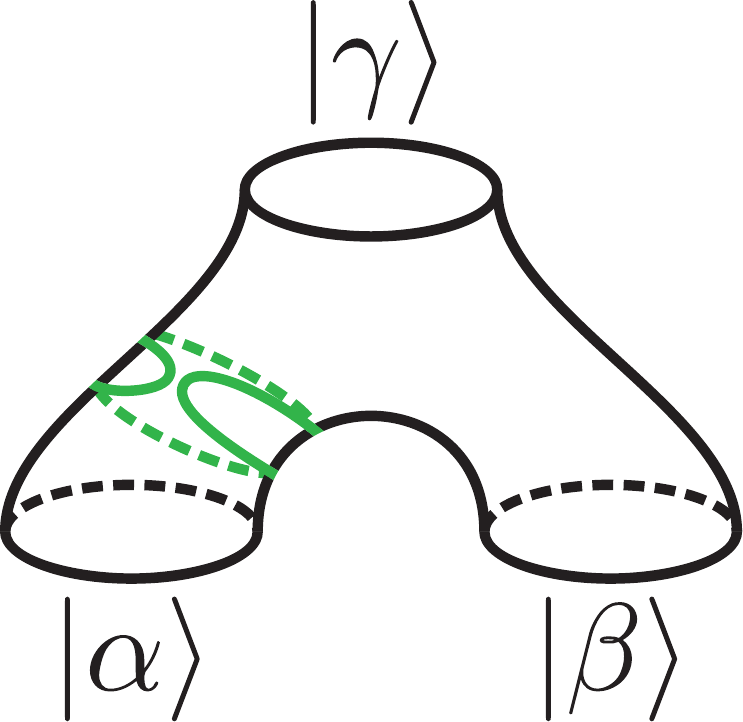}}}}}
\newcommand{\ZPantsc}{\mathord{\vcenter{\hbox{\includegraphics[scale=.25]{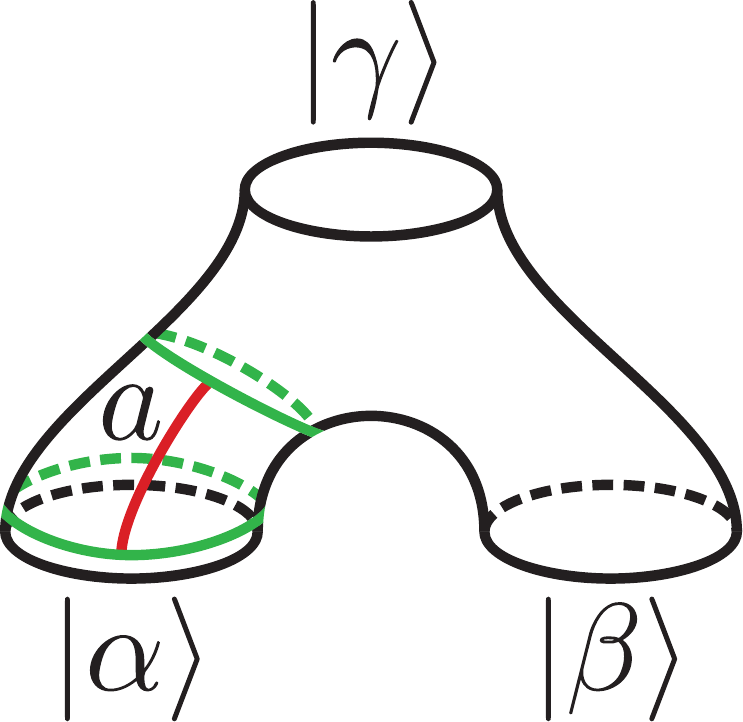}}}}}
\newcommand{\ZPantsd}{\mathord{\vcenter{\hbox{\includegraphics[scale=.25]{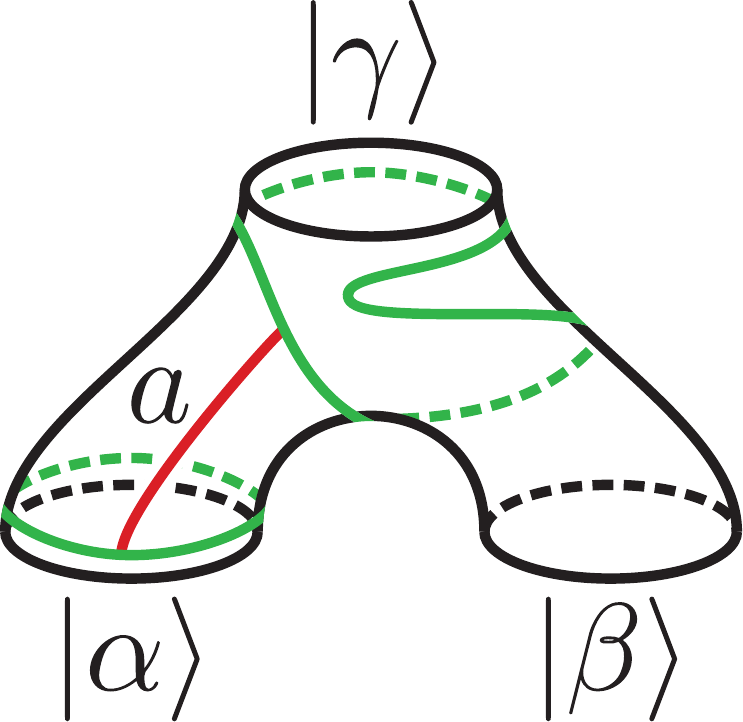}}}}}
\newcommand{\ZPantse}{\mathord{\vcenter{\hbox{\includegraphics[scale=.25]{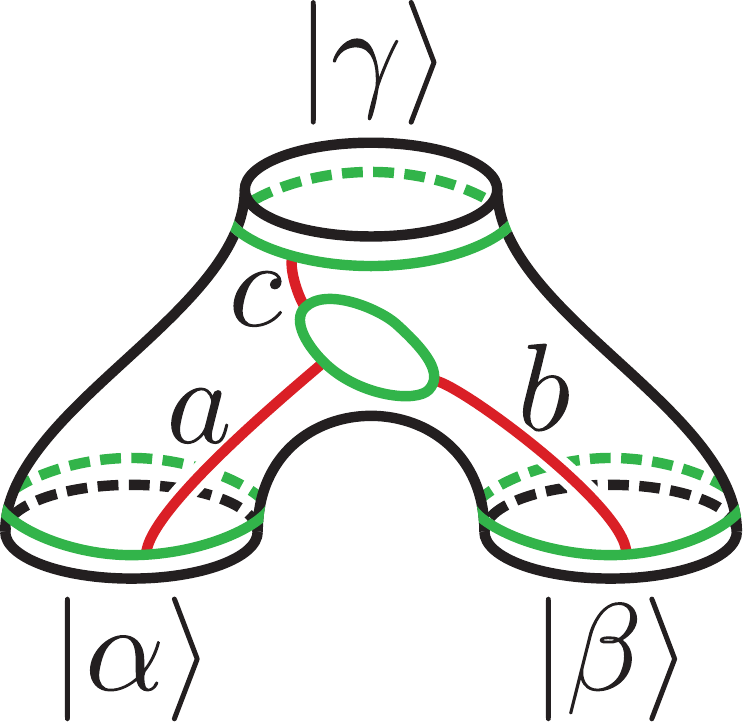}}}}}
\newcommand{\ZPantsf}{\mathord{\vcenter{\hbox{\includegraphics[scale=.25]{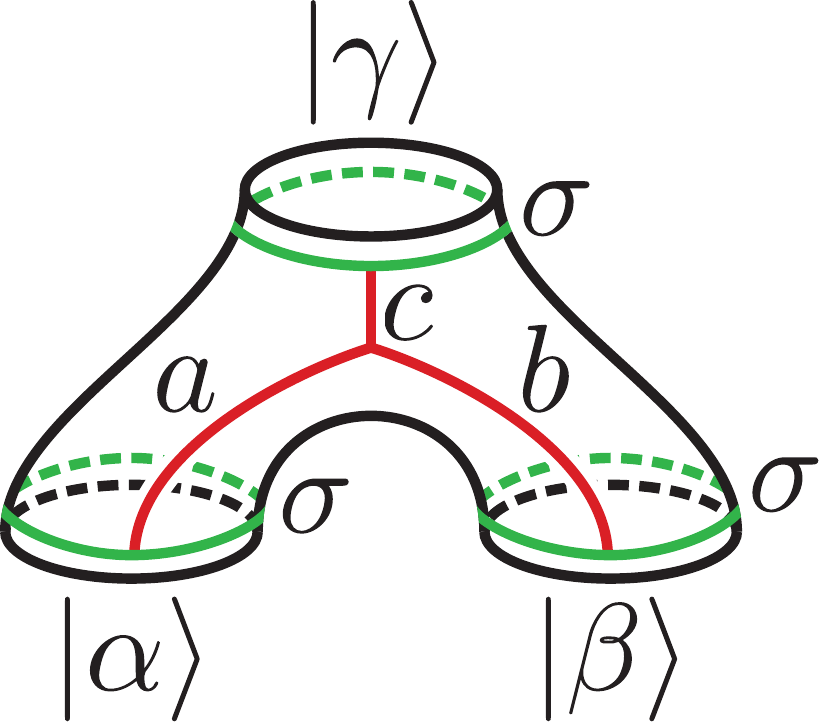}}}}}
\newcommand{\ZPuncturedTorus}{\mathord{\vcenter{\hbox{\includegraphics[scale=.25]{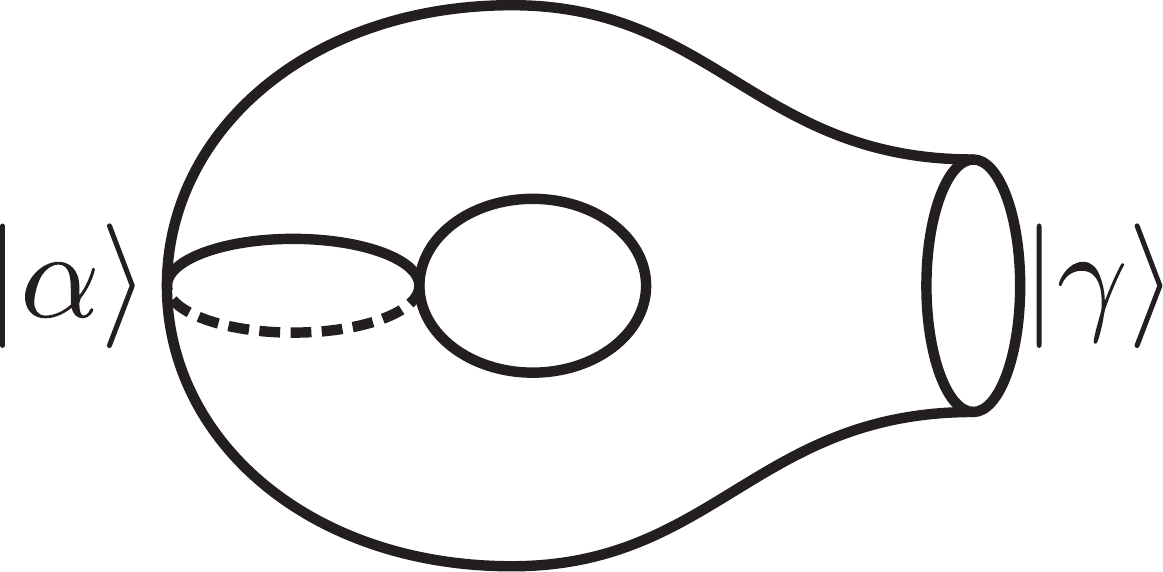}}}}}

\newcommand{\PantsGluea}{\mathord{\vcenter{\hbox{\includegraphics[scale=.25]{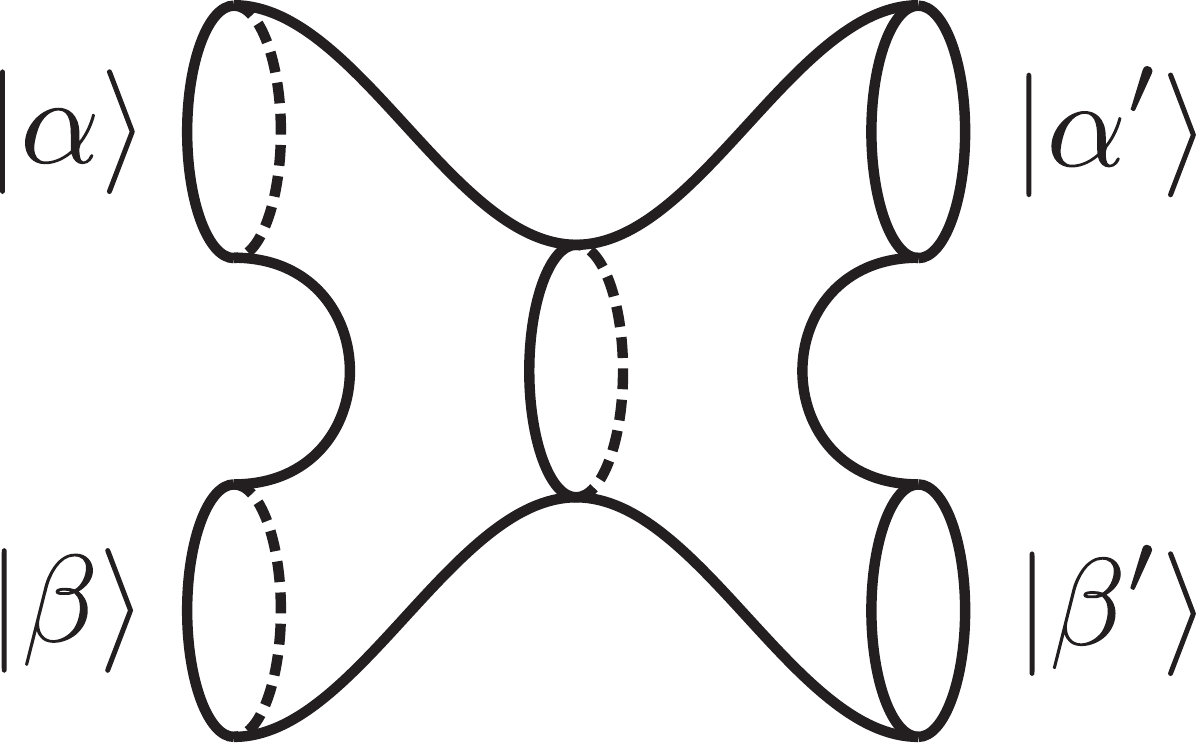}}}}}
\newcommand{\PantsGlueb}{\mathord{\vcenter{\hbox{\includegraphics[scale=.25]{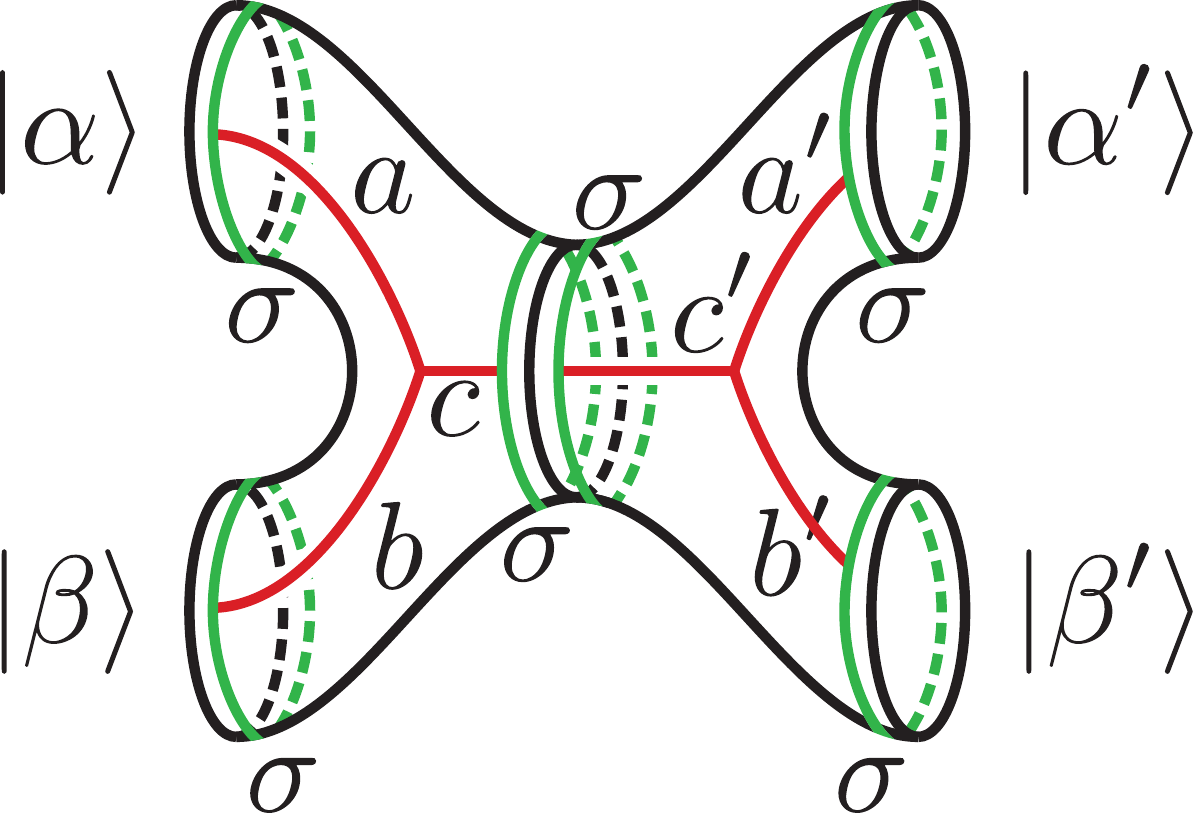}}}}}
\newcommand{\PantsGluec}{\mathord{\vcenter{\hbox{\includegraphics[scale=.25]{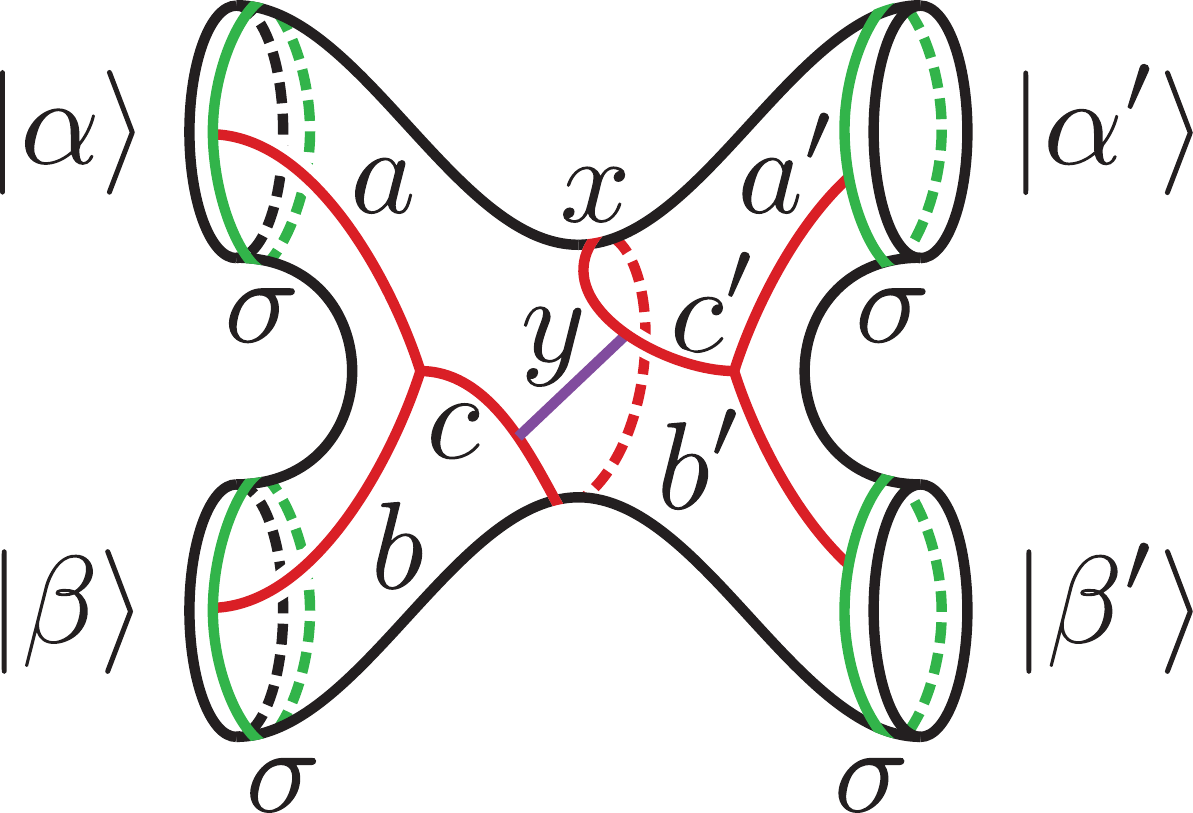}}}}}

\newcommand{\ZRGGw}{\mathord{\vcenter{\hbox{\includegraphics[scale=1]{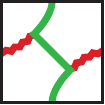}}}}}

\newcommand{\ZGGRw}{\mathord{\vcenter{\hbox{\includegraphics[scale=1]{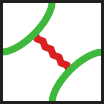}}}}}

\newcommand{\ZDisc}{\mathord{\vcenter{\hbox{\includegraphics[scale=1]{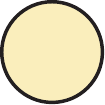}}}}}
\newcommand{\ZDiscG}{\mathord{\vcenter{\hbox{\includegraphics[scale=1]{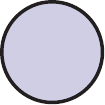}}}}}
\newcommand{\ZDiscGLoop}{\mathord{\vcenter{\hbox{\includegraphics[scale=1]{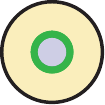}}}}}
\newcommand{\ZDiscGDual}{\mathord{\vcenter{\hbox{\includegraphics[scale=1]{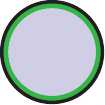}}}}}

\newcommand{\ZAnnulus}{\mathord{\vcenter{\hbox{\includegraphics[scale=1]{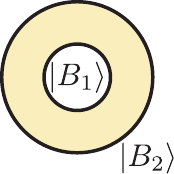}}}}}
\newcommand{\ZAnnulusDual}{\mathord{\vcenter{\hbox{\includegraphics[scale=1]{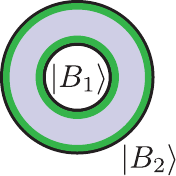}}}}}
\newcommand{\ZAnnulusDualPsi}{\mathord{\vcenter{\hbox{\includegraphics[scale=1]{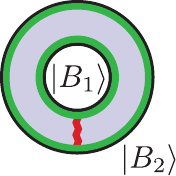}}}}}

\newcommand{\ThetaDiagram}{\mathord{\vcenter{\hbox{\includegraphics[scale=1]{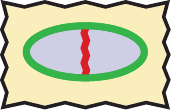}}}}}
\newcommand{\Barbell}{\mathord{\vcenter{\hbox{\includegraphics[scale=1]{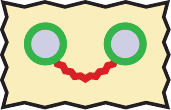}}}}}
\newcommand{\SigmaBubble}{\mathord{\vcenter{\hbox{\includegraphics[scale=1]{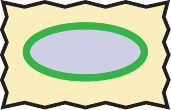}}}}}

\newcommand{\OxGG}{\mathord{\vcenter{\hbox{\includegraphics[scale=1]{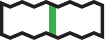}}}}}
\newcommand{\OGGx}{\mathord{\vcenter{\hbox{\includegraphics[scale=1]{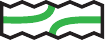}}}}}

\newcommand{\ORGGw}{\mathord{\vcenter{\hbox{\includegraphics[scale=1]{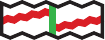}}}}}

\newcommand{\OxRGG}{\mathord{\vcenter{\hbox{\includegraphics[scale=1]{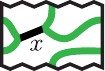}}}}}
\newcommand{\OGGxsquareed}{\mathord{\vcenter{\hbox{\includegraphics[scale=1]{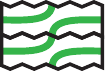}}}}}

\newcommand{\ORGGsquaredfuseFw}{\mathord{\vcenter{\hbox{\includegraphics[scale=1]{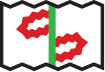}}}}}
 
\newcommand{\ORGGsquaredfusew}{\mathord{\vcenter{\hbox{\includegraphics[scale=1]{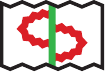}}}}} 
 
\newcommand{\ORGGsquaredw}{\mathord{\vcenter{\hbox{\includegraphics[scale=1]{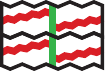}}}}}

\newcommand{\DualityGp}{\mathord{\vcenter{\hbox{\includegraphics[scale=1.0225]{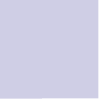}}}}}
\newcommand{\DualityGm}{\mathord{\vcenter{\hbox{\includegraphics[scale=1.0225]{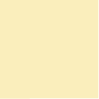}}}}}

\newcommand{\ZG}[1]{{
\mathord{\ooalign{ 
\vphantom{$\Big|^2$}\cr\hidewidth\ensuremath{#1}\hidewidth\cr$\vcenter{
\hbox{$\Z$}
}$\cr\vphantom{$\Big|_q$} }}}}
	
\newcommand{\ZRxRwG}[1]{{
\mathord{\ooalign{ 
\vphantom{$\Big|^2$}\cr\hidewidth\ensuremath{#1}\hidewidth\cr$\vcenter{
\hbox{$\ZRxRw$}
}$\cr\vphantom{$\Big|_q$} }}}}	

\newcommand{\ZxRRwG}[1]{{
\mathord{\ooalign{ 
\vphantom{$\Big|^2$}\cr\hidewidth\ensuremath{#1}\hidewidth\cr$\vcenter{
\hbox{$\ZxRRw$}
}$\cr\vphantom{$\Big|_q$} }}}}	

\newcommand{\ZRRxwG}[1]{{
\mathord{\ooalign{ 
\vphantom{$\Big|^2$}\cr\hidewidth\ensuremath{#1}\hidewidth\cr$\vcenter{
\hbox{$\ZRRxw$}
}$\cr\vphantom{$\Big|_q$} }}}}

\newcommand{\ZRRxpw}{\mathord{\vcenter{\hbox{\includegraphics[scale=1]{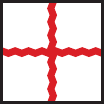}}}}}

\newcommand{\ZRRxpLocalw}{\mathord{\vcenter{\hbox{\includegraphics[scale=1]{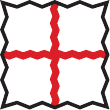}}}}}

\newcommand{\ZRRxLocalw}{\mathord{\vcenter{\hbox{\includegraphics[scale=1]{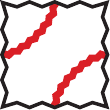}}}}}

\newcommand{\Zspsw}{\mathord{\vcenter{\hbox{\includegraphics[scale=1]{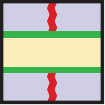}}}}}
\newcommand{\Zss}{\mathord{\vcenter{\hbox{\includegraphics[scale=1]{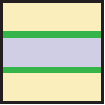}}}}}
\newcommand{\Zssdual}{\mathord{\vcenter{\hbox{\includegraphics[scale=1]{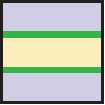}}}}}

\newcommand{\Zsspw}{\mathord{\vcenter{\hbox{\includegraphics[scale=1]{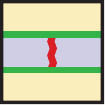}}}}}
\newcommand{\Zdsigmacycle}{\mathord{\vcenter{\hbox{\includegraphics[scale=1]{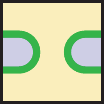}}}}}

\newcommand{\Zdsigmacyclepsiw}{\mathord{\vcenter{\hbox{\includegraphics[scale=1]{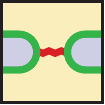}}}}}

\newcommand{\Zdsigma}{\mathord{\vcenter{\hbox{\includegraphics[scale=1]{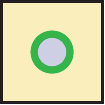}}}}}

\newcommand{\PartitionBasisabc}{\mathord{\vcenter{\hbox{\includegraphics[scale=1]{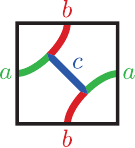}}}}}
\newcommand{\SPartitionBasisabc}{\mathord{\vcenter{\hbox{\includegraphics[scale=1]{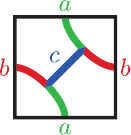}}}}}
\newcommand{\TPartitionBasisabc}{\mathord{\vcenter{\hbox{\includegraphics[scale=1]{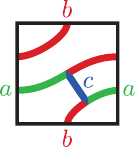}}}}}
\newcommand{\STPartitionBasisabc}{\mathord{\vcenter{\hbox{\includegraphics[scale=1]{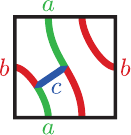}}}}}
\newcommand{\TSTPartitionBasisabc}{\mathord{\vcenter{\hbox{\includegraphics[scale=1]{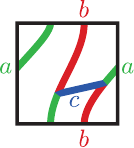}}}}}
\newcommand{\TSTPartitionBasisabcistopic}{\mathord{\vcenter{\hbox{\includegraphics[scale=1]{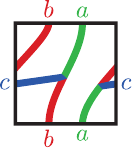}}}}}

\newcommand{\TinvSPartitionBasisabc}{\mathord{\vcenter{\hbox{\includegraphics[scale=1]{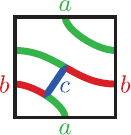}}}}}
\newcommand{\STinvSPartitionBasisabc}{\mathord{\vcenter{\hbox{\includegraphics[scale=1]{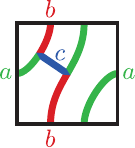}}}}}

\newcommand{\TransfMxIsingDots}{\mathord{\vcenter{\hbox{\includegraphics[scale=1.25]{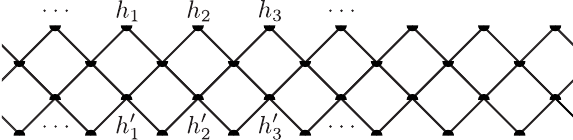}}}}}

\newcommand{\BWbare}{\mathord{\vcenter{\hbox{\includegraphics[scale=1.4]{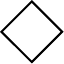}}}}}
\newcommand{\BWDotsbare}{\mathord{\vcenter{\hbox{\includegraphics[scale=1.4]{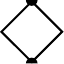}}}}}

\newcommand{\BWGraph}{\mathord{\vcenter{\hbox{\includegraphics[scale=0.9]{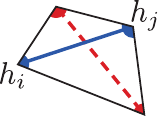}}}}}
\newcommand{\BWGraphDual}{\mathord{\vcenter{\hbox{\includegraphics[scale=0.9]{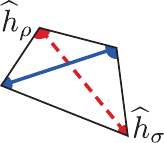}}}}}

\newcommand{\QD}{\mathord{\vcenter{\hbox{\includegraphics[scale=0.9]{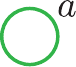}}}}}
\newcommand{\recouplinga}{\mathord{\vcenter{\hbox{\includegraphics[scale=0.9]{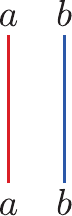}}}}}
\newcommand{\recouplingb}{\mathord{\vcenter{\hbox{\includegraphics[scale=0.9]{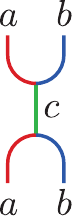}}}}}
\newcommand{\bubblea}{\mathord{\vcenter{\hbox{\includegraphics[scale=0.9]{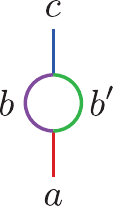}}}}}
\newcommand{\bubbleb}{\mathord{\vcenter{\hbox{\includegraphics[scale=0.9]{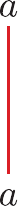}}}}}

\newcommand{\BWIsingH}{\mathord{\vcenter{\hbox{\includegraphics[scale=0.9]{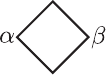}}}}}
\newcommand{\IsingClosedLoop}{\mathord{\vcenter{\hbox{\includegraphics[scale=0.9]{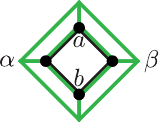}}}}}

\newcommand{\BWbarex}[1]{{\mathord{\ooalign{ \vphantom{$\Big|^2$}\cr\hidewidth\ensuremath{#1}\hidewidth\cr$\vcenter{\hbox{\includegraphics[scale=1.4]{BW.pdf}}}$\cr\vphantom{$\Big|_q$} }}}}		
\newcommand{\BWx}[5][]{\mathord{ \raisebox{0.2ex}{$\scriptstyle{#2}$}\mkern2mu\overset{#3}{\underset{#4}%
	{\ifthenelse{\isempty{#1}}{\BWbare}{\BWbarex{#1}}}%
	}\mkern2mu\raisebox{0.2ex}{$\scriptstyle{#5}$} }}

\newcommand{\BWDotsbarex}[1]{{\mathord{\ooalign{ \vphantom{$\Big|^2$}\cr\hidewidth\ensuremath{#1}\hidewidth\cr$\vcenter{\hbox{\includegraphics[scale=1.4]{BWDots.pdf}}}$\cr\vphantom{$\Big|_q$} }}}}	
\newcommand{\BWDotsx}[5][]{\mathord{ \raisebox{0.2ex}{$\scriptstyle{#2}$}\mkern2mu\overset{#3}{\underset{#4}%
	{\ifthenelse{\isempty{#1}}{\BWDotsbare}{\BWDotsbarex{#1}}}%
	}\mkern2mu\raisebox{0.2ex}{$\scriptstyle{#5}$} }}

\newcommand{\DefectSquareLabel}{\mathord{\vcenter{\hbox{\includegraphics[scale=0.3]{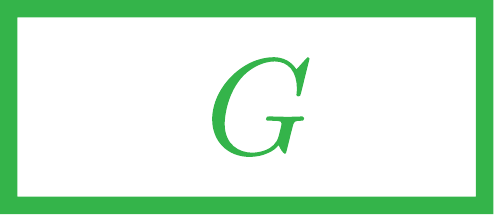}}}}}
\newcommand{\DefectSquare}{\mathord{\vcenter{\hbox{\includegraphics[scale=1.5]{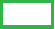}}}}}
\newcommand{\DefectSquareRed}{\mathord{\vcenter{\hbox{\includegraphics[scale=1.5]{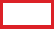}}}}}
\newcommand{\DefectSquarex}[4]{\mathord{\sideset{^{#1}_{#3}}{^{#2}_{#4}}{\mathop{\DefectSquare}}}}
\newcommand{\DefectSquareLabelx}[4]{\mathord{\sideset{^{#1}_{#3}}{^{#2}_{#4}}{\mathop{\DefectSquareLabel}}}}

\newcommand{\DefectCommuteaH}{\mathord{\vcenter{\hbox{\includegraphics[scale=0.8]{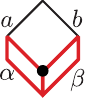}}}}}
\newcommand{\DefectCommutebH}{\mathord{\vcenter{\hbox{\includegraphics[scale=0.8]{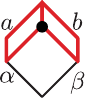}}}}}
\newcommand{\DefectCommuteaV}{\mathord{\vcenter{\hbox{\includegraphics[scale=0.8]{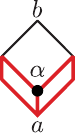}}}}}
\newcommand{\DefectCommutebV}{\mathord{\vcenter{\hbox{\includegraphics[scale=0.8]{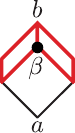}}}}}
\newcommand{\DefectCommutec}{\mathord{\vcenter{\hbox{\includegraphics[scale=0.8]{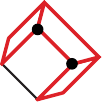}}}}}
\newcommand{\DefectCommuted}{\mathord{\vcenter{\hbox{\includegraphics[scale=0.8]{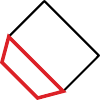}}}}}
\newcommand{\DefectCommuteaG}{\mathord{\vcenter{\hbox{\includegraphics[scale=0.8]{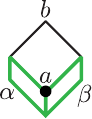}}}}}
\newcommand{\DefectCommutebG}{\mathord{\vcenter{\hbox{\includegraphics[scale=0.8]{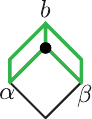}}}}}
\newcommand{\DefectCommuteaGp}{\mathord{\vcenter{\hbox{\includegraphics[scale=0.8]{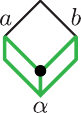}}}}}
\newcommand{\DefectCommutebGp}{\mathord{\vcenter{\hbox{\includegraphics[scale=0.8]{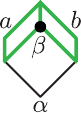}}}}}

\newcommand{\Wpsi}{\mathord{\vcenter{\hbox{\includegraphics[scale=0.8]{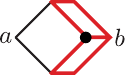}}}}}
\newcommand{\WpsiOp}{\mathord{\vcenter{\hbox{\includegraphics[scale=0.8]{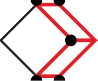}}}}}
\newcommand{\Wsigma}{\mathord{\vcenter{\hbox{\includegraphics[scale=0.8]{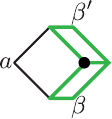}}}}}

\newcommand{\DefectCommuteaprime}{\mathord{\vcenter{\hbox{\includegraphics[scale=1.2]{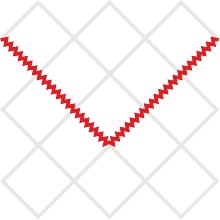}}}}}
\newcommand{\DefectCommutebprime}{\mathord{\vcenter{\hbox{\includegraphics[scale=1.2]{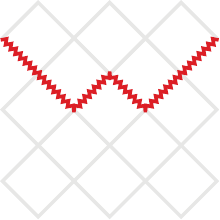}}}}}
\newcommand{\DefectCommutecprime}{\mathord{\vcenter{\hbox{\includegraphics[scale=1.2]{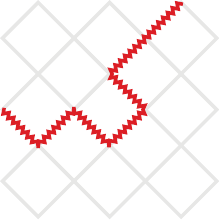}}}}}
\newcommand{\DefectCommutedprime}{\mathord{\vcenter{\hbox{\includegraphics[scale=1.2]{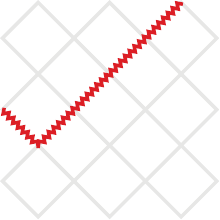}}}}}

\newcommand{\Trivalent}{\mathord{\vcenter{\hbox{\includegraphics[scale=0.3]{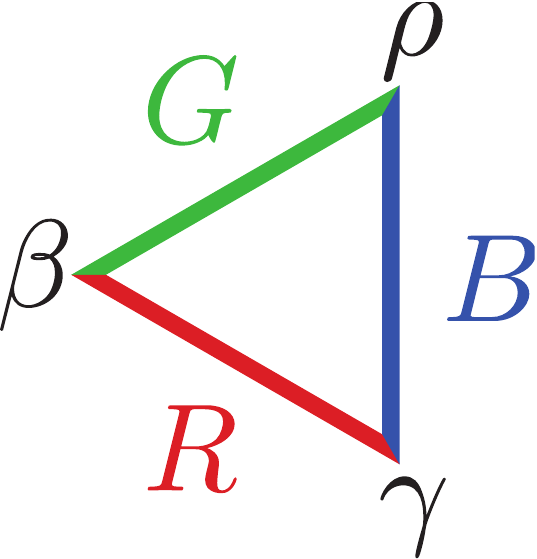}}}}}

\newcommand{\IsingFLeft}{\mathord{\vcenter{\hbox{\includegraphics[scale=1.6]{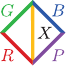}}}}}
\newcommand{\IsingFRight}{\mathord{\vcenter{\hbox{\includegraphics[scale=1.6]{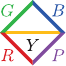}}}}}

\newcommand{\FmoveLSD}{\mathord{\vcenter{\hbox{\includegraphics[scale=1.6]{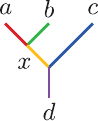}}}}}
\newcommand{\FmoveRSD}{\mathord{\vcenter{\hbox{\includegraphics[scale=1.6]{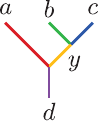}}}}}

\newcommand{\FmoveMacroLeft}{\mathord{\vcenter{\hbox{\includegraphics[scale=1.3]{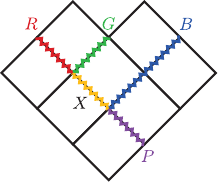}}}}}
\newcommand{\FmoveMacroRight}{\mathord{\vcenter{\hbox{\includegraphics[scale=1.3]{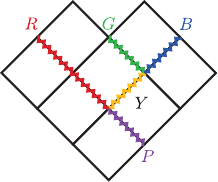}}}}}

\newcommand{\BubbleMacroLeft}{\mathord{\vcenter{\hbox{\includegraphics[scale=1.3]{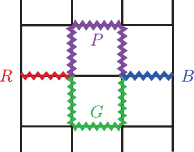}}}}}
\newcommand{\BubbleMacroRight}{\mathord{\vcenter{\hbox{\includegraphics[scale=1.3]{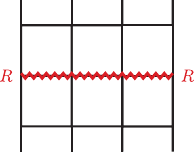}}}}}

\newcommand{\Trivalenta}{\mathord{\vcenter{\hbox{\includegraphics[scale=.7]{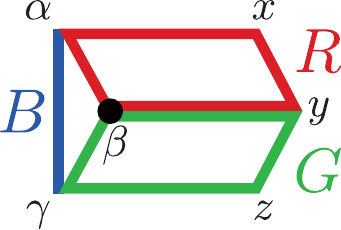}}}}}
\newcommand{\Trivalentb}{\mathord{\vcenter{\hbox{\includegraphics[scale=.7]{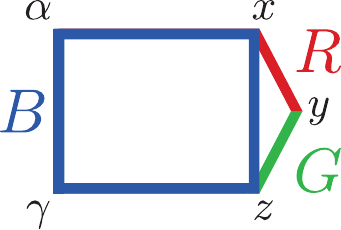}}}}}

\newcommand{\Trivalenth}{\mathord{\vcenter{\hbox{\includegraphics[scale=.25]{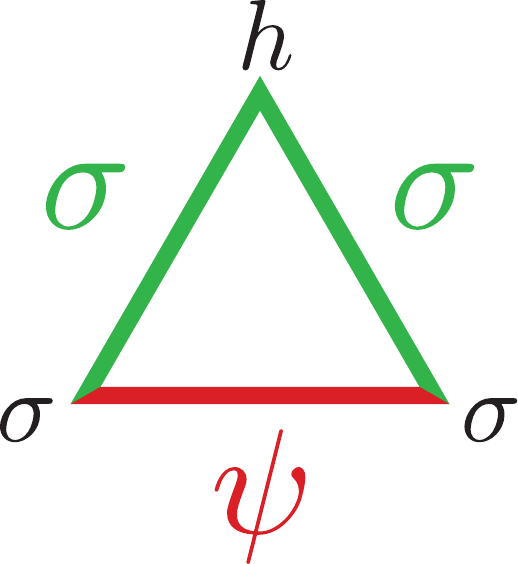}}}}}
\newcommand{\Trivalenthh}{\mathord{\vcenter{\hbox{\includegraphics[scale=.25]{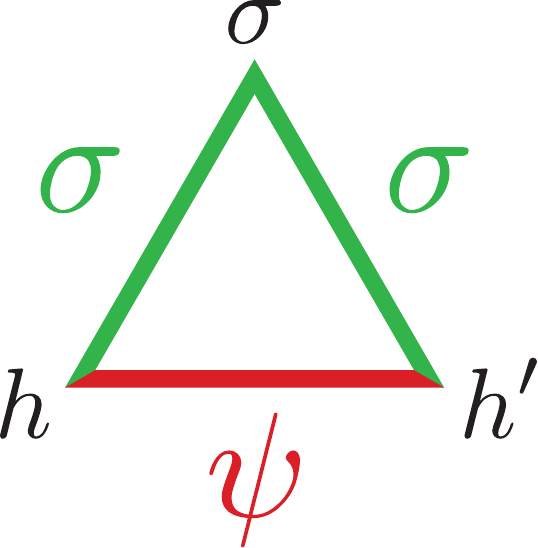}}}}}
\newcommand{\TrivalentIdh}{\mathord{\vcenter{\hbox{\includegraphics[scale=.25]{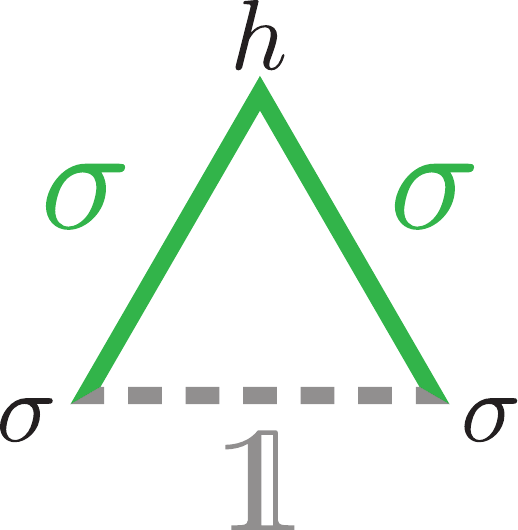}}}}}
\newcommand{\TrivalentIdhh}{\mathord{\vcenter{\hbox{\includegraphics[scale=.25]{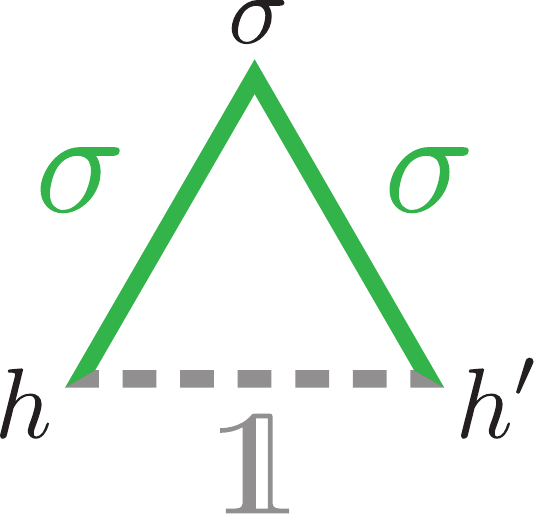}}}}}
\newcommand{\TrivalentIdpsipsih}{\mathord{\vcenter{\hbox{\includegraphics[scale=.25]{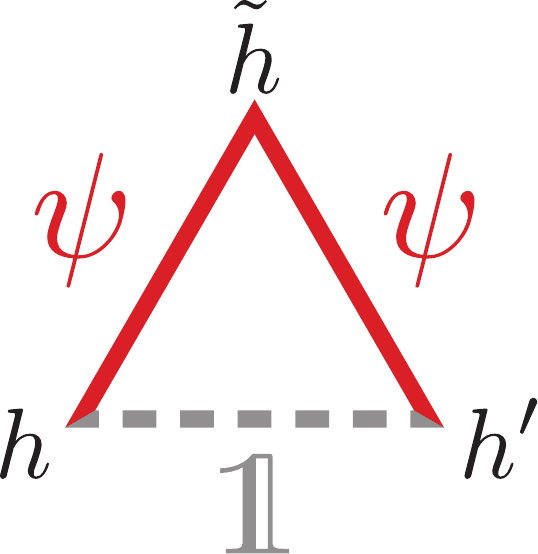}}}}}
\newcommand{\TrivalentIdpsipsihh}{\mathord{\vcenter{\hbox{\includegraphics[scale=.25]{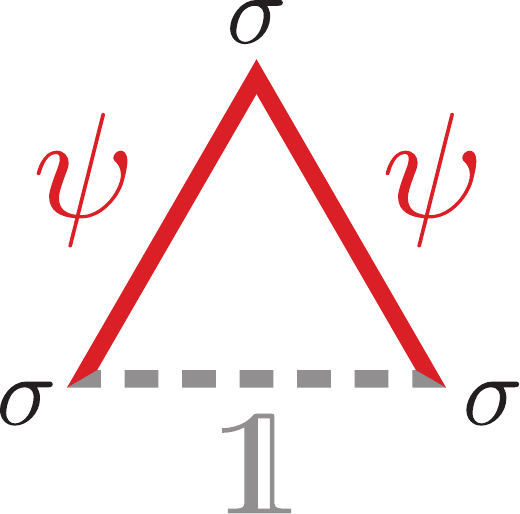}}}}}

\newcommand{\RedCrossLeft}{\mathord{\vcenter{\hbox{\includegraphics[scale=.25]{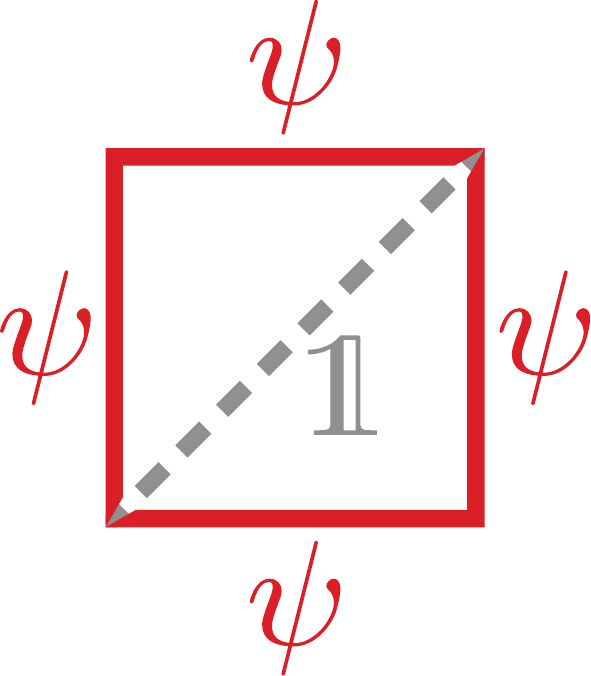}}}}}\newcommand{\RedCrossRight}{\mathord{\vcenter{\hbox{\includegraphics[scale=.25]{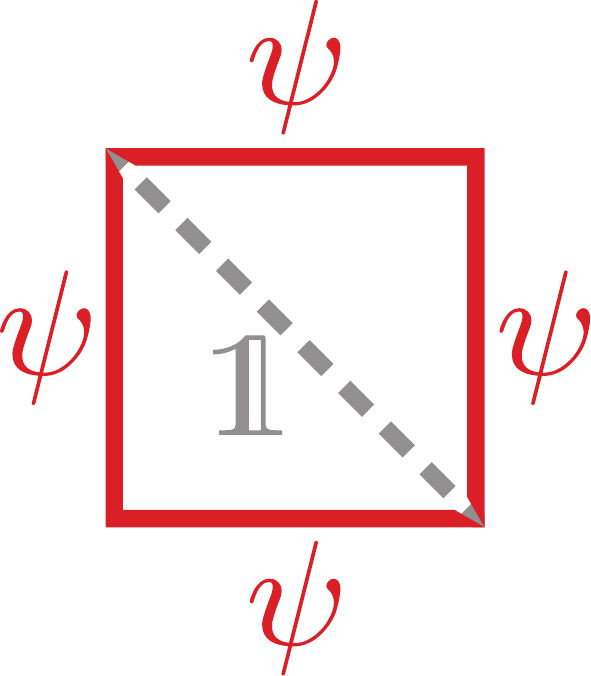}}}}}
\newcommand{\RedCross}{\mathord{\vcenter{\hbox{\includegraphics[scale=.25]{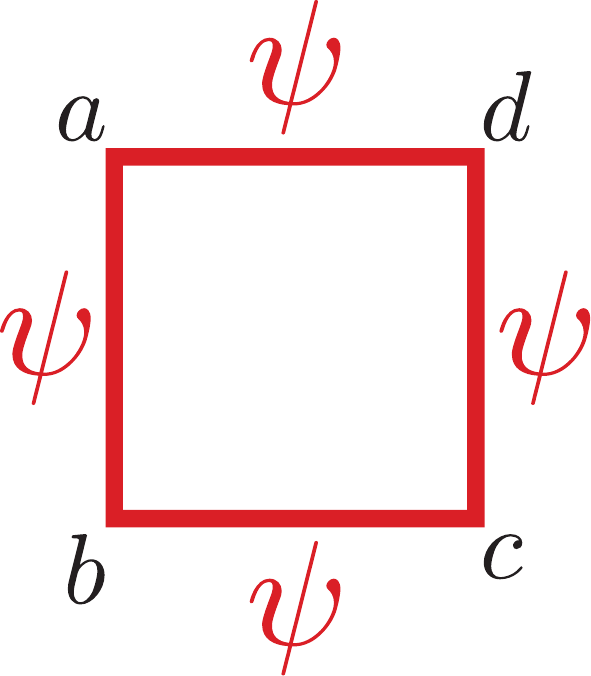}}}}}
\newcommand{\RedCrossNoHeights}{\mathord{\vcenter{\hbox{\includegraphics[scale=.25]{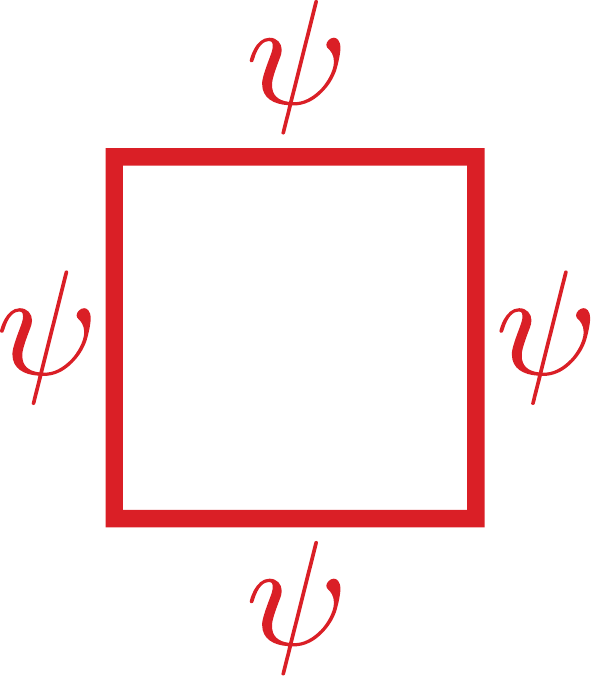}}}}}

\newcommand{\IsingFIdH}{\mathord{\vcenter{\hbox{\includegraphics[scale=.25]{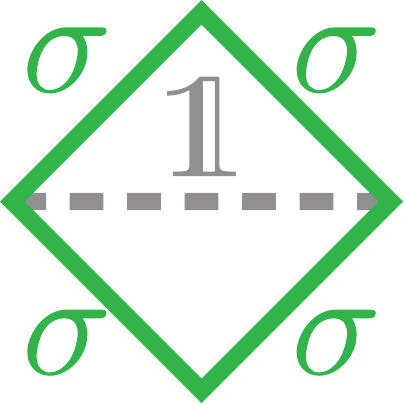}}}}}
\newcommand{\IsingFPsiH}{\mathord{\vcenter{\hbox{\includegraphics[scale=.25]{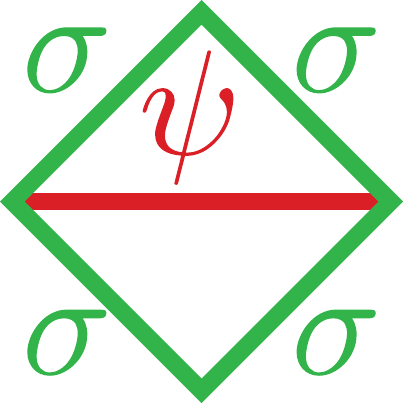}}}}}
\newcommand{\IsingFIdV}{\mathord{\vcenter{\hbox{\includegraphics[scale=.25]{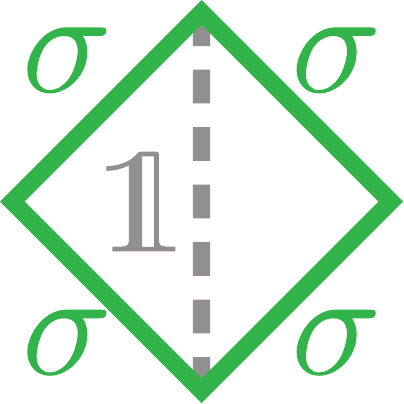}}}}}
\newcommand{\IsingFPsiV}{\mathord{\vcenter{\hbox{\includegraphics[scale=.25]{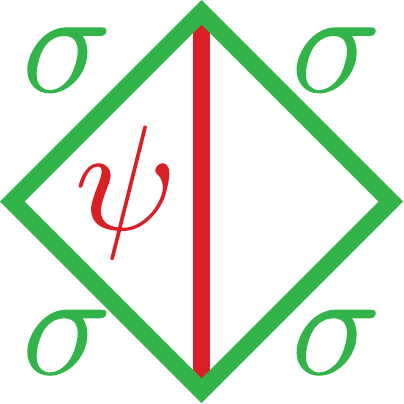}}}}}

\newcommand{\IsingFIdHMacro}{\mathord{\vcenter{\hbox{\includegraphics[scale=1]{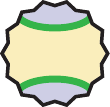}}}}}
\newcommand{\IsingFPsiHMacro}{\mathord{\vcenter{\hbox{\includegraphics[scale=1]{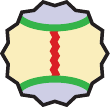}}}}}
\newcommand{\IsingFIdVMacro}{\mathord{\vcenter{\hbox{\includegraphics[scale=1]{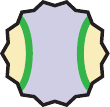}}}}}
\newcommand{\IsingFPsiVMacro}{\mathord{\vcenter{\hbox{\includegraphics[scale=1]{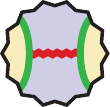}}}}}

\newcommand{\IsingFIdHp}{\mathord{\vcenter{\hbox{\includegraphics[scale=2]{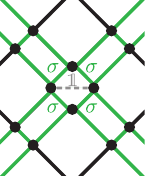}}}}}
\newcommand{\IsingDefectSplitp}{\mathord{\vcenter{\hbox{\includegraphics[scale=2]{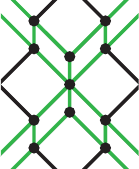}}}}}

\newcommand{\IsingBubbleLeftId}{\mathord{\vcenter{\hbox{\includegraphics[scale=.3]{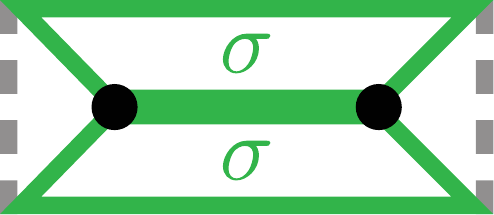}}}}}
\newcommand{\IsingBubbleLeftPsi}{\mathord{\vcenter{\hbox{\includegraphics[scale=.3]{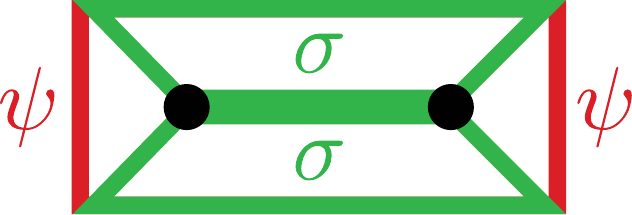}}}}}
\newcommand{\IsingBubbleRightId}{\mathord{\vcenter{\hbox{\includegraphics[scale=.3]{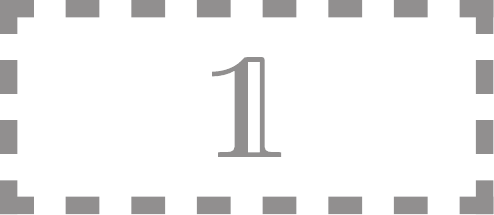}}}}}
\newcommand{\IsingBubbleRightPsi}{\mathord{\vcenter{\hbox{\includegraphics[scale=.3]{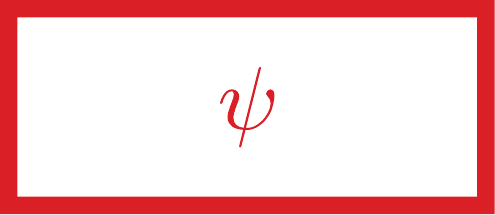}}}}}
\newcommand{\IsingBubbleLeftSigma}{\mathord{\vcenter{\hbox{\includegraphics[scale=.3]{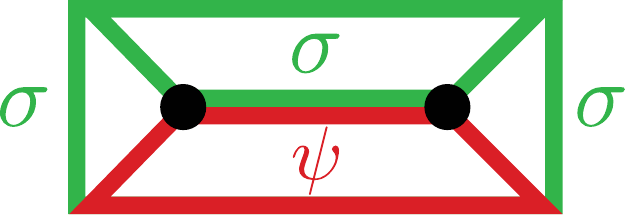}}}}}
\newcommand{\IsingBubbleRightSigma}{\mathord{\vcenter{\hbox{\includegraphics[scale=.3]{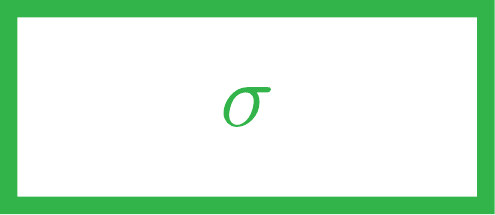}}}}}

\newcommand{\IsingBubbleMicroLeft}{\mathord{\vcenter{\hbox{\includegraphics[scale=.3]{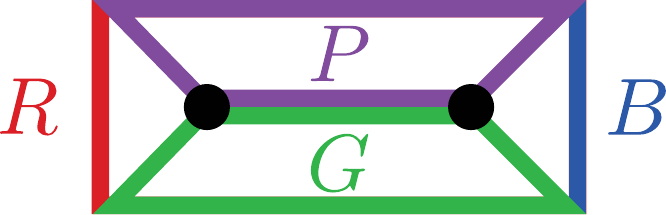}}}}}

\newcommand{\SpinFlip}{\mathord{\vcenter{\hbox{\includegraphics[scale=1]{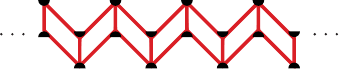}}}}}
\newcommand{\DualtyDefect}{\mathord{\vcenter{\hbox{\includegraphics[scale=1]{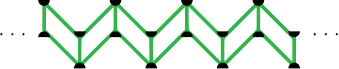}}}}}

\newcommand{\IsingDualityTranslationOperator}{\mathord{\vcenter{\hbox{\includegraphics[scale=.4]{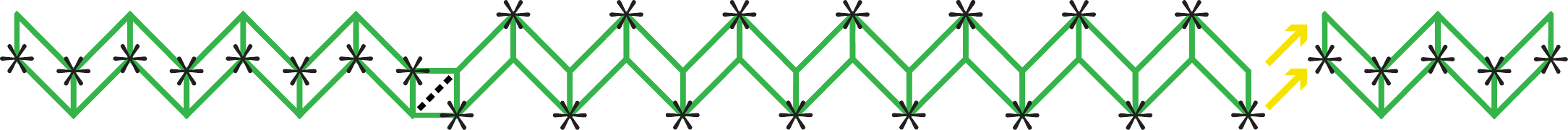}}}}}
\newcommand{\IsingDualityTranslationOperatorDual}{\mathord{\vcenter{\hbox{\includegraphics[scale=.4]{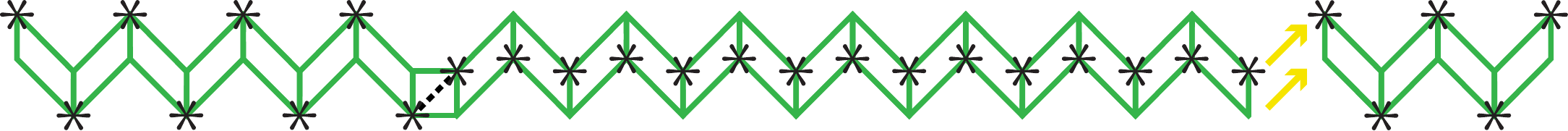}}}}}

\newcommand{\IsingDualityTranslate}{\mathord{\vcenter{\hbox{\includegraphics[scale=.4]{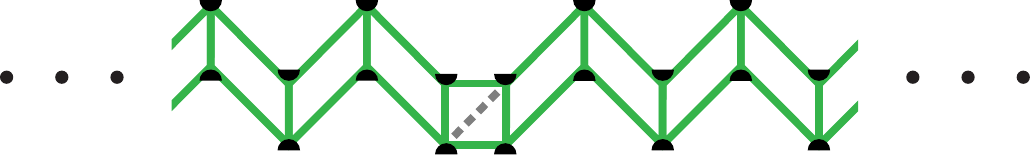}}}}}
\newcommand{\IsingDualityTranslateDagger}{\mathord{\vcenter{\hbox{\includegraphics[scale=.4]{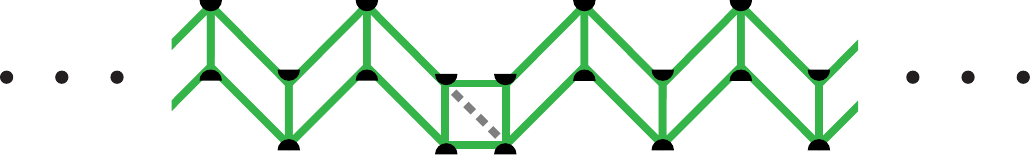}}}}}
\newcommand{\IsingIdentitytranslate}{\mathord{\vcenter{\hbox{\includegraphics[scale=.4]{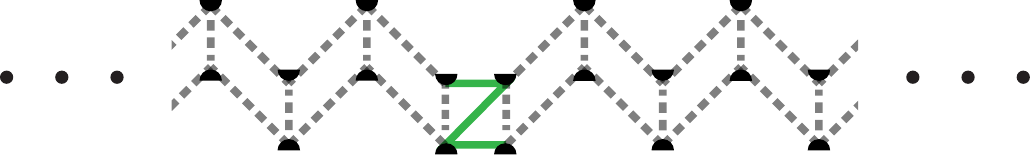}}}}}
\newcommand{\IsingSpinFlipTranslate}{\mathord{\vcenter{\hbox{\includegraphics[scale=.4]{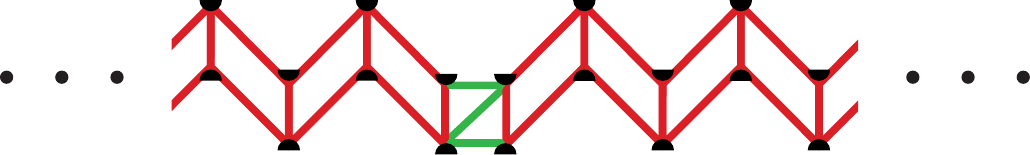}}}}}

\newcommand{\TDualityTwisted}{\mathord{\vcenter{\hbox{\includegraphics[scale=.4]{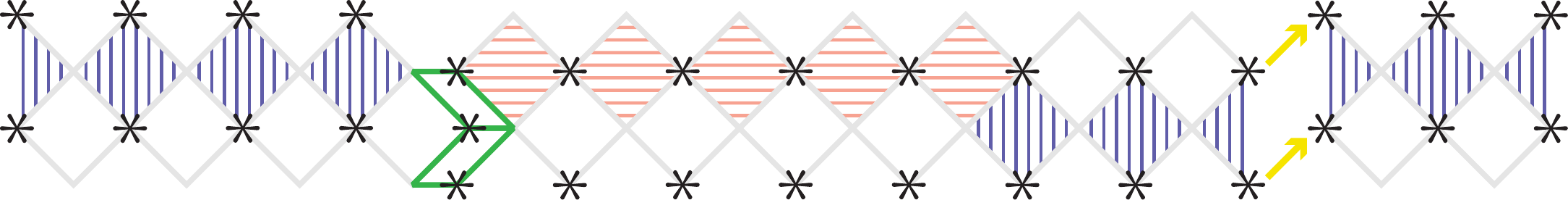}}}}}
\newcommand{\TDualityTwistedDual}{\mathord{\vcenter{\hbox{\includegraphics[scale=.4]{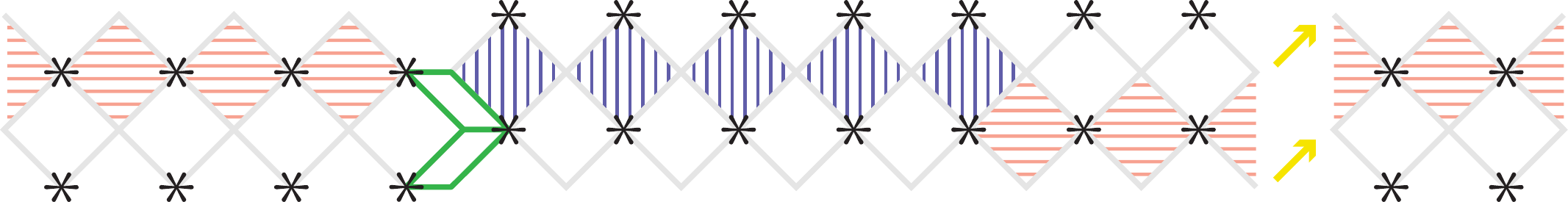}}}}}
\newcommand{\TDualityTwistedDots}{\mathord{\vcenter{\hbox{\includegraphics[scale=.4]{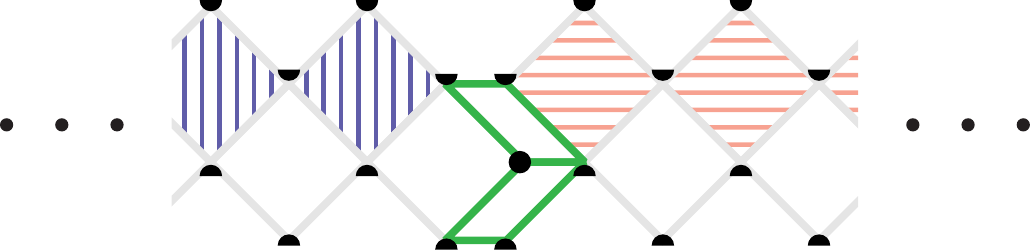}}}}}

\newcommand{\TrivalentConsistencya}{\mathord{\vcenter{\hbox{\includegraphics[scale=1.7]{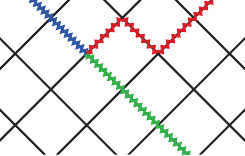}}}}}
\newcommand{\TrivalentConsistencyb}{\mathord{\vcenter{\hbox{\includegraphics[scale=1.7]{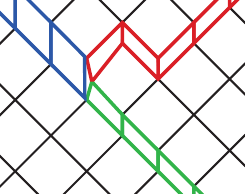}}}}}
\newcommand{\TrivalentConsistencyc}{\mathord{\vcenter{\hbox{\includegraphics[scale=1.7]{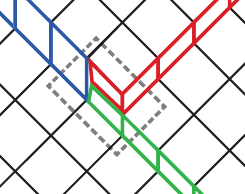}}}}}
\newcommand{\TrivalentConsistencyd}{\mathord{\vcenter{\hbox{\includegraphics[scale=1.7]{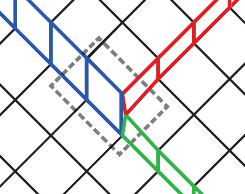}}}}}
\newcommand{\TrivalentConsistencye}{\mathord{\vcenter{\hbox{\includegraphics[scale=1.7]{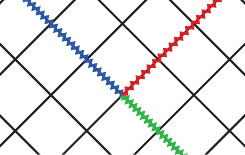}}}}}

\newcommand{\TrivalentJunctionInserta}{\mathord{\vcenter{\hbox{\includegraphics[scale=2.1]{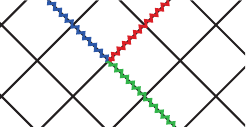}}}}}
\newcommand{\TrivalentJunctionInsertb}{\mathord{\vcenter{\hbox{\includegraphics[scale=2.1]{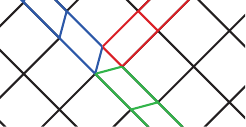}}}}}

\newcommand{\SpinFlipDefectInserta}{\mathord{\vcenter{\hbox{\includegraphics[scale=.85]{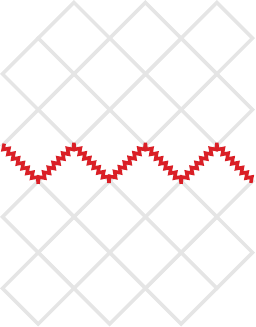}}}}}
\newcommand{\SpinFlipDefectInsertb}{\mathord{\vcenter{\hbox{\includegraphics[scale=.85]{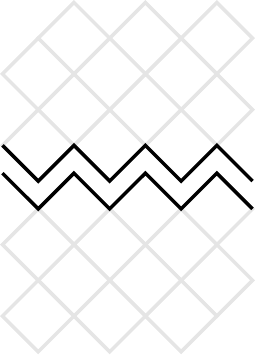}}}}}
\newcommand{\SpinFlipDefectInsertc}{\mathord{\vcenter{\hbox{\includegraphics[scale=.85]{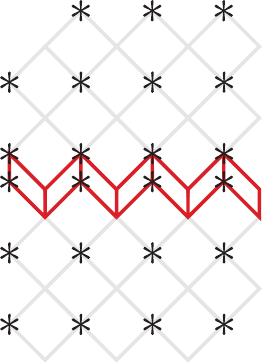}}}}}
\newcommand{\SpinFlipDefectInsertd}{\mathord{\vcenter{\hbox{\includegraphics[scale=.85]{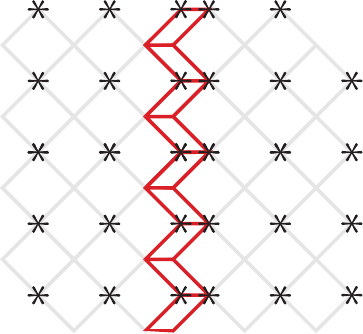}}}}}

 \newcommand{\SigmaDefecta}{\mathord{\vcenter{\hbox{\includegraphics[scale=.85]{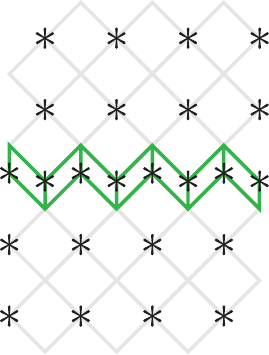}}}}}
  \newcommand{\SigmaDefectb}{\mathord{\vcenter{\hbox{\includegraphics[scale=.85]{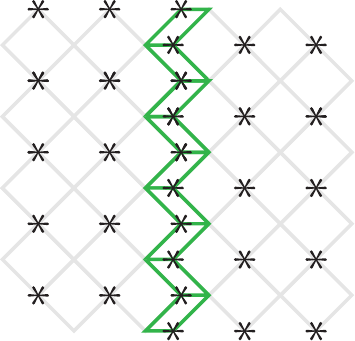}}}}}
   \newcommand{\SigmaDefectc}{\mathord{\vcenter{\hbox{\includegraphics[scale=.85]{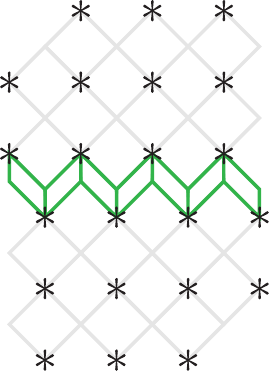}}}}}
    \newcommand{\SigmaDefectd}{\mathord{\vcenter{\hbox{\includegraphics[scale=.85]{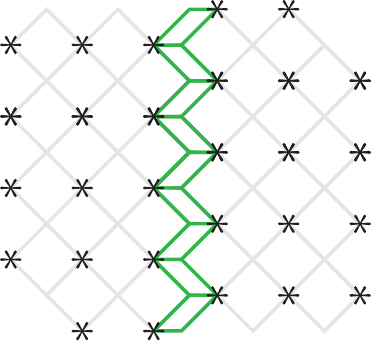}}}}}

\newcommand{\Grapha}{\mathord{\vcenter{\hbox{\includegraphics[scale=.85]{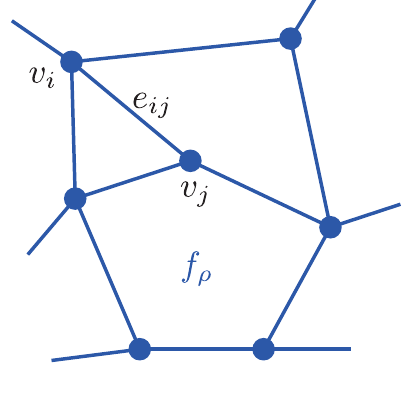}}}}}
\newcommand{\Graphb}{\mathord{\vcenter{\hbox{\includegraphics[scale=.85]{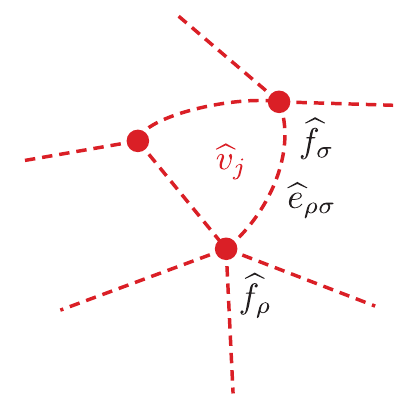}}}}}
\newcommand{\Graphc}{\mathord{\vcenter{\hbox{\includegraphics[scale=.85]{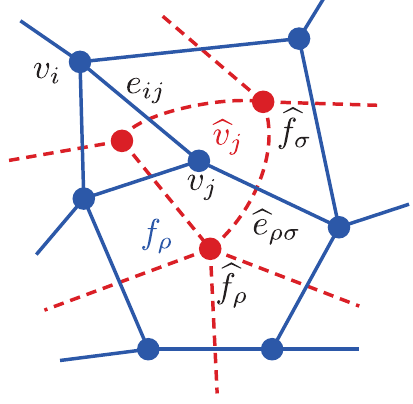}}}}}
\newcommand{\Graphd}{\mathord{\vcenter{\hbox{\includegraphics[scale=.85]{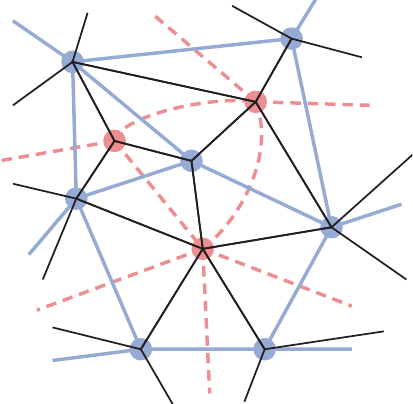}}}}}

\newcommand{\GraphCuta}{\mathord{\vcenter{\hbox{\includegraphics[scale=.85]{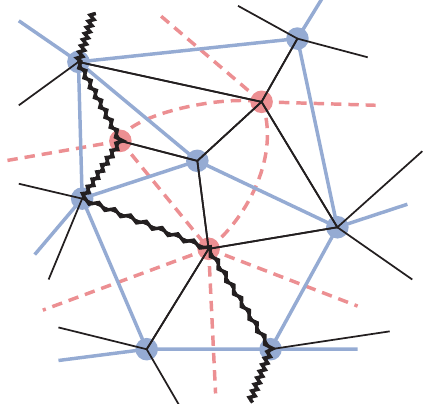}}}}}
\newcommand{\GraphCutb}{\mathord{\vcenter{\hbox{\includegraphics[scale=.85]{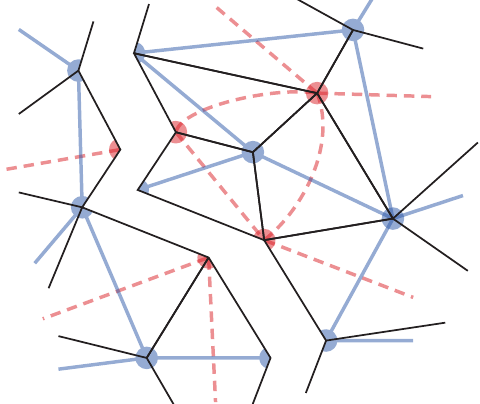}}}}} 
 
\usepackage{color}
\definecolor{purple}{rgb}{0.5,0,0.5}
\definecolor{dkgreen}{rgb}{0,0.5,0}
\definecolor{orange}{rgb}{1,0.5,0}

\newcommand{\comment}[1]{{}}

\numberwithin{equation}{section}		
\title{Topological Defects on the Lattice I: The Ising model }

\author{David Aasen$^1$, Roger S. K. Mong$^{2}$, Paul Fendley$^3$}
\affil{
$^1$ Department of Physics and Institute for Quantum Information and Matter, California Institute of Technology, Pasadena, CA 91125, United States}
\affil{
$^2$ Department of Physics and Astronomy, University of Pittsburgh, PA 15260, United States}
\affil{
$^3$ All Souls College and Rudolf Peierls Centre for Theoretical Physics,  University of Oxford, 1 Keble Road, Oxford, OX1 3NP, United Kingdom}

\begin{document}

\maketitle
\begin{abstract}
In this paper and its sequel, we construct topologically invariant defects in two-dimensional classical lattice models and quantum spin chains. We show how defect lines commute with the transfer matrix/Hamiltonian when they obey the {\em defect commutation relations}, cousins of the Yang-Baxter equation. These relations and their solutions can be extended to allow defect lines to branch and fuse, again with properties depending only on topology. In this part~I, we focus on the simplest example, the Ising model. We define lattice spin-flip and duality defects and their branching, and prove they are topological. One useful consequence is a simple implementation of Kramers-Wannier duality on the torus and higher genus surfaces by using the fusion of duality defects. We use these topological defects to do simple calculations that yield exact properties of the conformal field theory describing the continuum limit. For example, the shift in momentum quantization with duality-twisted boundary conditions yields the conformal spin $1/16$ of the chiral spin field. Even more strikingly, we derive the modular transformation matrices explicitly and exactly. 
\end{abstract}

%
%
%
%
%

\tableofcontents
\section{Introduction}

\label{sec:Intro}

The Yang-Baxter equation plays a fundamental role in the study of integrable models and their mathematics \cite{Perk2006}. Lattice Boltzmann weights satisfying the Yang-Baxter equation can be used to construct a family of commuting transfer matrices, yielding the conservation laws necessary for integrabilty \cite{Baxter1982}.
This has resulted not only in countless physical insights, but deep mathematics as well. For example, taking a limit of these Boltzmann weights yields topological invariants for knots and links generalizing the Jones polynomial \cite{Wadati1989}. 

In this paper and its sequel we explain how cousins of the Yang-Baxter equation allow these studies to be extended into further realms. We show how solving the {\em defect commutation relations} gives
topological defects in many lattice models. The partition function is independent of the path a topological defect follows except for topological properties, e.g.\ if and how the path wraps around a system with periodic boundary conditions. 
In this part~I, we study solutions of these defect commutation relations for the Ising model. In part~II, we explain how solutions follow from a general mathematical structure known as a fusion category, familiar from studies of anyons, rational conformal field theory, and topological quantum field theory. 

Magic follows. Much can be learned even in the case of a single defect line, where our construction gives a simple and systematic way to generalize defects in lattice models previously constructed by exploiting integrability \cite{Pearce2001,Pearce2003}. Solving the defect commutation relations allows a uniform way of understanding seemingly different observations, for example showing that Kramers-Wannier duality \cite{Kramers1941} can be extended to many other models in a generalisation known as ``topological symmetry'' \cite{Feiguin2007}. 
It provides a systematic way to define twisted boundary conditions, where a modified translation symmetry holds despite the presence of a defect seemingly breaking it. Understanding such translation invariance allows us 
to compute the spin of operators in conformal field theory  {\em exactly} in this lattice model, without using the full apparatus of integrability. For example, we find here in part~I the conformal spin 1/16 of the chiral spin field in the Ising model simply by computing the eigenvalues of the operator for translating around one cycle (i.e., a Dehn twist). 

We show how to include multiple topological defects by giving a precise method for understanding how they fuse together. Kramers-Wannier duality \cite{Kramers1941} becomes easily implemented with the transfer matrix formulation. This clarifies greatly a major subtlety, that duality is {\em not} a symmetry in the traditional sense \cite{Gaiotto}; while a duality transformation can be implemented by an operator commuting with the transfer matrix/quantum Hamiltonian, this operator is not unitary or even invertible. Defining this operator via an insertion of a ``duality defect'' line \cite{Schutz1993,Affleck1997,Grimm2003,Frohlich2004} makes it easy to derive what happens when two duality transformations are performed.

We show how to allow topological defect lines to branch in a topologically invariant fashion. Namely, we define a trivalent vertex where defect lines meet such that the partition function is independent of the location of the vertex. This results in a precise lattice definition of operations familiar in topological field theory, for example the ``$F$-move'' \cite{MSReview89} that in our context relates the partition functions of systems with different defect branching. This in turn allows 
some of the profound consequences of modular invariance, famed from conformal field theory \cite{Cardy1986}, to be derived and utilised directly on the lattice. It also allows a straightforward method for implementing duality when space is a higher-genus surface.

One of the messages of our papers is that topological defects are an absolutely fundamental characteristic of many two-dimensional lattice models, especially but not exclusively at their critical points.  At a crude level, it is because defects are effectively external probes, and so a system's response to their insertion gives useful information.  
However, the relation is much more profound, both physically and mathematically. Topological defects are intimately related to the symmetry structure of the theory, and so deeply ingrained. Indeed, this has long been known in conformal field theory: Cardy's seminal work on boundaries (in particular \cite{Cardy1989}) led to among other things an understanding of how defects play a fundamental role; see e.g.\  \cite{Petkova2001,Petkova2001b,Runkel2002,Runkel2004,Runkel2004b,Runkel2005,Runkel2006,Frohlich:2006}. We find similar structure in many lattice models.  For example, we show how ratios of the non-integer `ground-state degeneracy' \cite{AffleckLudwig} are given easily and exactly by fusing a topological defect with a boundary; they are simply the eigenvalues of the defect creation operators.

The defect commutation relations are typically a {\em milder} constraint than the Yang-Baxter equation: we will show how topological defects occur even when the Boltzmann weights are staggered or spatially varying. In the Ising model, this makes the model non-critical. In the more general models discussed in part~II, staggering the Boltzmann weights not only takes the model away from criticality, but breaks the integrability.

In this part~I, we downplay the general structure in favour of describing a particular model in depth, the Ising model. This we hope provides a very concrete realisation of the more general approach of part~II.  
Even though the two-dimensional Ising model has been studied for the best part of a century \cite{mccoywubook}, we believe our approach not only illuminates greatly some known results such as duality, but adds some new ones to the canon. For example, we compute the exact modular transformation matrices directly on the lattice. 

In section \ref{sec:Ising}, we review the basics of the two-dimensional Ising model. In section \ref{sec:defcomm}, we introduce topological defects and the defect commutation relations. We describe in depth two types, the spin-flip defect and the duality defect. In section \ref{sec:trivalent}, we show how defects can branch and join in a topologically invariant fashion. Even more remarkably, in the presence of multiple defect junctions,  we derive how to reshuffle the junctions, giving linear identities for partition functions. We  show how useful these identities are in section \ref{sec:torus}, by giving a simple implementation of duality on the torus, as well as deriving the aforementioned $1/16$. In section \ref{sec:partition}, we go further and derive the full set of modular transformation matrices for the Ising conformal field theory from purely lattice considerations. In section \ref{sec:highergenus} we find similar relations for partition functions on the disc and on higher genus surfaces. This in particular allows us to derive the ratios of ground-state degeneracies, and give a straightforward procedure for defining duality on any orientable surface.

\section{The Ising model}
\label{sec:Ising}

\begin{figure}[htb]
	\begin{center}
	 \includegraphics[width=2.75in]{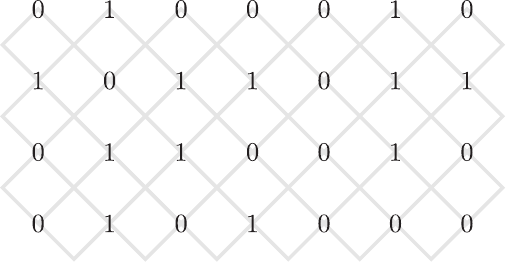}
	\end{center}
	\caption{%
		Ising model with a sample configuration on the square lattice.
		The height variables $h = 0,1$, corresponding to spins $\sigma = +1,-1$ respectively, live on what we call the ``original'' lattice. The original lattice and its dual also form a square lattice drawn in the figure.
	}
	\label{fig:IsingLattice}
\end{figure}

The degrees of freedom of the classical Ising model are spins $\sigma_j=\pm 1$ at each site $j$ of some graph, which for this section we take to be the square lattice.
To allow for general types of defects  we illustrate the configurations in a somewhat unconventional fashion in Fig.~\ref{fig:IsingLattice}. 
We draw both the square lattice and its dual, leaving the dual empty; we will see that duality maps the original model to one on the dual lattice. We also label the degrees of freedom by heights $h_j=0,1$, related to the customary variables by $\sigma_j = (-1)^{h_j}$. These conventions also have the advantage of making the generalisation to height models in part~II more transparent.

The Ising energy and partition function are given by
\begin{align}
	-\beta E[\{\sigma\}] & \;\; =
		\newcommand{\ColTwo}[2]{\renewcommand{\arraystretch}{0.4}
		\begin{tabular}{@{} c @{}}{#1}\\{#2}\end{tabular}}
		\mkern-5mu \sum_{x,y \,\, \in \, \text{\ColTwo{active}{sites}}} \mkern-5mu
		\big( J_x \sigma_{x,y}\sigma_{x+1,y} + J_y \sigma_{x,y}\sigma_{x,y+1} \big) ,
&	{Z} &= \sum_{ \{\sigma \} } e^{-\beta E[\{ \sigma \}]} .
\end{align}
where ($x,y$) labels the points on the original square lattice.
The critical and self-dual line for the square lattice is given by couplings $J_x$ and $J_y$ obeying
\begin{align}
	(\sinh 2J_x) (\sinh 2J_y) = 1\ .
\label{Isingcritline}
\end{align}
A convenient reparametrization of this line is given by introducing the ``anisotropy parameters'' $u_H$ and $u_V$ such that  \cite{Baxter1982}
\begin{align}
	e^{2J_y} = \cot u_V, \qquad\quad e^{2J_x} = \cot (\tfrac{\pi}{4}-u_H)\ .
\end{align}
The critical line (\ref{Isingcritline}) then corresponds to $u_H=u_V$. The couplings are isotropic when $u_H=\pi/4 - u_V$, so that $u_H=u_V=\frac{\pi}{8}$ is the critical and isotropic point.
Each link of the original lattice corresponds to a plaquette on the lattice comprised of the original and its dual, with the four sites comprised of two adjacent ones on the original lattice along with two on the dual.
Letting $a$ and $b$ be the values of the spins on the former two, the vertical and horizontal Boltzmann weights for each plaquette are 
\begin{subequations}
\begin{align}
	&\BWx{}{a}{b}{}  =\;
	\begin{cases}		
		\cos{u_V}	&	a=b , \\
		\sin{u_V}	&	a\ne b, 
		\end{cases}
		\label{eq:Ising_WV}
\\ &\BWx{a}{}{}{b}  =\;
	\begin{cases}		
		 \cos\big(\tfrac{\pi}{4}-u_H\big)	&	a=b , \\
		 \sin\big(\tfrac{\pi}{4}-u_H\big)	&	a\ne b ,
		\end{cases}
	\label{eq:Ising_WH}
\end{align}
\label{eq:Ising_W}\end{subequations}
In this paper we focus mainly on the square lattice (plus defects), but all of our results can be generalized greatly, as we explain in section \ref{sec:highergenus}. In general, one simply associates a parameter $u_p$ for each plaquette made from the combination of the lattice and its dual.

In the absence of defects, the partition function can be defined directly in terms of
(\ref{eq:Ising_W}). It turns out to be much more convenient once defects are introduced to complicate this definition by an additional weight per site of the lattice. This additional factor arises naturally in the closely related theories of anyons/topological field theory, where it is known as the ``quantum dimension''. Here it is simply
\begin{align}
	d_v=
	\begin{cases}
	1&\qquad\text{if there is a spin on site $v$},\\
	\sqrt{2}&\qquad\text{if site $v$ is empty}.
	\end{cases}
\label{vertexweights}
\end{align}
We thus define the Boltzmann weights as
\begin{align}
	e^{-\beta E[\{ \sigma \}]} = \prod_{\text{plaquettes $p$}} \BWx{\alpha_p}{\beta_p}{\delta_p}{\gamma_p} \times \prod_{\text{sites $v$}} d_v\;.
	\label{IsingConfigurationWeight}
\end{align}
Here we explicitly label the spins on the lattice and the dual, but each plaquette has only two active degrees of freedom. It is apparent from this definition that in the absence of defects, the extra site weight results in an unimportant overall constant.

 

The partition function is conveniently written in terms of the transfer matrix $T$. The operator $T$ acts on the vector space ${\cal H}$, whose orthonormal basis states are the spin/height configurations along a row of $L$ sites in the original lattice. Each basis state is labelled by $\ket{h_1 h_2 \cdots h_L}$, where each $h_j \in \{0,1\}$. Acting with $T$ represents one step of evolution in the $y$ direction (Euclidean ``time''), taking one row to the one above it; drawing both the lattice and the dual lattice makes the row zig-zag. The on-site weights are essentially gauge factors, and so there exist many ways of shuffling them around so that one reproduces the same partition function. We have found the least offensive way to include them is to split the weight per site evenly across the vertices in the vertical direction. To illustrate this splitting in the diagrams we put a half-dot at each of the vertices which serves to multiply the diagram by a $\sqrt{d_v}$. Explicitly, the Boltzmann weights with and without the half-dots are related by,
\begin{align}
\BWDotsx{\alpha}{\beta}{\delta}{\gamma} = \BWx{\alpha}{\beta}{\delta}{\gamma} \times\sqrt{d_{\beta} d_{\delta}}.
\label{halfdots}
\end{align}
With this in mind, the matrix elements of $T$ are illustrated by 
\begin{align}
	\bra{\{h\}}T \ket{\{h'\}} = \TransfMxIsingDots .
\label{Tdef}	
\end{align} 
To write $T$ explicitly, we define
\begin{align}
	\sigma^c_j = \mathds{1}\otimes\mathds{1}\otimes\dots\otimes\mathds{1}\otimes \sigma^c\otimes \mathds{1} \otimes \cdots~,
\end{align}
where the Pauli matrix $\sigma^c$ on the right-hand side acts on the $j$\textsuperscript{th} spin. We then define the operators $W^H_j$ and $W^V_j$ with matrix elements
\begin{equation}
\begin{aligned}
\label{IsingWVH}	\bra{\cdots h_j \cdots }W^V_j  \ket{\cdots h_j' \cdots} &=  \BWDotsx{}{h_j}{h_j'}{} 
		 = \big[ \cos u_V +  \sigma^x_j \sin u_V \big]_{h_j h_j'} ~, \\
	\bra{\cdots h_j h_{j+1}\cdots}W^H_{j+1/2} \ket{\cdots h_{j}' h_{j+1}' \cdots}&= \BWDotsx{h_{j}}{}{}{h_{j+1}}=  \big[ \cos u_H
			+  \sigma^z_j \sigma^z_{j+1}\sin u_H \big]_{h_j h_{j+1}; h_{j}' h_{j+1}'} ,
\end{aligned}
\end{equation}
The transfer matrix is given by 
$T^H T^V$, where
\begin{equation}
	T^H=\prod_{j=1}^L W^H_{j+1/2}\ ,\qquad T^V=\prod_{j=1}^LW^V_j\ .
	\label{IsingTHTV}
\end{equation}
For periodic boundary conditions in the horizontal direction, we let $\sigma^z_{L+1}\equiv \sigma^z_1$.
Putting periodic boundary conditions around the vertical direction as well makes two-dimensional space a torus.  The toroidal partition function for $r$ rows and $L$ columns is then
\begin{equation}
	{Z}_{1,1}(L,r) = \Tr\big[ (T^H T^V)^r\big] ~;
\end{equation}
the reason for the subscripts will be apparent shortly. 


Duality maps the Ising model on any planar lattice to one on the dual lattice. The partition functions of the two are the same when the coupling $J$ in the original model on a given link is related to the coupling $\widehat{J}$ on the corresponding link in the dual by \cite{Kramers1941}
\begin{align}
\sinh(J)\sinh(\widehat{J})=1\ .
\label{Isingduality}
\end{align}
Since the dual of a square lattice is also a square lattice, it is natural to expect that the critical point in this case is the self-dual point, $J =\widehat{J}$, and indeed this is so; note (\ref{Isingcritline}). 
The transfer matrix for the dual model can be defined using the {\em same} picture (\ref{Tdef}), but here the heights/spins $\hat{h}_{j-1/2}$ live on the dual lattice. We denote this vector space by $\widehat{\cal H}$, and its orthonormal basis elements are labelled  by $\ket{\hat{h}_{\frac{1}{2}}\hat{h}_{\frac{3}{2}}\cdots \hat{h}_{L-\frac{1}{2}}}$. The matrix elements are then given by replacing $h_j$ with $\hat{h}_{j-1/2}$ in (\ref{IsingWVH}).

In the following, we will describe how dualities of this sort can be derived in a very simple way by understanding the topological defects associated with the global symmetries of the model.
However, studying the Ising model on graphs with non-trivial topology and boundary segments will require further restrictions. Each boundary condition on the original graph will result in a change or possibly a sum over boundary conditions on the dual lattice. In the following, we provide a geometric way of performing the duality that provides a direct and simple way of mapping the boundary conditions between the model and its dual.


\section{The defect commutation relations and their solutions}
\label{sec:defcomm}

We now introduce defect lines, and show that when they and the Boltzmann weights obey certain relations, the partition function is independent of the deformations of the defect's path.  We focus on the two main types of topological defect lines in the Ising model, the spin-flip defect and the duality defect.

\subsection{The spin-flip defect}
\label{subsec:spinflip}

The Ising model has a ${\mathbb Z}_2$ symmetry given by flipping the spin at every site. In the transfer matrix formulation, this is generated by the operator $\D_{\psi} = \prod_n {\sigma_n^x}$, which obviously commutes with $T^H$ and $T^V$. The reason for the mysterious subscript will be explained later.
A straight spin-flip defect stretching in the horizontal direction is then created by inserting $\D_\psi$ between two of the rows of the transfer matrix. The toroidal partition function in the presence of such a defect line is then
\begin{align}
	{Z}_{1,\psi}(L,r) = \Tr \left[ \cdots T^HT^V \mathcal{D}_{\psi} T^HT^V \cdots \right].
\end{align}
We sketch one way to do this diagrammatically in Fig.~\ref{fig:IsingDefect-psi}(a), with the defect line corresponding to the sequence of parallelograms. 
%
The defect spin flip Boltzmann weights are given by:
\begin{align}
	\mathord{\sideset{^a_b}{}{\mathop{\DefectSquareRed}}}
	= \mathord{\sideset{}{^a_b}{\mathop{\DefectSquareRed}}}
	= 2^{-\frac14}\big[ \sigma^x \big]_{a,b}
	= \begin{cases} 0 & a = b, \\ 2^{-\frac14} & a \neq b. \end{cases}
\end{align}
Here we draw the parallelograms as rectangles.
As Boltzmann weights, $\DefectSquareRed$ simply enforces that the spins on either side are opposite. In terms of the half-dot notation introduced in (\ref{halfdots}), the spin-flip defect creation operator is
\begin{align}
\mathcal{D}_\psi = \SpinFlip = \prod_{j} \sigma_j^x\ ,
\end{align}
taking ${\cal H}\to {\cal H}$, and $\widehat{\cal H}\to \widehat{\cal H}$.
The semi-circles have been introduced so that the operator can be inserted into the partition function between rows of the transfer matrix without changing the overall constant in front of the partition function. 
Because $(\D_\psi)^2=1$, inserting two horizontal defects successively is equivalent to inserting none.
\begin{figure}[h] \centerline{%
 \xymatrix @!0 @M=2mm @R=19mm @C=38mm{
 &&(a)&(b) \\
 \SpinFlipDefectInserta \ar[r] &\SpinFlipDefectInsertb \ar[r] & \SpinFlipDefectInsertc & \SpinFlipDefectInsertd
 }}
\caption{A defect line by choosing a closed path along the edges of the lattice, and then inserting a seam of parallelograms. (a) and (b) are horizontal and vertical spin-flip defects respectively; the stars label the locations of the spins. We have omitted the dots (weight per site) for clarity.}	
\label{fig:IsingDefect-psi}
\end{figure}

The spin-flip defect line can be moved around without changing the partition function $\mathcal{Z}_{1,\psi}$. It is therefore a {\em topological defect}.
This follows from the fact that the $\D_{\psi}$ commutes with each individual plaquette operator used to build the transfer matrix:
\begin{align}
	[\D_\psi\,,W^H_{j+1/2}] = [\D_\psi\,,W^V_j] = 0
\end{align}
for all $j$. 
Pictorially, these can be represented as {\em defect commutation relations}
\begin{align}
	\DefectCommuteaH\ =\  \DefectCommutebH\ ,  \qquad\quad  \sum_\alpha \DefectCommuteaV\ =\ \sum_\beta \DefectCommutebV
	\label{Isingdefcomm}
\end{align}
where the two cases correspond to the different possible locations of the spins.
Similarly,
\begin{align}
	\sum_{\rm internal\ labels}\DefectCommutec\ =\ \DefectCommuted\ ,
	\label{Isingdefcomm2}
\end{align}
where here we do not show the (fixed) external labels and the (summed over) internal labels, and so only need write the diagram once.
The partition function remains invariant under these moves when the factor of $d_v$ per site, denoted by the full circle, is included. These relations mean that the defect line can be deformed without modifying the partition function; for example
\begin{equation}
	\DefectCommuteaprime \leftrightarrow \DefectCommutebprime \quad\text{ and }\quad \DefectCommutecprime \leftrightarrow \DefectCommutedprime\ .
\label{symmcomm}
\end{equation}
Thus $Z_{1,\psi}(L,r)$ is independent of any local deformations of the defect line.
This implies that in the critical and continuum limits, the partition function remains conformally invariant.


Inserting a vertical defect as shown in Fig.~\ref{fig:IsingDefect-psi}b can be interpreted as changing the boundary condition on the transfer matrix. Such boundary conditions defined using a topological defect  are called {\em twisted}.  
Twisted boundary conditions are often associated with a discrete symmetry like spin-flip one here \cite{Pearce1989}. The spin-flip defect effectively changes the interaction bridging the defect line from ferromagnetic to antiferromagnetic, modifying the Boltzmann weight and corresponding operator to
\begin{align}
	\Wpsi &\ =\  \begin{cases}		
		 2^{-\frac12}\sin\big(\tfrac{\pi}{4}-u_H\big)	&	a=b  \\
		 2^{-\frac12}\cos\big(\tfrac{\pi}{4}-u_H\big)	&	a\ne b
		\end{cases}\ , \\
		W^H_{L+1/2} &\ =\  \WpsiOp = \cos u_H-  \sigma^z_L \sigma^z_{1}\sin u_H
\end{align}
respectively. Correspondingly the transfer matrix is modified to give the partition function 
\begin{equation}
Z_{\psi,1}(L,r)= \Tr\, (T^VT^H_\psi)^r\ ,
\end{equation}
where
\begin{equation}
	T^H_\psi = (\cos u_H - \sigma_L^z\sigma^z_1\sin u_H) \prod_{j=1}^{L-1}W^H_{j+1/2}  \ .
\end{equation}
The path independence of the defect can be seen directly in the transfer matrix formulation by utilizing a unitary transformation. For example, taking $T^H_\psi \to U T^H_\psi U^{-1}$ with $U= \sigma_L^x$ moves a defect between sites $L$ and $1$ to one between sites $L-1$ and $L$ (or vice versa) without changing the partition function.

Twisted boundary conditions are particularly interesting because the model still obeys a modified version of translation invariance. This follows from the path independence of the defect; it can be placed between any two sites without changing the partition function. To find the modified translation operator, we first define the translation operator $\T_1$, the translation operator for untwisted periodic boundary conditions. Precisely, $ \T_1$ has matrix elements
\begin{align}
\bra{ \{ h_i\}} { \T}_{\mathds 1} \ket{\{h_i'\}}=  \left( \prod_{i=1}^{L} \delta_{h_{i+1},h_i'} \right)\ .
\label{T1def}
\end{align}
It is simple to check that  ${\cal T}_{\mathds{1}}$ indeed obeys ${\cal T}_{\mathds{1}} T^V {\cal T}_{\mathds{1}}^{-1}=T^V$ and ${\cal T}_{\mathds{1}} T^H {\cal T}_{\mathds{1}}^{-1}=T^H$. However, ${\cal T}_{\mathds{1}}$ does not commute with $T^{\psi}$, because translating moves the defect over to between sites $1$ and $2$. However, the defect can be moved back by the unitary transformation $U=\sigma^x_1$. Equivalently, one can first shift the defect to the left, and then translate it back.
Thus the modified translation operator 
\begin{align}
{\T}_\psi = \sigma^x_1 { \T}_{\mathds 1}={ \T}_{\mathds 1}\sigma^x_L
\label{Tpsidef}
\end{align}
does indeed commute with the transfer matrix in the presence of these twisted boundary conditions:
${ \T}_\psi T^V {\T}_\psi^{-1}=T^V$ and ${ \T}_\psi T^H_\psi {\T}_\psi^{-1}=T^H_\psi$. 
Thus in the continuum and critical limits, the model remains conformally invariant even in the presence of a defect.

This simple result has another interesting consequence.
Because $({ \T}_\psi)^2=\sigma^x_1{\T}_{\mathds 1} \sigma^x_1 {\T}_{\mathds 1} = \sigma^x_1\sigma^x_2({\T}_{\mathds 1})^2$, 
\begin{align}
({\T}_{\psi})^L=\prod_{j=1}^L \sigma^x_j = \D_\psi\ .
\label{TpsiL}
\end{align}
Thus translating around a cycle in the presence of a vertical spin-flip defect is equivalent to inserting a horizontal defect! 
As a map between partition functions, it can be pictured schematically by
\begin{align}
\ZxRRw\ \xrightarrow{({\T}_{\psi})^L}\ \ZRRxpw\ \equiv\ \ZRRxw\ .
\label{Tpsi}
\end{align}
In section \ref{sec:trivalent} we define precisely the quadrivalent defect Boltzmann weight at the point where the vertical defect meets the horizontal defect, and  (see equation \ref{redcross}) derive the equality between partition functions on the right-hand-side of (\ref{Tpsi}).
Intuitively, this is a consequence of the fact that any path traversing either cycle of the torus will always cross a spin flip defect. Indeed we also have the equality
\begin{align}
\newcommand{\ZRRxwS}{\mathbin{\rotatebox[origin=c]{90}{$\ZRRxw$}}}
\ZRRxw\ =\ \ZRRxwS\ .
\end{align}
This can also be verified by explicit calculation using the defect commutation relations.

Inserting the operator $\modT_\psi\equiv ({\T}_{\psi})^L$ into the toroidal partition function can be thought of as  cutting the torus into a cylinder, twisting one end of the cylinder around a full cycle, and gluing it back together.
This process in topology is known as a Dehn twist, and in geometry as the modular transformation ${\bf T}$. (Sadly, we now have three objects conventionally labelled by the letter T.) In the absence of vertical defects, doing a Dehn twist in the horizontal direction leaves the toroidal partition function invariant. In the presence of a vertical defect, it is non-trivial.

\subsection{The duality defect}
\label{subsec:dualitydefect}

We now move on to a more intricate example, the duality defect line \cite{Schutz1993,Frohlich:2006}. 
This defect line is associated with the Kramers-Wannier duality of the Ising model described at the end of section \ref{sec:Ising}. We will explain here precisely how to define this defect. In subsequent sections we will show how implementing duality via this topological defect proves a powerful tool in understanding duality, both on and off the critical point. 


\begin{figure}[h] \centerline{%
 \xymatrix @!0 @M=2mm @R=19mm @C=38mm{
 (a1)&(b1)&(a2)&(b2) \\
 \SigmaDefecta &\SigmaDefectc & \SigmaDefectb & \SigmaDefectd
 }}
\caption{Illustrations of a horizontal duality defect line in (a1) and (a2) and a vertical duality defect line in (b1) and (b2), with the spins denoted by stars. The duality defect line splices together the lattice with the dual lattice. The only difference between (a1) and (a2) is the position of the spins, and likewise for (b1) and (b2).}	
\label{fig:IsingDefect-sigma}
\end{figure}
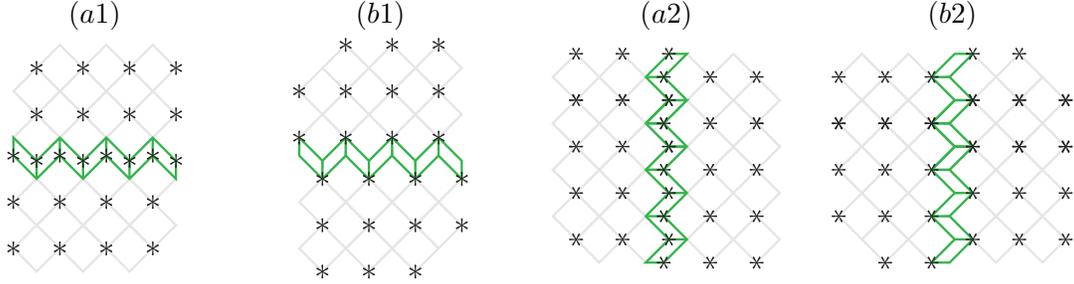

Since duality replaces a model on any lattice with one on the dual lattice, the parallelogram underlying a duality defect line needs to implement this. Thus a duality defect must join a lattice and its dual, as illustrated for the square lattice and the dual square lattice in figure \ref{fig:IsingDefect-sigma}. Note that the spins are located on opposite corners of the parallelogram, in constrast with the spin-flip defect. 

We saw that the ${\mathbb Z}_2$ spin-flip symmetry generator commutes with the individual Boltzmann weights, the relation pictorially presented in (\ref{symmcomm}). This guaranteed that the partition function was independent of continuous deformations of the defect path. 
This will also hold for the duality defect if the analogous relations are satisfied, namely
\begin{equation}
\sum_a \DefectCommuteaG = \DefectCommutebG\ ,  \qquad \DefectCommuteaGp =  \sum_\beta \DefectCommutebGp \ 
\label{dualitydefcomm}
\end{equation}
as well as these relations rotated by 90 degrees, along with (\ref{Isingdefcomm2}).
Commuting a duality defect line through a plaquette with its two spins along the vertical axis turns it into one with spins along the horizontal, and vice versa. Thus dragging a duality defect through the lattice using the commutation relations (\ref{dualitydefcomm}) relates a model on one lattice to one on the dual lattice.

The solution of the duality defect commutation relations is less obvious than in the spin-flip case.
The building block is given by
\begin{align}
	\DefectSquarex{a}{}{}{b} = \DefectSquarex{}{a}{b}{} = 2^{-\frac12} (-1)^{ab}\ ,
	\label{dualitydefectdef}	
\end{align}
where $a$ and $b$ are heights (related to the traditional spins by $\sigma = (-1)^h$). To satisfy (\ref{dualitydefcomm}) one needs to include the internal vertex weight as well (recall: a factor of $1$ for each Ising spin and $\sqrt{2}$ for the blank vertices). Multiplying (\ref{dualitydefectdef}) by a factor
$(-1)^{\mu} (-1)^{\nu(a+b)}$ still solves the defect commutation relations, but these cancel out in the string of defect Boltzmann weights. We thus set $\mu=\nu=0$ in the following.

It is instructive to work out explicitly the weights for a single closed duality defect loop. 
For example, consider the duality defect loop surrounding a vertical Boltzmann weight, 
\begin{equation}
\begin{aligned}
	\IsingClosedLoop &= \sum_{a,b} \underbrace{1\times\sqrt{2}\times1\times\sqrt{2}}_{\text{internal vertex weights}}
		\times \underbrace{ 2^{-2} (-1)^{a\alpha} (-1)^{a\beta} (-1)^{b\beta} (-1)^{b\alpha} }_{\text{duality defect}}
		\times W_V(a,b;u) 
\\	&= \cos{u} + (-1)^{\alpha+\beta}\sin{u} = \sqrt{2}\ \BWIsingH\ .
	\label{eq:IsingClosedLoop}
\end{aligned}
\end{equation}
The partition function with a single closed loop not wrapped around a cycle is therefore simply that without the loop, multiplied by the ``quantum dimension'' $d_\sigma=\sqrt{2}$. 
Because of the defect commutation relations, this statement holds for any topologically trivial loop. 
Thus we can map a model on to its dual by nucleating a loop (and dividing the partition function by $\sqrt{2}$) and then growing it. This is obvious for a planar graph, but by introducing trivalent vertices in section \ref{sec:trivalent}, we will show how to implement duality on the torus and higher genus surfaces as well.

Note that in (\ref{eq:IsingClosedLoop}), the factor $u$ is associated with a particular plaquette; thus inside the closed loop the couplings $J_x$ and $J_y$ are interchanged from those outside.  This can be made clear by utilising the operator formulation  \cite{KadanoffCeva}. The operator $\D_\sigma$ creating a duality defect takes a spin/height configuration on the original lattice to one on the dual, i.e.\ takes ${\cal H}\to \widehat{\cal H}$, or vice versa. The graphical picture applicable to both is
\begin{align}
\D_{\sigma} = \DualtyDefect
\label{Dsigmadef}
\end{align}
so that the matrix elements are
\begin{equation}
\bra{\{\hat{h}\}} \mathcal{D}_{\sigma} \ket{\{h\}} =
\prod_{j=1}^L \frac{1}{\sqrt{2}}(-1)^{\hat{h}_{j-1/2}h_j } (-1)^{h_j\hat{h}_{j+1/2}}\\
=2^{-L/2}(-1)^{\sum_{j=1}^L (h_{j-1}+h_{j})\widehat{h}_{j-1/2}}
\end{equation}
where $h_{0}\equiv h_L$ and $h_{L+1/2}\equiv h_{1/2}$. 
It is easy to check that $  \bra{\{h\}} \mathcal{D}_{\sigma} \ket{\{\hat{h}\}}=
\bra{\{\hat{h}\}} \mathcal{D}_{\sigma} \ket{\{h\}}$.
Following the standard convention that the Pauli matrices acting on sites in the new vector space are denoted by $\mu^r_{j+1/2}$, the operator version of (\ref{dualitydefcomm}) is
\begin{eqnarray}
\mathcal{D}_\sigma \sigma_j^z \sigma_{j+1}^z = \mu_{j+1/2}^x \mathcal{D}_\sigma\ ;\qquad
\mathcal{D}_\sigma \sigma_j^x &=& \mu_{j-1/2}^z \mu_{j+1/2}^z\mathcal{D}_\sigma \ .
\end{eqnarray}
Letting $\widehat{W}^{V,H}_{j+1/2}$ be the operators acting on the new vector space analogous to ${W}^{V,H}_{j}$ from (\ref{IsingWVH}) on the original, these relations then yield
\begin{align}
\mathcal{D}_\sigma W^H_{j+1/2} = \widehat{W}^V_{j+1/2} \mathcal{D}_\sigma\ ; \qquad
\mathcal{D}_\sigma W^V_j = \widehat{W}^H_{j} \mathcal{D}_\sigma\ .
\end{align}
where $\widehat{W}^{H,V}$ are given by $W^{H,V}$ with $\sigma^r$ replaced by $\mu^r$. Thus the duality relates the corresponding transfer matrices by
\begin{align}
\mathcal{D}_\sigma T^H = \widehat{T}^V \mathcal{D}_\sigma\ ; \qquad
\mathcal{D}_\sigma T^V= \widehat{T}^H \mathcal{D}_\sigma\ .
\end{align}
Thus if a Boltzmann weight associated with a given plaquette on the square lattice is $W^H(u_H)$, commuting the duality defect gives a plaquette described by Boltzmann weight $W^V(u_H)$. Likewise it changes $W^V(u_V)\to W^H(u_V)$. 

Commuting a duality defect through the lattice thus not only changes the degrees of freedom from the original lattice to the dual lattice, it interchanges $J_x\leftrightarrow J_y$ as well. This indeed is the familiar statement of Ising duality. Our definition of defect lines can be utilized for any lattice or graph; one must just keep track of which factors $u_p$ are associated with each plaquette, i.e., each link of the original lattice or graph. Commuting a duality defect through the lattice then changes the spins from the original lattice to the dual lattice, but leaves the $u_p$ for each plaquette the same.


Since the $\mu_j^r$ satisfy the same algebra as the $\sigma_j^r$, $\mathcal{D}_\sigma$ preserves the algebra of Pauli matrices, and all local physical properties will be invariant under duality. However, non-local properties such as ground-state degeneracy can and typically do change.
The reason is that duality is not really a symmetry in the traditional sense: ${\mathcal D}_\sigma$ is not unitary, or even invertible. An intuitive way of understanding why the duality transformation is not invertible is to note that because duality interchanges $u_H$ and $u_V$, it interchanges the ordered phase ($u_H>u_V$) with the disordered phase ($u_H<u_V$). There are two minima of the free energy in the former case (or equivalently, a two-fold degenerate ground state in the quantum Hamiltonian limit). In the latter case, there is a unique minimum. Thus the duality map is not one-to-one and cannot be unitary.
The way this has been traditionally remedied in the operator formulation is to project onto sectors of the theory for which duality is invertible, and consider the appropriate boundary conditions in each sector \cite{KadanoffCeva,Pasquier1990,Levy1991}. In such constructions, duality includes a translation by half a site (to go effectively from lattice to dual), which meant that within one of these sectors, doing duality twice gives a translation, not the identity.  Our scheme provides a different perspective where we explicitly see the sectors over which $\D_\sigma$ is invertible and make use of both the lattice and dual lattice all along. This will prove particularly useful in section \ref{sec:sixteenth}.

Thus, despite the name, the duality defect operator $\mathcal{D}_\sigma$ does {\em not} square to the identity. In fact, it has a zero eigenvalue. To see this, we work out the defect {\em fusion algebra}, which describes what happens when two horizontal defects are combined. 
Putting two duality defects together gives
\begin{align}
\bra{\{ h_{j}' \}} \mathcal{D}_\sigma^2   \ket{\{ h_j \}} = 2^{-L}\sum_{\{\hat{h}_{j+1/2}=0,1\}} (-1)^{\sum_j \hat{h}_{j+1/2}(h_j+h_j' +h_{j+1}+h_{j+1}')}.
\end{align}
Each sum over the hatted variables imposes the constraint $h_j+h_j' +h_{j+1}+h_{j+1}' =0 \ {\rm mod}\ 2$. The constraint is satisfied only if $h_j' = h_j$ for all $j$ or $h_j' = 1+h_j$ for all $j$. Thus two duality defects either gives the identity, or a spin-flip defect:
$\D_\sigma^2=\mathds{1}+\D_\psi$.
The spin-flip defect operator is invertible: $\D_\psi^2=\mathds{1}$, so the duality-defect operator is not: ${\cal D}_\sigma^4=2{\cal D}_\sigma^2.$ Thus $\mathcal{D}_\sigma^2/2$ is a projection operator onto the sector where $\D_\psi$ has eigenvalue $1$. 
Putting a duality defect and a spin-flip defect together gives just a duality defect:
\begin{align}
\bra{\{ \hat{h}_{j+1/2} \}} \mathcal{D}_\sigma \mathcal{D}_{\psi}  \ket{\{ h_j \}}= 2^{-L/2}(-1)^{\sum_j (\hat{h}_{j-1/2 }+\hat{h}_{j+1/2})(1-h_j)} = \bra{ \{ \hat{h}_{j+1/2} \} } \mathcal{D}_\sigma   \ket{ \{ h_j \}} \ ,
\end{align}
since $(-1)^{\sum_j \hat{h}_{j-1/2}+\hat{h}_{j+1/2}}=1$. 
Thus $\mathcal{D}_\sigma(\mathcal{D}_{\sigma}^2-2)=0$, and so the eigenvalues of $\D_\sigma$ are $0,\pm\sqrt{2}$.

The fusion algebra for the topological defects in the Ising model is therefore
\begin{align}
{\mathcal D}_\psi^2=\mathds{1}\ ;\qquad\quad  \mathcal{D}_\sigma \mathcal{D}_{\psi} =\mathcal{D}_\psi \mathcal{D}_{\sigma} =  \mathcal{D}_\sigma\ ;\qquad\quad
 \mathcal{D}_{\sigma}^2 = \mathds{1} + \mathcal{D}_{\psi}
\label{fusionalgebra}
\end{align}
Fusion algebras will play an important role in our analysis. In general,
the fusion algebra can be be expressed in a compact notation
\begin{align}
\D_a \D_b = \sum_c N_{ab}^c \D_c\ .
\label{Ndef}
\end{align}
The $\D_a$ are normalized so that the  $N_{ab}^c $ are non-negative integers. Remarkably, for all the examples discussed in part~II as well, it is always possible to find a topological defects obeying (\ref{Ndef}). This is an explicit and exact correspondence between defects on the lattice and a rational conformal field theory, because chiral vertex operators in the latter obey such a fusion algebra. This suggests strongly that defect lines in the lattice can be associated directly with such operators. For example, in the Ising conformal field theory the chiral operators are $\psi$, the Majorana fermion,  and $\sigma$ the chiral part of the spin field. These obey the fusion algebra $\psi^2=\mathds{1}$, $\sigma\times\sigma = \mathds{1} +\psi$, and $\sigma\psi=\sigma$, just as their namesake defects do. This association of chiral operators with defects explains our naming of the defects, but is much more than a coincidence: it will allow us to generalise Kramers-Wannier duality to the height models in part~II.

\subsection{Duality-twisted boundary condition}
\label{sec:dualitybc}

\begin{figure}[bht]
	\begin{center}
	\includegraphics[width=5.5in]{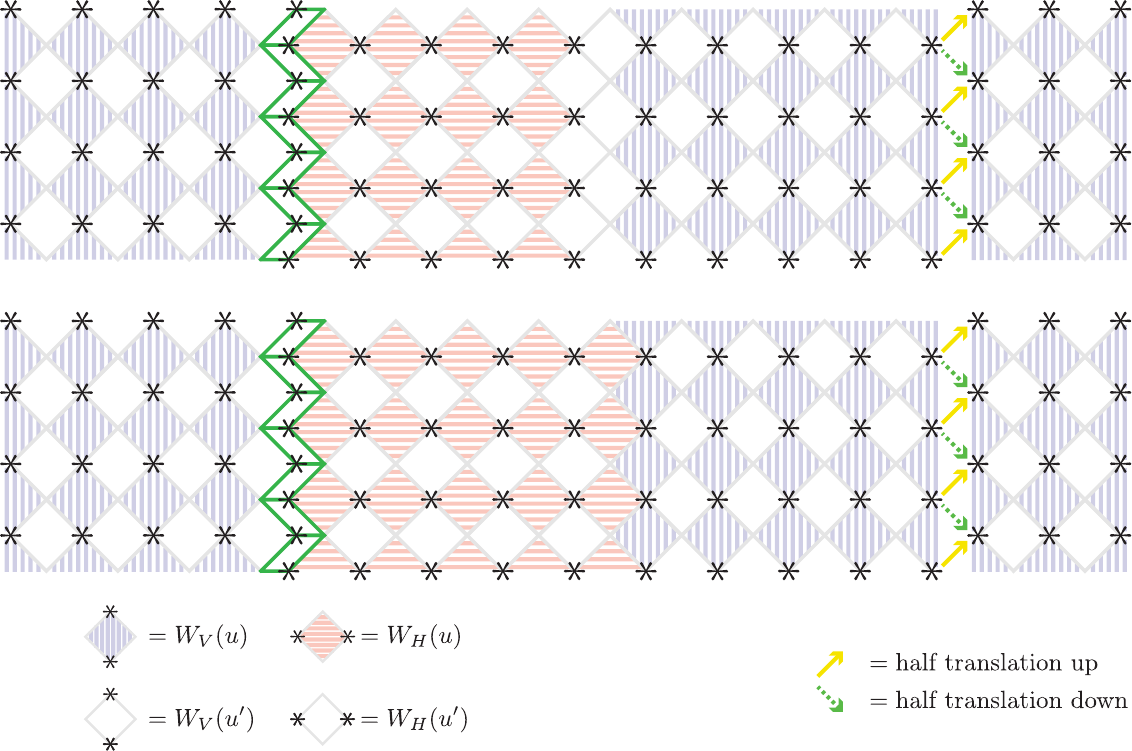}
	\end{center}
	\caption{The two types of domain walls present with periodicity in the horizontal direction and a vertical defect. The duality defect and the domain wall each separate the ordered phase with the disordered phases.
	 Squares with hatching have spectral parameter $u$ while blank ones have $u'$; the direction of hatching indicates the locations of the spins.  Top and bottom differ by the locations of the spins, and dots are omitted for clarity.}
	\label{Gluing}
\end{figure}

Since the duality defect shifts the spins from one sub-lattice to the other, care is needed when considering a single duality defect on a manifold with non-trivial topology. Consider a torus with a single vertical duality defect illustrated in Fig.~\ref{Gluing}.
The weights on the hatched and unhatched plaquettes are defined using parameters $u$ and $u'$ respectively, so that when crossing the duality defect on the square lattice, the vertical and horizontal Boltzmann weights change roles. Away from the critical point, this glues together the ordered and disordered phases in a seamless way -- the duality defect is mobile. However, in the presence of a single duality defect (or any odd number), the only way to sew together left and right sides is include an immobile, physical domain wall between the ordered and disordered phases. At the domain wall, the two types of hatching meet so that now the two boundaries can be sewed together with a half vertical translation in either direction. The domain wall breaks translation invariance except at the critical point $u=u'$, where the hatching is unnecessary and the system becomes translation invariant again. 

Including a single vertical duality defect is equivalent to a {\em duality twisted boundary condition} \cite{Levy1991,Schutz1993,GrimmSchutz1993,Affleck1997,Pearce2001,Grimm2002,Grimm2003,Armen2015}.
Due to the presence of the domain wall,  there are two transfer matrices one can define in the presence of a vertical duality defect, depending on which sites are occupied by heights/spins. 
The one arising from the bottom part of figure \ref{Gluing} has matrix elements pictured as
\begin{align}
&T_{\sigma} \ =\  \TDualityTwisted \ , 
\end{align}
where the one spin in the middle of the defect is summed over.
We have left the dots implicit in favour of displaying where the Ising spins reside.
We denote by ${\cal H}_\sigma$ the vector space comprised of the spin configurations on a row, so that $T_\sigma$ takes ${\cal H}_\sigma\to {\cal H}_\sigma$ .
We denote by $\widehat{\cal H}_\sigma$ the other vector space on which $T_\sigma$ acts, with
\begin{align}
&{T}_{\sigma} \ =\  \TDualityTwistedDual \ .
\end{align}
The two transfer matrices differ only by where the Ising degrees of freedom are placed, and
the two vector spaces ${\cal H}_\sigma$  and $\widehat{\cal H}_\sigma$  are isomorphic. They are related by a half translation, just as ${\cal H}$ and  $\widehat{\cal H}$ are.  We can thus adopt a unified convention for the $T_\sigma$ acting on the full vector space ${\cal H}_\sigma\oplus \widehat{\cal H}_\sigma$ just as for $T$ in (\ref{Tdef}): 
\begin{align}
T_\sigma \ =\  \TDualityTwistedDots
\end{align}

To be more explicit, consider $T_\sigma$ acting on ${\cal H}$, and let the duality defect be between sites $L$ and $1$, while the domain wall is at site $r$. The duality defect modifies the adjacent Boltzmann weight to
\begin{align}
\Wsigma \;=\; 2^{-\frac{1}{2}} \big[ \cos u
			+  \sigma^z_L \otimes \sigma^x_{1}\sin u \big]_{a\beta;a\beta'}\ .
\end{align}
The transfer matrices in the presence of a vertical duality defect are therefore
\begin{align}
T^{V}_\sigma = \big[ \cos u +  \sigma^z_L \otimes \sigma^x_{1}\sin u \big]\prod_{j=2}^{r-1} W^V_j(u')\,
\prod_{j=r}^{L} W^V_j(u) \ , \qquad T^{H}_\sigma =  \prod_{j=1}^{r-1}W_{j+1/2}^H(u)  
\prod_{j=r}^{L-1}W_{j+1/2}^H(u')\ .
\label{Tsigmadef}
\end{align}
The partition function is then
\begin{align}
Z_{\sigma,1} = \Tr(T^{V}_\sigma T^{H}_\sigma)\ .
\label{Zsigma}
\end{align}
The Boltzmann weights $W^V_1$ and $W^H_{L+1/2}$ do not appear, and there is only one new weight associated with the defect. Thus only $2L-1$ different weights are involved in the transfer matrix here.

 

Since the duality defect is topological, there exists a local unitary transformation $U_\sigma$ that shifts it over one site \cite{Levy1991,GrimmSchutz1993,Schutz1993}. The way to find it in our approach is to
drag the duality defect over the Boltzmann weight using the defect commutation relations. 
For the defect between $L$ and $1$, it  is 
\begin{align}
U_\sigma = {\cal H}_L\otimes CZ_{L,1}
\label{DualityTranslate}
\end{align}
where ${\cal H}_L = \frac{1}{\sqrt{2}} (\sigma^x_L + \sigma^z_L)$ and $CZ_{L,1} = \frac{1}{2} (1+\sigma_L^z +\sigma_1^z - \sigma_L^z \sigma_1^z)$ are the Hadamard and control-Z gates. 
However, the domain wall is {\em not} movable by a unitary transformation. Thus only at the critical point $u=u'$ does the transfer matrix commute with the modified translation operator ${\cal T}_\sigma= {\cal T}U_\sigma$. Relating ${\cal T}_\sigma$ to the corresponding Dehn twist is subtle, and so we defer this analysis to section \ref{sec:sixteenth}.

\subsection{A Majorana zero mode in the Hamiltonian limit}
\label{sec:Majorana}

Here we consider the Hamiltonian limit and study twisted boundary conditions in the Ising quantum spin chain. 
The critical case has been investigated thoroughly \cite{Schutz1993, Affleck1997,Grimm2002}, so we focus mainly on a very interesting piece of physics in the gapped phase. In the presence of a duality defect, a Majorana zero mode occurs, localized on the domain wall separating ordered and disordered regions. Such zero modes are currently of great current interest in the study of topological order \cite{Aliceareview}. An interesting feature is that this zero mode is unpaired -- the ``branch cut'' or ``string'' associated with the zero mode terminates on the defect instead of another zero mode.

Including no defect or a vertical spin defect results in periodic or anti-periodic boundary conditions on the spin chain. The Hilbert space on which the quantum Hamiltonian $H$ acts is then the vector space ${\cal H}$ on which the transfer matrix acts.  In the limit $u_H,u_V\to 0$, while keeping $u_H/u_V=J$ finite, the transfer matrix approaches the identity matrix. The order $u$ corrections are local, and yield the Hamiltonian: $T-\mathds{1} \propto u H + O(u^2)$, where
\begin{align}
H^{\pm} = -\sum_{j=1}^L \sigma^x_j - J \sum_{j=1}^{L-1}\sigma_j^z \sigma_{j+1}^z 
\mp J\sigma^z_L\sigma^z_{1}\ .
\label{HamiltonianLimit}
\end{align}
The top sign is for periodic boundary condition, while the bottom sign is for antiperiodic, a spin-flip defect between sites $L$ and $1$. 
Both Hamiltonians commute with the global spin-flip $\mathbb{Z}_2$ symmetry generated by ${\cal D}_\psi=\prod_j \sigma_j^x$.
The spectrum is independent of the location of the defect, since the defect can be moved at will by a unitary transformation. Multiple vertical spin-flip defect lines simply result in minus signs in the corresponding terms $J\sigma^z_k\sigma^z_{k+1}$ in the Hamiltonian. However, any even number can be removed by a unitary transformation; e.g.\ spin-flip defects at $(k,k+1)$ and $(L,1)$ are removed by the unitary transformation $\prod_{j=1}^k\sigma^x_j$. This is the vertical version of the horizontal defect fusion $\D_\psi\D_\psi =\mathds{1}$.



The Hamiltonian for a duality-twisted boundary conditions includes a domain wall at site $r$, and so acts on the Hilbert space ${\cal H}_\sigma$. Taking $u,u'\to 0$ with $u'/u=J$ gives 
\begin{align}
H^{d} =& - J\sum_{j=1}^{r-1}\sigma^z_j \sigma^z_{j+1} -\sum_{j =2}^{r-1}  \sigma^x_j
 -\sum_{j=r}^{L-1} \sigma^z_j \sigma^z_{j+1} - J\sum_{j=r}^{L} \sigma^x_j - \sigma_{L}^z \sigma_{1}^y\ .
\label{Hduality}
\end{align}
The final term is the effect of the duality defect, and is unitarily equivalent to the one discussed in \cite{Schutz1993} and \cite{Grimm2002}. For later convenience we have done a unitary transformation that changes $\sigma^x_1$ to $\sigma^y_1$ in this term, leaving the others untouched.
A straightforward calculation shows that if we wrap the defect line around the boundary differently, i.e., shifted by half a unit cell so that it wraps around $W_L^V$, we find the last term in (\ref{Hduality})
is replaced by $J\sigma_L^x-\sigma^x_1 -J\sigma^x_L\sigma_1^z$. 
The spectra for these three spin-chain boundary conditions in the critical case has been found explicitly \cite{Grimm2002,OBrien1996,Chui2002}, but we will show in sec.\ \ref{sec:partition} that the explicit expressions are not needed to relate the critical partition functions to those in conformal field theory.

The spin-flip symmetry remains, still generated by ${\D}_\psi$.  With a duality defect, as opposed to other boundary conditions, the spectrum is identical in both sectors labelled by eigenvalues $\pm 1$ of $\D_\psi$. This double degeneracy is a consequence of Kramers pairing. 
Precisely, note that conjugating $H^d$ and $\D_\psi$ by the operator $\sigma^z_1$ gives 
\begin{align}
\sigma^z_1 H^d \sigma^z_1\,=\, (H^d)^*\ , \qquad\quad \sigma^z_1\, \D_\psi\,\sigma^z_1\, =\, - \D_\psi\ .
\end{align}
Since $H^d$ is Hermitian, $(H^d)^*$ has the same eigenvalues. Hence the pairing: for each eigenstate $|\psi\rangle$ of $H^d$ and $\D_\psi$ with eigenvalues $E$ and $\omega$ respectively, $\sigma_z^1|\psi\rangle^*$ has  eigenvalues $E$ and $-\omega$ respectively. This degeneracy has an interesting consequence for partition functions. Because the spectra with $\D_\psi=1$ and $-1$ are the same,
\begin{align}
\hbox{tr }\left(\frac{1+\D_\psi}{2} e^{-\beta H_d} \right) = \hbox{tr }\left(\frac{1-\D_\psi}{2} e^{-\beta H_d} \right)
\quad\Rightarrow\quad \hbox{tr }\left(\D_\psi e^{-\beta H_d} \right)\ =\ 0\ .
\label{ZDpsi}
\end{align}
The analogous statement holds in the classical model on the torus as well; see (\ref{ZDpsi2}). 

Unless $J=1$ so that the system is critical, the domain wall illustrated in figure \ref{Gluing} breaks translation invariance.  In (\ref{Hduality}), the coupling $J$ changes roles at the domain wall. Thus on one side the system is ordered, and on the other it is disordered; which is which depends on whether $|J|<1$ or $|J|>1$. As a consequence, a  zero mode operator ``localised'' on the domain wall appears. This is straightforward to understand by using a Jordan-Wigner transformation of the Hamiltonian into free fermions; see Appendix \ref{zero-mode} for details.  In the gapped phase, the Hamiltonian includes a mass term for the fermions that changes sign at the location of the domain wall. Thus one would expect a Majorana zero mode localised there \cite{Aliceareview}, an expectation confirmed in the appendix. 

Its construction is similar to that of the zero mode occurring at the edge of the Ising/Kitaev chain with open boundary conditions \cite{Kitaev2001}, and utilises the iterative procedure described in \cite{Fendley2015}. We consider the case $|J|<1$, so that the region from $r$ to $L$ is ordered (i.e.\ the two-point function $\langle\sigma_j^z\sigma_k^z\rangle$ in the ground state approaches a non-zero constant in the limit $r\ll j\ll k\ll L$ ), while the region from $1$ to $r$ is disordered. The starting point for the iteration is the operator 
\begin{align} \Psi_0=i\sigma^z_1\sigma^z_r \prod_{k=1}^{r-1} \sigma^x_k\ ,
\label{psi0}
\end{align}
which commutes with the Hamiltonian in the extreme limit $J=0$. The string of spin-flip operators terminating at the duality defect means that the only terms in the full Hamiltonian not commuting with $\Psi_0$ are in the region around the duality defect. Moreover, these terms are of order $J$. The resulting commutator can be written as a commutator of $H^d$ with an order $J$ operator, up to corrections of order $J^2$:
\begin{align}
\left[H^d,\Psi_0\right]\ =\ -2J(\sigma^x_r+\sigma^z_{r-1}\sigma^z_{r})\Psi_0
 \ =\ -\left[H^d,\,\Psi_1\right]\ +\ {\cal O}(J^2)\ .
\end{align}
where
\begin{align}
\Psi_1=  iJ\sigma^z_1\left(\sigma^z_{r+1} \prod_{k=1}^{r} \sigma^x_k\  +\ \sigma^z_{r-1}\prod_{k=1}^{r-2} \sigma^x_k\right)\ .
\end{align}
Thus $[H^d,\Psi_0+\Psi_1]$ contains only terms of order $J^2$. These terms can then be written as $-[H^d,\Psi_2]$ and the iteration continues. The analysis in the appendix shows that that the iteration can stop, giving an operator $\Psi=\Psi_0+\Psi_1+\Psi_2+\dots+\Psi_{L}$ that commutes with the Hamiltonian. 
For $|J|>1$, there is an analogous construction, given in the appendix.

All the terms in the zero mode have strings of spin flips terminating at the duality defect. This string is familiar from the Jordan-Wigner transformation: it means the zero mode is non-local in terms of the spins, but local in terms of the fermions. Thus if the fundamental degrees of freedom are fermions, $\Psi_0$ and $\Psi_1$ are localised at the domain wall. As the operators get farther from $r$,  the norm of each term falls off exponentially, up to one (important) subtlety discussed in the appendix. One other feature worthy of note is that there is just one zero mode for a given value of $J$, as opposed to the case of an open Ising chain, where there is a zero mode at each end \cite{Kitaev2001}. Here the duality defect plays the role of the other end, but there is no zero mode localised there. The operator $\sigma^z_1$ does result in the Kramers pairing, but it is not a zero mode, as it does not commute with $H^d$.  

A nice picture also emerges in terms of the spins: the zero mode contains a spin-flip defect stretching from the domain wall to the duality defect. As we have repeatedly emphasised, the duality defect is topological and so can be moved without changing the energy. This suggests that in the full 2d classical model, there is a trivalent junction of defects where a spin-flip defect terminates at the duality defect, and that its location can be changed without changing the free energy/partition function. In the next section \ref{sec:trivalent} we define such a junction precisely.

\section{Trivalent junctions of defects}
\label{sec:trivalent}

Having described two different types of defects, our next step is to show how they can intersect. In this section we show how to branch and fuse defect lines at intersections. We show that in our construction, it is quite straightforward to find intersections that can be moved around without changing the partition function. Understanding these intersections is essential for defining and analysing partition functions on the torus twisted along multiple cycles.

\subsection{The triangular defect}

\begin{figure}[h] \centerline{%
 \xymatrix @!0 @M=8mm @C=85mm{
 \TrivalentJunctionInserta\ar[r] & \TrivalentJunctionInsertb
 }}
\caption{On the left, a schematic view of a trivalent junction of vertex lines. On the right, how this is defined precisely by introducing a triangular face.}
\label{fig:Trivalent}
\end{figure}

The basic object is a trivalent junction of defect lines. The three lines meeting need not be of the same type (and in Ising, cannot be). In our construction of defects using quadrilaterals, a trivalent junction requires introducing a triangular face, as illustrated in figure \ref{fig:Trivalent}. 
A nontrivial statement we elaborate on below is that any higher valency junction of defect lines can always be decomposed into trivalent junctions.
Each side of the triangle is labelled by the type of defect coming into that side of the junction, as well as a height label at each vertex:
\begin{align}
\Trivalent 
\label{eq:triangle}
\end{align}

\begin{figure}[h] \centerline{%
 \xymatrix @!0 @M=2mm @R=42mm @C=62mm{
 {\quad \TrivalentConsistencya \quad } \ar[rr]\ar[d]& & {\quad \TrivalentConsistencye \quad}  \\
  {\quad \TrivalentConsistencyb \quad } \ar[r]&{\quad  \TrivalentConsistencyc \quad} \ar[r]       & {\quad \TrivalentConsistencyd \ar[u] \quad}
}}
\caption{
Moving a trivalent junction on the lattice is schematically illustrated in the top row. Doing this precisely requires two steps, shown in the bottom row.
		The first step uses the defect commutation relations to move the red defect line around a plaquette. 
		The second step 
uses	 the triangle defect commutation relation (\ref{eq:defcomm3}) inside the boxed region. The dots are omitted for clarity. }
		\label{fig:TrivalentConsistency}
\end{figure}
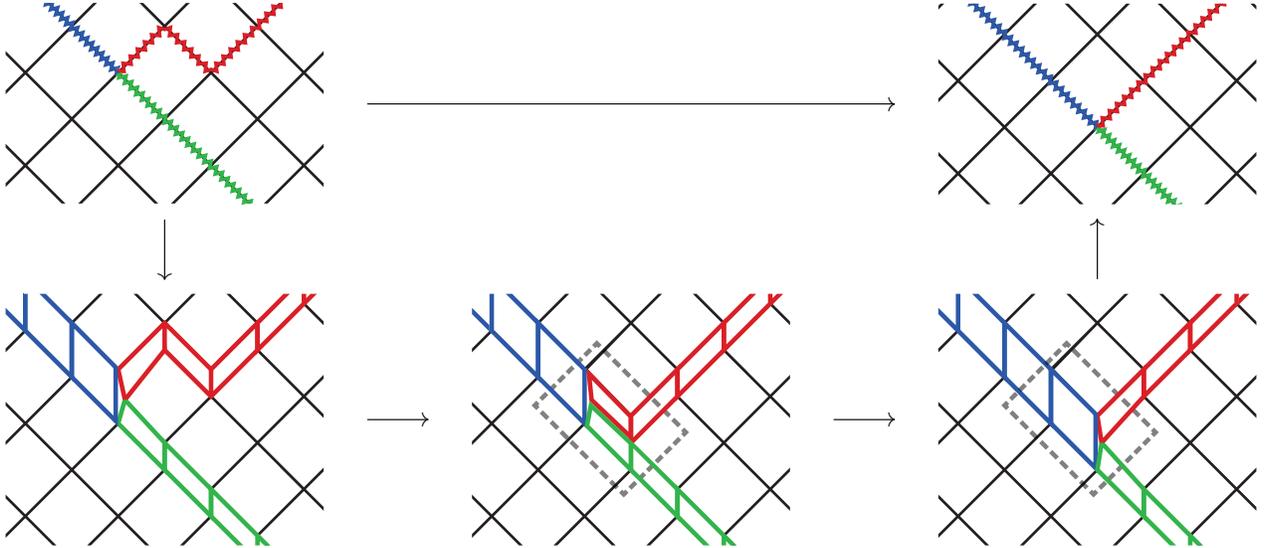

In the interest of studying topological defects we will also enforce the condition that the triangle defect is free to move without changing the value of the partition function. We thus
require that the triangle defect commutes locally with the transfer matrix. The analogous triangle defect commutation relations to be solved are given by
\begin{align}
\sum_\beta \Trivalenta =  \Trivalentb,
\label{eq:defcomm3}
\end{align}
When this is satisfied, triangle defects indeed can be moved around without changing the partition function, as illustrated in figure \ref{fig:TrivalentConsistency}.

Finding solutions to (\ref{eq:defcomm3}) appears to be quite an imposing problem, since it involves the defect Boltzmann weights and the triangle weights.  However, as we will explain in detail in part~II, there is a procedure for finding solutions.  The general structure developed there requires that the weight of the triangle is non-vanishing only when the labels obey certain conditions.
These conditions work in much the same way as the defect lines themselves -- they occur at a termination point of three defect lines with a rule on how to fuse them together. 
We used the word ``fuse'' in the previous section to describe for example how creating two successive horizontal duality defects is equivalent to the sum of the identity operator and a horizontal spin-flip defect, i.e.\ $\D_\sigma^2=\mathds{1} + \D_\psi$ in operator language.
This same algebra (\ref{fusionalgebra}) is utilised to constrain the allowed triangles at a given trivalent junction.  For this to make sense, the heights themselves must be labelled by elements of the fusion algebra. In the Ising model, the heights $0$ and $1$ (i.e.\ spin up and down) correspond respectively to $\mathds{1}$ and $\psi$, while the empty site corresponds to $\sigma$. The condition that the triangle be non-vanishing is that at each vertex, the two defect labels must fuse to the height variable at that vertex. 

Precisely, this means that for the weight of (\ref{eq:triangle}) not to vanish, $N_{\beta \rho}^G N_{\rho \gamma}^B N _{\gamma \beta}^R \ne 0$, where the $N_{ab}^c$ are the non-negative integers defined by the fusion algebra (\ref{Ndef}). Since the fusion algebra is associative, this implies that $N_{RG}^B  >0$ as well. This correspondence between heights and defect labels seems rather ad hoc and coincidental. However, as will be apparent in the more general height models of part~II, it is not, but instead a fundamental part of the definitions of the theory itself. This indicates rather strongly that the defect lines we study are themselves a fundamental part of the theory, and gives excellent intuition into why their topological properties are so remarkable.


We therefore need to introduce the identity ``defect'', which  leaves the partition function invariant, even if wrapped around a cycle.
We draw it with dashed lines, and its Boltzmann weights are
\begin{equation}
\mathord{\sideset{^{a}_{\alpha}}{^{b}_{\beta}}{\mathop{\IsingBubbleRightId}}}\ =\  2^{-\tfrac14}\,\delta_{a\alpha}\delta_{b\beta}\  .
\end{equation}
The factor of $2^{-\tfrac14}$ is needed to cancel the weights (\ref{vertexweights}) associated with the extra sites arising from the defect insertion.
The identity defect obviously satisfies the defect commutation relations.

The identity defect also can terminate along another defect at a trivalent junction.
It is easy to find the corresponding solution of (\ref{eq:defcomm3}); the only non-trivial part is to again compensate for the weights from the extra sites. 
In our conventions the triangle weights that occur at the termination of an identity defect are then:
\begin{equation}
\begin{aligned}
\TrivalentIdh&\ =\ 2^{-\frac14}\ ,&&\qquad  \TrivalentIdhh\ =\  \delta_{hh'},\\
\TrivalentIdpsipsihh& \ =\  2^{-\frac14}\ ,  &&\qquad\TrivalentIdpsipsih \ =\  [\sigma^x]_{h\tilde{h}} \delta_{h h'},
\end{aligned}
\label{identitytri}
\end{equation}
where $h$ can be the height $0$ or $1$ (i.e.\ $\mathds{1}$ or $\psi$, i.e.\ spin up or down), and the $\sigma$ height corresponds to an empty site. We can then insert an identity defect between say two duality defect lines that touch at a single site. At that site, we can insert two triangle defects using (\ref{identitytri}):
\begin{align}
\IsingDefectSplitp \quad \ =\  \quad \IsingFIdHp
\label{insertidentity}
\end{align}
A possible factor of $\sqrt{2}$ coming from (\ref{identitytri}) is cancelled by the extra weight coming from increasing the number of sites by one.

The Ising triangle defect commutation relations are fairly straightforward to solve in general, since the only non-vanishing $N_{RG}^B$ in the Ising model not involving the identity defect are $N_{\sigma\sigma}^\psi=N_{\sigma\psi}^\sigma=N_{\psi\sigma}^\sigma=1$. Thus the only non-trivial trivalent junction here involves two duality defects and one spin flip defect.  There are two distinct types of labelings of the heights around the triangle, and the weights are
\begin{align}
\Trivalenth &=2^{-\frac14} (-1)^h, & \Trivalenthh &= \big[ \sigma^x \big]_{hh'} .
\label{psitri}
\end{align}
It is straightforward to check directly that indeed these triangle weights satisfy (\ref{eq:defcomm3}). 

\subsection{Partition function identities}

The fact that the partition function is independent of the locations of trivalent junctions has a number of remarkable consequences. 
Here we show that there are {\em local} relations between different ways defect lines can join, resulting in linear relations among partition functions with different defects present.
We develop a toolset that relates microscopic properties of the defect lines to macroscopic pictures, making deriving these linear relations very easy.

We showed in  (\ref{eq:IsingClosedLoop}) that the partition function including a closed isolated defect loop not wrapped around any cycle is simply $\sqrt{2}$ times that without the loop. Let us now consider a more general situation with a closed loop, still topologically trivial, but with two defect lines terminating at it.
A concrete example is two spin-flip defects terminating at a duality defect loop.
To express such configurations in a uniform fashion, we use the trivalent defect commutation relations (\ref{eq:defcomm3}) to push the two trivalent junctions so that they are both on top of one defect Boltzmann weight. Furthermore, using the standard defect commutation relations (\ref{dualitydefcomm}) we rearrange the picture so that the defect loop is in line with the defect lines that terminate at it. In a picture,
\begin{align}
\xymatrix @!0 @M=3mm @C=38mm{
 \BubbleMacroLeft \ar[r] &\IsingBubbleMicroLeft 
 }
 \label{IsingBubbleMicro}
\end{align}
It applies for any allowed external heights, while the internal heights (the dots) are summed over.

In part~II we explain a general method for simplifying the right-hand-side of (\ref{IsingBubbleMicro}). 
The result is very elegant.
First, (\ref{IsingBubbleMicro}) is zero unless $R=B$. One can check this in Ising by direct computation; more generally it follows from an orthogonality relation satisfied by the trivalent junction weights. 
If $P = \mathds{1}$ then obviously (\ref{IsingBubbleMicro}) is simply $\delta_{RB}\delta_{GB}$, 
and likewise with $P\leftrightarrow G$.
We can explicitly compute the non-trivial ones to give
\begin{align}
\IsingBubbleLeftId \ =\  \sqrt{2}\  \IsingBubbleRightId\ , \qquad \IsingBubbleLeftSigma \ =\  
\IsingBubbleRightSigma\ , \qquad \IsingBubbleLeftPsi \ =\ \sqrt{2}\ 
 \IsingBubbleRightPsi
\label{MicroBubble}
\end{align}
where these apply independently of the heights at the corners. 
This special property means the macroscopic defect lines can be manipulated independently of the microscopic configurations, making the analysis quite tractable. Indeed, 
all these rules are summarized with the following \emph{macroscopic} picture,
\begin{align}
\BubbleMacroLeft \ =\ \delta_{RB} \sqrt{\frac{d_G d_P}{d_R}} \BubbleMacroRight
\label{BubbleMacro}
\end{align}
where $d_\sigma = \sqrt{2}$ and $d_0 = d_\psi = 1$.


This relation, and its successors, are local, meaning that they can be applied to any subgraph of defects: (\ref{BubbleMacro}) applies no matter what the lines going out are connected to. Any loop with zero, one or two external defects attached can thus be removed. In fact, setting  $B = \mathds{1}$ and $R=\psi$ shows that any partition function containing a ``tadpole'' vanishes, e.g.\
\begin{align}
\Barbell  \ =\ 0 \ .
\label{barbell}
\end{align}
The zigzag around the edge illustrates that this could be a sub-part of any partition function. 
The two colours of shading correspond to the two types of staggering of the couplings. These necessarily change across the duality defect, as indicated by the types of hatching in figure \ref{Gluing}. At the critical point $u\ =\ u'$, so there is no staggering and no need for the shading.

We already came across a simple example of such a partition function identity when implementing translation invariance in the presence of a vertical spin-flip defect: translating fully around a cycle results in the 
 crossing red lines appearing in (\ref{Tpsi}). The junction is given by two spin-flip defects meeting at a point and then leaving. 
The termination of the spin-flip defect lines at that point defines a quadrivalent defect Boltzmann weight:
\begin{align}
\RedCross\ =\ \begin{cases}		
		 2^{-\frac12} 	&	a=b=c=d = \sigma,  \\
		\sigma^x_{ab}\sigma^x_{bc} \sigma^x_{cd} \sigma^x_{da}	&	\text{otherwise}
		\end{cases}
\end{align}
Either the defect lines terminate at a dual lattice point shown in the first line or a point on the lattice shown in the second line. The quadrivalent defect of spin flips can be decomposed into two trivalent junctions in two distinct ways. Using the explicit weights (\ref{identitytri}) shows that the two are the same
for any labelling of the heights:
\begin{align}
\RedCrossNoHeights\ =\ \RedCrossLeft\ =\  \RedCrossRight\ .
\label{redcross}
\end{align}
The trivalent junctions give a linear relation between these. Since the trivalent junctions in (\ref{redcross}) are free to move we indeed have
\begin{align}
\newcommand{\ZRRxLocalwS}{\mathbin{\rotatebox[origin=c]{90}{$\ZRRxLocalw$}}}
\ZRRxpLocalw\ =\ \ZRRxLocalw\ =\ \ZRRxLocalwS
\label{redcross2}
\end{align}
as promised.

The analogous situation for duality defects is even more interesting.
Here the relations between pairs of triangle defects is more intricate; the corresponding identity is
\begin{align}
\IsingFIdH \ =\ \frac{1}{\sqrt2} \Bigg( \IsingFIdV+ \IsingFPsiV \Bigg),
\label{reshuffle1}
\end{align} 
as is easily verified from (\ref{identitytri},\ref{psitri}).
This identity also holds rotated by 90 degrees, so
\begin{align}
\IsingFPsiH \ =\ \frac{1}{\sqrt2} \Bigg( \IsingFIdV- \IsingFPsiV \Bigg),
\label{reshuffle2}
\end{align} 
for any labelling of the external vertices. 
The trivalent defect commutation relations (\ref{eq:defcomm3}) allow the junctions to be moved apart, connected by a defect of the type stretched across. The ordinary defect commutation relations allow the individual lines to be moved at will. 

All these manoeuvres leave the partition function invariant. Therefore (\ref{reshuffle1}) and (\ref{reshuffle2})  can be recast as linear identities for various partition functions under local rejoining. Schematically,
\begin{align}
\IsingFIdHMacro \ &=\ \frac{1}{\sqrt2} \Bigg( \IsingFIdVMacro+ \IsingFPsiVMacro \Bigg),\\
\IsingFPsiHMacro \ &=\ \frac{1}{\sqrt2} \Bigg( \IsingFIdVMacro- \IsingFPsiVMacro \Bigg).
\label{reshuffleMacro1}
\end{align} 
Again, the two colours of shading correspond to the two types of staggerings and the zig-zag boundary signifies that it could be a sub-part of any partition function.
As a consistency check, we can rederive the vanishing of the `barbell'' graph from (\ref{barbell}):
\begin{align}
\Barbell  \ =\  \frac{1}{\sqrt{2}} \Bigg( \SigmaBubble - \ThetaDiagram \Bigg) \ =\  
\frac{1}{\sqrt{2}} \Bigg( \SigmaBubble - \SigmaBubble \Bigg) =
0\ .
\end{align}
using first the reshuffling and then (\ref{BubbleMacro}) with $P=\psi$ and $G=R=B=\sigma$.

\subsection{Rearranging intersections: $F$ moves}
\label{sec:Fmoves}

We have shown how to manipulate defect lines at the macroscopic level without any reference to the microscopic details of how the defect lines fuse together. 
What is even more remarkable is that the macroscopic rules they obey are precisely those of the underlying microscopic degrees of freedom. This of course is not a coincidence, but rather both are consequences of a deep mathematical structure. These linear relations are called in the context of topological field theory {\em $F$-moves}. These are precisely the rules obeyed by defect lines in the conformal field theory describing the continuum limit of the critical model \cite{Frohlich:2006}. The rules we derive here apply off the critical point as well. Here we say a little about their general structure. In part~II we will go much further.

Whenever four defect line segments meet at a point, the intersection can be decomposed into trivalent vertices in two distinct ways. However, as apparent in (\ref{redcross}, \ref{reshuffle1}, \ref{reshuffle2}), there are linear relations between the two ways the lines join. The $F$ moves governing this reshuffling are in general
\begin{align}
	 \IsingFLeft \;=\; \sum_Y \Big[F^{RGB}_P \Big]_{XY} \IsingFRight \;.
	\label{MicroF}
\end{align}
where the coefficients $\Big[F^{RGB}_P \Big]_{XY}$ are called $F$-symbols.
Conjoined with the trivalent defect commutation relations (\ref{eq:defcomm3}) and the standard duality defect relations (\ref{dualitydefcomm}), the analogous statement to (\ref{BubbleMacro}) holds:
\begin{align}
\FmoveMacroLeft \ =\  \sum_{Y} \big[ F_{P}^{RGB} \big]_{XY}\FmoveMacroRight
\label{MacroF}
\end{align}
The $F$-symbols form a unitary matrix for fixed external legs $R,G,B,P$, and in simple examples like Ising are independent of the heights at the corners.

We have already worked out most of the $F$-symbols for Ising. For example, the relation (\ref{redcross}) or (\ref{redcross2}) arises because spin flip defects can only fuse to the identity defect.
The corresponding $F$-symbol is thus $\Big[F_\psi^{\psi \psi \psi} \Big]_{\mathds{1} \mathds{1}}=1$.
The nontrivial $F$-symbols arise when the external legs have duality defects, and follow from (\ref{reshuffle1}) and (\ref{reshuffle2}):
\begin{align}
	\left[F_{\sigma}^{\sigma \sigma \sigma} \right]_{ab} \ &=\  \frac{1}{\sqrt2}(-1)^{ab},
	& F_{a}^{ \sigma b \sigma } \ &=\  F_{ \sigma }^{a \sigma b} \ =\   (-1)^{ab}.
\label{FSymbol}
\end{align}
where $a$ or $b=0$ for the identity defect $\mathds{1}$, and $1$ when it is the spin-flip defect $\psi$ (the same notation as for the height labels). 
The other non-zero $F$-symbols are equal to one when allowed by the fusion rules. If not allowed by the fusion rules, they of course give zero.

The $F$-moves allow one to decompose any higher-valency junction of defect lines into trivalent junctions, as promised. 
Crucially, the $F$-symbols are independent of the microscopic height configurations at the corners of the square. This is why we have always been able to omit writing them, and why the schematic pictures contain all the information necessary. The $F$ moves are macroscopic operations; they only depend on the types of defect lines present and how they fuse. However, they are uniquely determined by the microscopic rules we have used to define the Boltzmann and defect weights. 
This beautiful relation between microscopic and macroscopic properties illustrates how defect lines and junctions are fundamental objects in the theory. The relations between them and the continuum results we derive in subsequent sections are natural consequences of this common structure. 


This common structure is known as a {\em fusion category}, and arises in studies of rational conformal field theory, anyonic particles, and topological quantum field theory \cite{MooreSeiberg:89}; for reviews relevant to our story, see \cite{MSReview89,Bondersonthesis,Kitaev2006}.
The fusion category provides a general set of rules for manipulating these lines.
Written in abstract form, these are summarized as 
\begin{equation}
\begin{aligned}
\QD = d_a\ , \qquad \bubblea &= \delta_{ac}\sqrt{\frac{d_b d_{b'}}{d_a}}\ \bubbleb\ , \qquad  \recouplinga = \sum_c N_{ab}^c \sqrt{\frac{d_c}{d_a d_b}}\ \recouplingb\ ,\\
&\FmoveLSD = \sum_y \Big[ F^{abc}_d \Big]_{xy} \FmoveRSD\ .
\label{threerules}
\end{aligned}
\end{equation}
These should be understood as diagrammatic rules for the defect lines -- i.e. each line in the picture corresponds to some defect line with the corresponding label in the Ising model.
The first two generalize (\ref{BubbleMacro}) and (\ref{MacroF}), while the third is a special case of the $F$-symbol, whose general form is given in the last.
In fact, the triangle and defect weights themselves are expressed in terms of the $F$-symbols:
\begin{align}
\Trivalent = (\frac{d_R d_G} {d_B d_\beta^2})^{\frac14}\Big[ F^{RG \rho}_\gamma \Big]_{ B \beta}, \quad \DefectSquareLabelx{a}{b}{\alpha}{\beta} = \frac{1}{\sqrt{d_{G}d_\sigma}} \Big[F^{a\alpha \beta}_ b \Big]_{G\sigma}
\end{align}

The specific rules analysed in this paper are those of the Ising fusion category.\footnote{It turns out there are two solutions of \eqref{threerules} compatitble with the Ising fusion rules $\sigma \times \sigma = \mathds{1} + \psi$. Another is known as $\SU2_2$ and involves a sign known as the Frobenius-Schur indicator for the $\sigma$ label. We have chosen the positive sign here; in 
part~II will elaborate.}
In part~II we will extend this correspondence to a large class of lattice models, including for example the height models of Andrews, Baxter and Forrester \cite{Andrews1984}. 
In our lattice defects,  we have only required the rules of a fusion category; we have not utilised the braiding appearing in the additional structure present in a modular tensor category. It thus would be quite interesting to see if lattice models could be defined using fusion categories that cannot be extended to allow braiding.


\section{Duality on the torus}
\label{sec:torus}
 
With these diagrammatic rules we can prove some remarkable facts about the Ising model on the torus with relatively little amounts of work. One particularly useful application is a very straightforward way  to implement Kramers-Wannier duality on the torus. Our analysis makes it possible to do all the manipulations using macroscopic pictures that only depend on the topology of the defects. Using such manipulations, we show how to easily derive identities between toroidal partition functions on the lattice and those on the dual lattice. Even more strikingly, by studying the translation operator in the presence of duality defects, we derive the conformal dimension 1/16 of the chiral spin field directly in the critical lattice model.

\subsection{Identities for toroidal partition functions}
\label{sec:partitionidentities}


A simple example is two defect loops wrapped around the same cycle of a torus.
Applying (\ref{reshuffle1}) gives
\begin{align}
 \Zss \ =\   \frac{1}{\sqrt{2}} \Bigg( \Zdsigmacycle + \Zdsigmacyclepsiw   \Bigg).
 \label{id1}
\end{align}
The fact that the trivalent junction is free to move allows us to pull the duality defect loop around the torus and find a linear combination of two duality defect bubbles, one having no defect loops terminating at it, the other having two spin-flip defects terminating at it. Using equation (\ref{BubbleMacro}) allows these loops to be removed at the expense of some quantum dimensions:
\begin{align}
 \Zss =  \frac{1}{\sqrt{2}} \Bigg( \Zdsigma+ \sqrt{2}\,\ZRxRwG{\DualityGm} \Bigg)=  \ZG{\DualityGm} +  \ZRxRwG{\DualityGm}\ .
 \label{id2}
\end{align} 
We thus recover $\D_\sigma \D_\sigma = \mathds{1} + \D_\psi$ from section \ref{subsec:dualitydefect}
by purely local moves, and without any need to manipulate the explicit definitions of the operators.

Since (\ref{id2}) shows the operator creating a duality defect wrapped around a cycle is not invertible, it is not immediately obvious how to define a duality transformation on the torus. If it were invertible, the procedure would be to create such a line and its inverse, move one around the other cycle, and then annihilate the two. This would leave behind couplings with the other staggering on the dual lattice, and so show that the partition function for the model and its dual on the torus are the same. The fact that $D_\sigma$ is not invertible means that they are not. We can however use (\ref{id2}) to derive a first identity between toroidal partition functions. We move one of the duality defect lines on the left-hand side around the torus in the vertical direction before fusing it with the other. This then leaves the dual lattice behind, so that
\begin{align}
\ZG{\DualityGm} + \ZRxRwG{\DualityGm} \ =\   \ZG{\DualityGp} +\ZRxRwG{\DualityGp}\ .
\label{zzdual1}
\end{align}
In words, the partition function of the model and its dual are the same, when restricted to the even sector of the ${\mathbb Z}_2$ symmetry generated by $\D_\psi$ \cite{KadanoffCeva}.

Using our defect lines, there is a simple way to derive a less obvious relation between the partition functions of the Ising model and its dual. This exploits the fact that 
creating a topologically trivial defect loop is invertible:
\begin{align}
\ZG{\DualityGm} \ =\  \frac{1}{\sqrt{2}}\,\Zdsigma \ =\  \frac{1}{\sqrt{2}}\,\Zdsigmacycle\ .
\label{nucleate1}
\end{align}
The first relation follows from (\ref{eq:IsingClosedLoop}), while the second follows from using the defect commutation relations to move the loop. 
Using the $F$-moves derived above we also have the equality
\begin{align}
  \frac{1}{\sqrt{2}}\,\Zdsigmacycle  \ =\  \frac{1}{2} \Bigg ( \Zss +\Zsspw \Bigg )\ .
\label{nucleate2}  
\end{align} 
All spins between the defect lines reside on the dual lattice, as indicated by the darker shading. Next we drag one $\sigma$ defect line around the vertical cycle, giving
\begin{align}
\ZG{\DualityGm} \ =\ \frac{1}{2} \Bigg( \Zssdual +\Zspsw \Bigg).
\label{eq:Z_dd}
\end{align}
Finally, using another $F$-move and (\ref{threerules}) we annihilate these duality defect lines to find
\begin{align}
\ZG{\DualityGm} \ =\ \frac{1}{2} \Bigg ( \ \ZG{\DualityGp} +\ZRxRwG{\DualityGp} +\ZxRRwG{\DualityGp}  +\ZRRxwG{\DualityGp}\ \Bigg )\ .
\label{z-to-zdual}
\end{align}
All the partition functions on the right-hand side are defined solely on the dual lattice with the opposite staggering. A more traditional way of deriving this would be to rewite the Ising model in terms of dimers \cite{mccoywubook} and utilise the analogous result there \cite{Tesler}. Here we have a direct and simple proof valid both off the critical point and on the lattice. This linear constraint is also familiar in conformal field theory \cite{Ginsparg},  
and indeed it will prove quite useful in section \ref{sec:partition} studying modular transformations on critical partition functions.

Combining (\ref{zzdual1}) with (\ref{z-to-zdual}) yields another relation between toroidal partition functions:
\begin{align}
\ZG{\DualityGm} - \ZRxRwG{\DualityGm} \ =\  \ZxRRwG{\DualityGp} + \ZRRxwG{\DualityGp}\ ,
\label{zzdualagain}
\end{align}
This provides a nice consistency check, since it also is obtained by directly annihilating the defects in first picture on the right-hand side of (\ref{eq:Z_dd}).

$F$-moves give an easy way of seeing that one particular toroidal partition function vanishes. An $F$ move relates two partition functions with a duality defect wrapped around one cycle and a spin-flip defect around the other: 
\begin{align}
{\mathbin{\rotatebox[origin=c]{90}{$\ZGRGw$}}}\ =\ [F^{\psi \sigma \psi}_\sigma]_{\sigma\sigma}\, \ZRGGw\ =\ -\ \ZRGGw\ ,
\end{align}
where we have left the domain wall unlabelled and so omitted the shading here.
However, shifting the defects part way around the vertical direction shows that the two partition functions are equal as well as opposite, and hence vanish:
\begin{align}
{\mathbin{\rotatebox[origin=c]{90}{$\ZGRGw$}}} \ =\ \ZRGGw\ =\  0\ .
\label{ZDpsi2}
\end{align}
This is the full 2d classical version of (\ref{ZDpsi}), a consequence of the  double degeneracy of the Hamiltonian $H^d$ in the presence of a vertical duality defect.

\subsection{Translation across a duality defect}
\label{sec:sixteenth}

We have shown in the previous section how $F$ moves of defect lines result in identities for toroidal partition functions. Here we extend this analysis further by studying the Dehn twist $\modT_\sigma$, the translation around a complete cycle of the torus in a direction orthogonal to a duality defect. 
We use this to compute the exact shift in the momenta, enabling us to give a direct and exact (mod 1/2) lattice derivation of the conformal spin 1/16 of the chiral spin field in the corresponding conformal field theory. 
This Dehn twist will also play a large role in section \ref{sec:partition} when we explore the modular transformations of the partition functions and their continuum limits. 

The presence of a conserved translation operator guarantees conservation of momentum, with the momentum operator ${\cal P}_{\varphi}$ in the presence of a twisted boundary condition defined via
 \begin{equation}
e^{\frac{2\pi i {\cal P}_\varphi}{L}} =\T_\varphi .
\label{Pdf}
\end{equation}
With periodic boundary conditions (no vertical defects), $({ \T}_\mathds{1})^L=1$, so the total momenta (the eigenvalues of ${\cal P}_1$) are constrained to be integers. With anti-periodic boundary conditions from a vertical spin-flip defect, this is modified. Here $({\T}_{\psi})^L=\D_\psi$, as detailed in (\ref{TpsiL}). 
Since ${\T}_\psi$ commutes with the transfer matrix, we can group the states into sectors labelled both by momentum and the eigenvalue $\pm 1$ of $\D_\psi$. The momenta in the sector with $D_\psi=-1$ and anti-periodic boundary conditions are therefore given by half integers. 

This seemingly innocuous half-integer shift in the momentum gives an exact result in the conformal field theory describing the continuum limit.  In a CFT, one can think of a conformally twisted boundary condition as being ``created'' by pair producing an operator and its conjugate, dragging one around the torus in the vertical direction, and then re-annihilating the two \cite{Cardy1989}. A classic result relates the eigenvalues of the energy and momentum operators in the CFT to the scaling dimensions of operators \cite{Blote1986,Affleck1986}.
Up to an integer, the conformal spin (the difference of its right and left scaling dimensions) of the operator creating the twisted boundary condition is then precisely the momentum shift. Thus in the Ising CFT there must be an operator with conformal spin $1/2$ modulo an integer. The fermion field $\psi$ indeed has conformal spin $1/2$. Thus simple manipulations on the lattice model in the presence of conformally twisted boundary conditions yield exact results for the boundary-condition changing operator in the corresponding conformal field theory!  


Finding the shift in momentum quantisation for duality-twisted boundary conditions requires more work, but ultimately becomes a calculation similar to those in the preceding subsection \ref{sec:partitionidentities}. 
Since a duality-twisted boundary condition requires a domain wall, only at the critical point is it possible to define a translation operator that commutes with the transfer matrix, as explained above by (\ref{DualityTranslate}).
However, a suitably defined Dehn twist $\modT_\sigma$ does commute for all $u$ and $u'$. The Dehn twist
$\modT_\psi$ in the presence of a vertical spin-flip defect was described at the end of section \ref{subsec:spinflip}. and qualitatively the effect is the same for $\modT_\sigma$: doing a Dehn twist in the presence of a vertical duality defect creates a horizontal defect. At the putative crossing of the vertical and horizontal lines in $\modT_\sigma$, the lines avoid, just as for $\modT_\psi$ in (\ref{redcross}). 
Therefore, schematically, the Dehn twist implements a map of partition functions
\begin{align}
\ZxGG \xrightarrow{\modT_\sigma} \ZGGx
\label{Dehn1}
\end{align}  
just as in (\ref{redcross2}).
For simplicity we specialise to the critical point, where we can omit the shading and the domain wall.

Precisely, the operator $\modT_\sigma$ takes 
${\cal H}_\sigma$ to $\widehat{\cal H}_\sigma$ and vice-versa; these vector spaces were defined in section \ref{sec:dualitybc}. 
The matrix elements acting on ${\cal H}$ are given graphically by
\begin{align}
\modT_\sigma^{l} \ =\  \IsingDualityTranslationOperator \ .
\label{dualitytranslation}
\end{align}
The superscript $l$ is to note that the spins have effectively shifted half a unit cell to the left at the defect. 
The analogous operator acting on ${\widehat{\cal H}_\sigma}$ has matrix elements
\begin{align}
{\modT}_\sigma^{r}\ =\  \IsingDualityTranslationOperatorDual \ .
\label{dualitytranslationdual}
\end{align}
The superscript $r$ indicates the half-unit-cell shift to the right of the spins. 
Similarly to the transfer matrix, the pictures (\ref{dualitytranslation}) and (\ref{dualitytranslationdual}) differ only in where the Ising spins are placed. A unified notation for the two in the same fashion as (\ref{Dsigmadef}) is therefore
\begin{align}
\modT_\sigma \ =\  \IsingDualityTranslate
\label{Tsigma}
\end{align}
Unlike the single insertion of a horizontal duality defect by $\D_\sigma$, the Dehn twist is invertible (in fact unitary).
Reversing the orientation of how the full twist is glued back together gives
\begin{align}
\modT_\sigma^{-1} \ =\  \IsingDualityTranslateDagger
\end{align}
Using the defect commutation relations gives
$\modT_\sigma^{-1} \modT_\sigma \ =\  \modT_\sigma \modT_\sigma^{-1}\ =\ \mathds{1}_\sigma$ 
where
\begin{align}
\mathds{1}_\sigma \ =\  \IsingIdentitytranslate\ ,
\end{align}
is the identity operator for ${\cal H}_\sigma$ and $\widehat{\cal H}_\sigma$.

It is now straightforward to use the defect commutation relations to show that the Dehn twist commutes with the transfer matrix:
\begin{align}
 \modT_\sigma T_\sigma \ =\  T_\sigma \modT_\sigma\ .
\end{align}
However, one interesting characteristic of $\modT_\sigma$ is that it is {\em not} a product of the translation operators ${\mathcal T}_\sigma$, even at the critical point. Instead, a tedious calculation gives
\begin{align}
\modT_\sigma^2\ =\ \T_\sigma^{2L-1}\ .
\label{TT}
\end{align}
The effective length of the system is therefore $L_{\text{eff}} = L-\frac{1}{2}$ for $L$ Ising spins. This is also apparent from (\ref{Hf}) in the appendix;  there are two fermions per site, but only $2L-1$ of them appear in the fermionic version of the Hamiltonian. 
As a consequence, the $L$-dependence of the momenta is in units of $2\pi/(L-\frac12)$ instead of the usual $2\pi/L$ \cite{Grimm2003}. 

To compute the shift in momenta, we need not diagonalize
the $2^L \times 2^L$ matrix.  Instead we derive identities for products of $\modT_\sigma$ using the $F$ moves, and so constrain the momenta. The product $\modT_\sigma^2$ is easy to compute by fusing together the defect lines locally, as done without the vertical duality defect in (\ref{id2}). Here this results in
\begin{align}
\modT_\sigma^2 \ =\  \frac{1}{\sqrt{2}} \Big( \IsingIdentitytranslate + \IsingSpinFlipTranslate \Big)
\label{MicroTSquared}
\end{align}
Away from the vertical duality defect, this resembles $(\D_\sigma)^2=1+\D_\psi$. However, the presence of the vertical duality defect results in the trivalent junctions at the end points of the spin-flip defect lines. These leave $\modT_\sigma$ invertible, unlike $\D_\sigma$. 

This operator identity also can be derived in the same spirit as those for the partition function in section \ref{sec:partitionidentities}. We label
\begin{align}
\modT_{\sigma} \ =\  \OGGx\ , \qquad \mathds{1}_\sigma \ =\  \OxGG\ , \qquad \psi_\sigma \ =\ \ORGGw\ .
\end{align}
The multiplication rule for these operators is simply to stack the first one on top of the second. Thus
(\ref{MicroTSquared}) can be rederived simply via
\begin{align}
\modT_{\sigma}^2 \ =\  \OGGxsquareed \ =\  \sum_x \frac{1}{\sqrt{2}} \OxRGG  \ =\  \frac{1}{\sqrt{2}} \big( \OxGG + \ORGGw \big)
\ =\ \frac{1}{\sqrt{2}} \big( \mathds{1}_\sigma+\psi_\sigma \big)\ .
\label{Tsquared}
\end{align}
The subtleties involving the half-lattice translations and dislocation are immaterial in this method, since this relies only on the topological properties of the defect lines.

We have just one more identity to derive. As opposed to $\D_\psi^2=\mathds{1}$, the operator $\psi_\sigma$ squares to $-\mathds{1}_\sigma$:
\begin{align}
\psi_\sigma^2 \ =\  \ORGGsquaredw \ =\  \ORGGsquaredfusew \ =\ \Big[ F_{\sigma}^{\psi \sigma \psi} \Big]_{\sigma \sigma}\ORGGsquaredfuseFw \ =\  -\, \OxGG \ =\  -\mathds{1}_\sigma
\end{align}
In the first step we fuse the spin flip defects locally, which gives the identity. In the second step we used another $F$-move defined in (\ref{FSymbol}) to pull the spin flip defects past each other on the duality defect. Lastly we used the relation for the bubble diagrams in (\ref{threerules}) to remove the spin flip defects from the duality defect.  
Therefore 
\begin{align}
\modT_\sigma^4\ =\ \frac{1}{2}\big(\mathds{1}_\sigma+\psi_\sigma\big)^2\ =\ \psi_\sigma\ .
\end{align}
Thus
\begin{align}
\modT_\sigma^4 \ =\  \sqrt{2}\,\modT_\sigma^2-\mathds{1}_\sigma\ , \qquad \modT_\sigma^8\ =\  -\mathds{1}_\sigma\ ,\qquad \modT_\sigma^{16} \ =\  \mathds{1}_\sigma\ .\
\label{idX}
\end{align}

The identities (\ref{idX}) strongly constrain the allowed momenta. The relation (\ref{TT})  means that when $e^{2\pi ip_\sigma/L_{\rm eff}}$ is an eigenvalue of the translation operator $\T_\sigma$,  then $e^{4\pi i p_\sigma}$ is an eigenvalue of $\modT_\sigma^2$. Therefore  (\ref{idX}) requires 
\begin{align}
p_\sigma \ =\  h_\sigma + n, \quad \text{where} \quad n \in \mathbb{Z} \quad \text{and} \quad h_\sigma \ =\ \pm \frac{1}{16}, \pm \frac{7}{16}\ . 
\end{align}
The momentum offset given by $h_\sigma$ is one of our main results. This discrete quantity cannot change as one takes the continuum limit, so it indicates that there is a field of conformal spin exactly $\pm 1/16$ in
the Ising conformal field theory, up to a half integer. It of course has long been known that the chiral part of the spin field indeed has conformal spin 1/16, but our derivation required none of the detailed apparatus of integrability, just simple manipulations using the $F$ moves.
We will explore this and other consequences in detail in the next section.

\section{Partition functions and modular invariance} 
\label{sec:partition}

By analysing the behaviour of the partition function in the presence of horizontal and vertical duality defects, we showed in the preceding section how to extract an exact continuum quantity, the conformal spin (mod half an integer) of the chiral spin field at criticality. In this section, we show that this correspondence between lattice and continuum results can be extended substantially. By including the appropriate defect lines, we find lattice analogs of the Ising conformal field theory partition functions in all sectors. 

The method is to exploit {\em modular transformations}. The partition functions of conformal field theories on a torus exhibit remarkable mathematical structure, and often can be computed exactly. A key tool in this analysis is modular invariance, a consequence of the fact that the physics must be invariant under reparametrizations of the torus \cite{Cardy1986}. 
We show that modular transformations are not solely a continuum property: we derive the complete and exact modular transformations for the critical Ising model purely by lattice considerations. Although the Ising lattice model is special in that partition functions can be computed exactly even in the presence of defects \cite{Grimm2003}, this knowledge is not necessary for our computation. Indeed, in part~II we will derive similar results for models where it is not possible to compute partition functions on the lattice.

\subsection{Modular transformations of lattice defect lines}
\label{subsec:latticemod}

We went to great lengths in preceding sections to show precisely how to define defect lines branching, fusing, and wrapping around cycles of the torus.  Therefore in this section we do all the calculations using schematic pictures, since all have precise lattice definitions, up to ambiguities of half-lattice translations that do not affect the partition functions in the continuum limit. 
We restrict to the critical point ($u_H=u_V$ or $u=u'$ on the square lattice), so we need no shading in the presence of the duality defect. The most general defects we need consider have toroidial partition functions
\begin{align} 
Z_{ac}^b \equiv \PartitionBasisabc \ \ :\ \  N_{ac}^b \neq0\ .
\end{align}
Any other toroidal partition functions involving defect lines can be reduced to sums over these by utilising the $F$ moves described in section \ref{sec:Fmoves}, summarized in  (\ref{threerules}).

Modular transformations form a group with two generators, both of which were discussed above. One generator, conventionally labelled $\modS$, simply exchanges the two cycles of the torus. 
The other is the Dehn twist $\modT$, discussed at length in section \ref{sec:sixteenth}, with
the precise lattice definitions of the particular cases $\modT_\sigma$ and $\modT_\psi$ given by (\ref{Tsigma}) and (\ref{TpsiL}) respectively. The modular transformations act on the defect partition functions as
\begin{align}
\PartitionBasisabc \ \xrightarrow{\ \modS\ } \SPartitionBasisabc\ ,\qquad\qquad
\PartitionBasisabc \ \xrightarrow{\ \modT\ }\ \TPartitionBasisabc\ .
\label{STgen}
\end{align}
Using the $F$ moves and other transformations described above, the right-hand side of the latter can be written as a linear combination of the $Z_{ab}^c$.

We now show explicitly how the lattice modular transformations act on all the states $Z_{ac}^b$, deriving a general formula for $\modS$ and $\modT$ in the critical Ising lattice model. 
It turns out to be slightly nicer to compute $\modS \modT \modS$ instead of $\modT$; since $\modS^2=\mathds{1}$,  $\modT$ of course can be extracted.
A critical toroidal partition function with no defects (i.e.\ periodic boundary conditions around both cycles) must be invariant under both modular transformations,  since they simply reparamaterise the torus: \begin{align}
\Z \ \xrightarrow{\modS} \Z && \Z \ \xrightarrow{ \modS \modT\modS} \Z
\end{align}
When defects are present, however, they need not be. 
Wrapping a spin-flip defect around one or both cycles results in anti-periodic boundary conditions. These three types of configurations close under the action of both $\modT$ and $\modS$:
\begin{equation}
\begin{aligned}
	\xymatrix @C=12mm {
		\ZxRRw \ar[r]^{\modS}
	&	\ZRxRw &
	\ZRxRw \ar[r]^{\modS}
	&	\ZxRRw &
	\ZRRxw \ar[r]^{\modS}
	&	\ZRRxw &
	\\	\ZxRRw \ar[r]^{\modS \modT \modS}
	&	\ZxRRw &
	\ZRxRw \ar[r]^{\modS \modT \modS}
	&	\ZRRxw &
	\ZRRxw \ar[r]^{\modS \modT \modS}
	&	\ZRxRw &
	}
\end{aligned}
\end{equation}

It is a little more work to find modular transformations involving duality-twisted boundary conditions. 
Obviously,
\begin{align}
\ZxGG \ \xrightarrow{\modS} \ZGxG \quad \text{and} \quad \ZGxG \ \xrightarrow{\modS} \ZxGG\ 
\end{align}
For a duality defect wrapping around both cycles of the torus,
\begin{align}
\newcommand{\ZGGxS}{\mathbin{\rotatebox[origin=c]{90}{$\ZGGx$}}}
\ZGGx \ \xrightarrow{\modS} \ZGGxS = \frac{1}{\sqrt{2}} \Bigg(\ZGGx + \ZGGRw \Bigg)\ ,
\end{align}
where the equality follows from using an $F$ move in the middle of the figure. Another configuration with duality-twisted boundary conditions around both cycles has a spin-flip defect beginning and ending on the duality defect. The analogous computation gives
\begin{align}
\newcommand{\ZGGRwS}{\mathbin{\rotatebox[origin=c]{90}{$\ZGGRw$}}}
\ZGGRw \ \xrightarrow{\modS} \ZGGRwS = \frac{1}{\sqrt{2}} \Bigg(\ZGGx - \ZGGRw \Bigg)\ .
\end{align}
The last ones to consider correspond to a mix of boundary conditions; duality twisted in one direction and anti-periodic in the other. These are given by
\begin{align}
\newcommand{\ZRGGwS}{\mathbin{\rotatebox[origin=c]{90}{$\ZRGGw$}}}
\newcommand{\ZGRGwS}{\mathbin{\rotatebox[origin=c]{90}{$\ZGRGw$}}}
\ZRGGw \ \xrightarrow{\modS} \ZRGGwS = -\,\ZGRGw \quad \text{and} \quad \ZGRGw \ \xrightarrow{\modS} \ZGRGwS = -\,\ZRGGw
\end{align}
where the right hand side of each equality follows again from (\ref{threerules}).
Finally, the action of $\modS \modT \modS$ on the partition functions involving duality defects  
is a fun exercise to compute, yielding
\begin{align}
\ZGGx \ \xrightarrow{\modS \modT \modS}\  \ZGxG \qquad \text{and} \qquad \ZGGRw  \ \xrightarrow{\modS \modT \modS}\  -\, \ZGRGw
\end{align}
as well as
\begin{align}
\ZGxG \ \xrightarrow{\modS \modT \modS}  \frac{1}{\sqrt{2}} \Bigg(\ZGGx + \ZGGRw \Bigg) \quad \text{and} \quad \ZGRGw \ \xrightarrow{\modS \modT \modS}  \frac{1}{\sqrt{2}} \Bigg(\ZGGx - \ZGGRw \Bigg)\ .
\end{align}

We now summarize the transformations in matrix form. To do so we pick the basis given by
\begin{align}
\Bigg\{\Z\,,\,\ZxRRw\,,\, \ZRxRw\,,\, \ZRRxw\,,\, \ZGGx\,,\, \ZGGRw\,,\, \ZGxG\,,\, \ZGRGw\,,\, \ZRGGw\,,\, \ZxGG \Bigg\}
\label{latticebasis}
\end{align}
so that
\begin{align}
\modS = 
\begin{bmatrix}
1&&&\\
&&1 & \\
&1&&\\
&&&1
\end{bmatrix} \oplus
\frac{1}{\sqrt{2}}\begin{bmatrix}
1&1\\
1&-1
\end{bmatrix} \oplus
\begin{bmatrix}
&&&1\\
&&-1& \\
&-1&&\\
1&&&
\end{bmatrix}
\end{align}
and
\begin{align}
\modS \modT \modS = 
\begin{bmatrix}
1&&&\\
&1& & \\
&&&1\\
&&1&
\end{bmatrix} \oplus
\begin{bmatrix}
&&1&\\
&&&-1 \\
\frac{1}{\sqrt{2}}&\frac{1}{\sqrt{2}}&&\\
\frac{1}{\sqrt{2}}&-\frac{1}{\sqrt{2}}&&
\end{bmatrix}\oplus
\begin{bmatrix}
-1&\\
&1
\end{bmatrix}.
\end{align}
As with all the results in this paper, these are special cases of the general results described in part~II. Indeed, all the above modular transformations are summarized in a remarkably elegant fashion:
\begin{align}
\modS(Z_{ac}^b) = \sum_{\gamma} \Big[F_{a}^{bab} \Big]_{c \gamma} Z_{b\gamma}^a\qquad \text{and} \qquad \modT(Z_{ac}^b) = \sum_{\gamma} \Big[F_{b}^{aba} \Big]_{c \gamma}Z_{\gamma a }^b\ .
\end{align}
These can be found by applying the transformation to the basis states $Z_{ac}^b$ as in (\ref{STgen}), and  then using (\ref{threerules}) and the $F$-moves to simplify the resulting expression.

We conclude this subsection by showing that indeed $\modS$ and $\modT$ acting on the $Z_{ac}^b$  provide a representation of the modular group. The group is determined by two constraints: $\modS^2=1$ and $(\modS\modT)^3=1$. 
The definition of $\modS$ given by (\ref{STgen}) makes it obvious that $\modS^2=\mathds{1}$. (More formally, it is because both kinds of defects are their own conjugates: $N_{\psi \psi}^0 = N_{\sigma \sigma}^0 = 1$.)
Hence we only need to verify that $(\modS \modT)^3 = \mathds{1}$. Because 
the $\modT$ transformation defined by (\ref{STgen}) is obviously invertible, this is equivalent to $\modT \modS \modT = \modS \modT^{-1} \modS$.  The latter is a consequence of the commutativity of the diagram 
\begin{align}
	\vcenter{\xymatrix @!0 @M=0mm @R=13mm @C=26mm{
		&	{\ \  \SPartitionBasisabc \ \  }
				\ar[rr]^{\modT^{-1}}
		&&	{\ \ \TinvSPartitionBasisabc \ \  } \ar[rd]^{\modS}
	\\	{\ \ \PartitionBasisabc \ \ }\ar[ru]^{\modS} \ar[rd]^{\modT}
		&&&&	{\ \ \STinvSPartitionBasisabc \ \ } 
	\\	&{\ \ \TPartitionBasisabc \ \ }\ar[rr]^{\modS}
		&&{\ \ \STPartitionBasisabc \ \ } \ar[ru]^{\modT}				
	}}
	\label{ST3}
\end{align}
The last step along the lower path is the only non-obvious identity. It follows by moving lines around the torus as
\begin{align}
\vcenter{\xymatrix{
\STPartitionBasisabc  \ar[r]^{\modT} &\TSTPartitionBasisabc \ar@{-->}[r]& \TSTPartitionBasisabcistopic \ar@{-->}[r]& \STinvSPartitionBasisabc}}
\end{align}
The proof that (\ref{ST3}) commutes is competely independent of any basis choice -- it only requires that the defect lines can move without changing the value of the partition function.

\subsection{The Ising CFT partition functions}
\label{subsec:IsingCFT}

We briefly review the Ising CFT and its partition functions on the torus \cite{BPZ:CFT:84,Cardy1986,Ginsparg}. 
The conformal field theory describing the continuum limit of the critical lattice Ising model has long been understood. More recently, it has been rigorously proven that the critical Ising lattice model does converge to this CFT upon taking the appropriate limit 
 \cite{Smirnov2012}.

Conformal symmetry in two dimensions is infinite dimensional, and so is quite powerful, often allowing the partition function to be computed exactly. For the most part, this is an exercise in the representation theory of the Virasoro algebra, the algebra of the generators of conformal symmetry. These generators split into left- and right-moving sectors, with those in each sector obeying the Virasoro algebra. 
Irreducible representations of the Virasoro algebra are called (chiral) Verma modules; every state in the Hilbert space of the 1+1 dimensional quantum theory belongs to some module which is a product of left and right Verma modules.
Each Verma module is labeled by a ``highest weight'' state, the state of lowest energy in that sector, created from the vacuum by an operator called a primary field. The infinite tower of states belonging to each Verma module can be found by acting with the generators 
of conformal symmetry. 
The Ising CFT is minimal, meaning there are only a finite number of Verma modules, and the three primary fields are the fermion $\psi$ of scaling dimension $h_\psi=1/2$, the chiral spin field $\sigma$ of dimension $h_\sigma=1/16$, as well as the identity of dimension $h_{\mathds{1}}=0$. 
A remarkable fact is that after an overall rescaling, 
each state in the Verma module of a primary field $\mathcal{A}$ has energy on the circle determined up to a positive integer $N$, namely $E = N + h_\mathcal{A}-c/24$, where $c$ is the central charge of the theory \cite{Blote1986,Affleck1986,Cardy1986}.
The exercise in representation theory is then to find the multiplicities for each $N$. 
A convenient way to package this information is through the Virasoro character for each module, given by summing $q^E$ over all states in the module, for some parameter $q$.
For the Ising model, the central charge $c=1/2$, and the chiral characters are \cite{Cardy1986,Ginsparg}
\begin{subequations}\begin{align}
	\chi_{\mathds{1}} &= \frac{1}{\eta(q)} \sum_{k \in \mathbb{Z}}\left( q^{(24k+1)^2/48}-q^{(24k+7)^2/48} \right) ,\\
	\chi_{\sigma} &= \frac{1}{\eta(q)} \sum_{k \in \mathbb{Z}}\left( q^{(24k+2)^2/48}-q^{(24k+14)^2/48} \right) ,\\
	\chi_{\psi} &=  \frac{1}{\eta(q)} \sum_{k \in \mathbb{Z}} \left(q^{(24k+5)^2/48}-q^{(24k+11)^2/48}\right) ,
\end{align}\end{subequations}
where $ \eta(q) = q^{\frac{1}{24}} \prod_{k = 1}^{\infty} (1 - q^k)$.
The $k=0$ terms in the summands indeed give the correct $h-c/24$ in each sector.
Because of the minus signs, it is not obvious that each term is $q^E$ times a non-negative integer coefficient, but expanding out $1/\eta(q)$ in powers of $q$, one finds it is indeed so. These all can be written in terms of Jacobi elliptic theta functions%
	\footnote{or directly as Weber modular functions: $\mathfrak{f}(\tau) = \chi_{\mathds{1}}+\chi_\psi$, $\mathfrak{f}_2(\tau) = \sqrt{2}\chi_\sigma$, $\mathfrak{f}_1(\tau) = \chi_{\mathds{1}}-\chi_\psi$.}; see e.g.~\cite{Ginsparg}.

The partition function of the Ising model with various boundary conditions can be built from these characters. We consider here a torus, which need not be square or even rectangular. In general, a torus is specified by two complex numbers $c_1$ and $c_2$ and identifying all points in the complex plane through the equivalence relations $z \sim z+ c_1$ and $z \sim z +c_2 $. Conformal symmetry allows us to rescale and rotate the torus so that one of the periods is unit norm and on the real axis. Each torus can thus be labelled by the modular parameter $\tau =c_2/ c_1$, the ratio of the two periods. 
On a torus with periodic boundary conditions around both cycles,  the left- and right- Virasoro algebras remain independent, so the states belong to Verma modules of both chiralities. 
The partition functions are given by sesquilinear forms defined on the characters, 
\begin{align}
	{\cal Z}_{M}(\tau) = \sum_{\alpha \beta}\bar{ \chi}_\alpha(\bar{q}) M_{\alpha \beta} \chi_{\beta}(q).
\label{ZM}
\end{align}
where $q = e^{2 \pi i \tau}$ is called the nome.
The $3\times3$ matrix $M$ thus completely specifies the partition function on the torus, which can be written in the form
\begin{align}
{\cal Z}_{M}(\tau) =\bar{v} Mv,\qquad v(\tau) &= \begin{pmatrix} \chi_{\mathds{1}} \\ \chi_\sigma \\ \chi_\psi \end{pmatrix} ,
&	\bar{v}(\bar{\tau}) &= \begin{pmatrix} \bar\chi_{\mathds{1}} & \bar\chi_\sigma & \bar\chi_\psi \end{pmatrix}\ .
\end{align}
For the partition function to be real as in the lattice model, we require $M=M^\dagger$. 

The characters transform nicely under the modular group. Since the modular $\modS$ transformation interchanges the two cycles, it corresponds to taking $\tau \rightarrow -1/\tau$. The Dehn twist implemented by the modular $\modT$ transformations corresponds to $\tau \rightarrow \tau +1$. They are linear transformations on the characters, and so take the form \cite{Cardy1986}
\begin{align}
{\cal Z}_{M}(-1/\tau) = \bar{v} s^{\dagger} M s v\ , \qquad\quad {\cal Z}_{M}(\tau+1)= \bar{v}t^{\dagger} M t v, 
\end{align}
where for Ising the matrices $s$ and $t$ are
\begin{align}
	s = \frac{1}{2} \begin{pmatrix}1&\sqrt{2}&1\\ \sqrt{2}&0&-\sqrt{2}\\1&-\sqrt{2}&1 \end{pmatrix},\qquad\quad
	t = \zeta_{48}^{-1} \begin{pmatrix}1&&\\ &\zeta_{16}&\\&& -1\end{pmatrix}, 
\label{stcft}	
\end{align}
where $\zeta_r = e^{2\pi i/r}$.
Using these relations partition functions with the desired modular transformation properties can be built quite easily.

\subsection{Identifying lattice partition functions with their continuum counterparts}


It is natural to expect that the topological defects we have defined in this paper will in the continuum limit turn into the topological defects studied in the CFT \cite{Frohlich:2006}.  The purpose of this section is to make this notion precise. Namely, for any configuration of topological defects in the critical lattice model, we identify the corresponding CFT partition function.

To do this, we utilise 
the modular transformation properties derived for the lattice and continuum in sections \ref{subsec:latticemod} and \ref{subsec:IsingCFT} respectively.
The algebra that the defect lines satisfy strongly constrains the kinds of partition functions appearing in the continuum limit. In particular, the modular transformations in the continuum limit must preserve the structure of the modular transformations on the lattice, i.e., the map from lattice to continuum partition functions must be equivariant. Since in both cases they are linear, these manipulations are not difficult.

\comment{

We can view the continuum limit as  a map from the lattice model boundary conditions to a particular matrix $M$ which describes the sesquilinear form on the character. Self consistency of the theory requires that this map commutes with the group action of the modular transformations:
\begin{align}
	\xymatrix @C=20mm {
		Z_{\text{lattice}} \ar[r]^{\phi} \ar[d]_{g}
	&	{\cal Z}_{\text{continuum}} \ar[d]_{g}
	\\	Z_{\text{lattice}} \ar[r]^{\phi}
	&	{\cal Z}_{\text{continuum}}
	}
\end{align}
Where we have defined the map from lattice to continuum as $\phi$ and $g$ is a modular transformation. In matrix elements, this map requires $Z(g.Z_{ac}^b) = g. \phi(Z_{ac}^b)$. Since all elements of the modular group are given by words in $\modS$ and $\modT$ it means that we can solve this constraint for $\modS$ and $\modT$ then we have solved it for the entire group of modular transformations.

 The fact that there are as many character by-liners as their are lattice partition functions may inspire one to believe that we can find lattice versions of all character. Indeed this generically won't be true, see \cite{PARTII}.
}

It is useful to start by counting all the inequivalent partition functions. The 10 lattice partition functions listed in (\ref{latticebasis}) form a closed set under modular transformations. However, using a duality defect  and the $F$-moves, in (\ref{z-to-zdual}) we found a relation between the partition function without defects and a linear combination of dual partition functions. At the critical point, the Ising model on the square lattice is self-dual, so the relation \ref{z-to-zdual} reduces the number of independent lattice partition functions on the torus to nine. In the continuum expression (\ref{ZM}) there are also nine different real partition functions arising from the independent numbers in $M=M^\dagger$. 
It is straightforward to check that all nine can be found by applying $\modS$ and $\modT$ to the three partition functions with twisted boundary conditions:
\begin{align}
\Z\ , \quad \ZxGG\ ,\quad \ZxRRw.
\end{align}
Hence, if we can find three matrices $M_\mathds{1}$, $M_\sigma$ and $M_\psi$ such that the corresponding ${\cal Z}_M$ transforms under modular transformations identically as these three lattice lattice partition functions, then the remaining identifications will follow automatically. 

The first thing to check is for modular invariants, combinations that are invariant under both $\modS$ and $\modT$. Of the 10 lattice partition functions in (\ref{latticebasis}), there only two modular invariants, and (\ref{z-to-zdual}) says they are same at the critical point: 
\begin{align}
\Z\ =\ \frac{1}{2} \Bigg(\, \Z + \ZRxRw + \ZxRRw + \ZRRxw\, \Bigg).
\end{align}
There is indeed only one modular-invariant CFT partition function, having $M_\mathds{1}$ the identity matrix.
Thus we identify
\begin{align}
\Z \longrightarrow{\cal Z}_{M_\mathds{1}}
=\bar{v}\begin{bmatrix}
1&& \\
&1&\\
&&1
\end{bmatrix} v
\end{align}
in the continuum limit.
Thus, not surprisingly, the lattice partition function with no defects present corresponds to the well-known Ising modular-invariant partition function.

Others are not quite as obvious. Consider now a single vertical duality defect line. As we showed at great length in section \ref{sec:sixteenth}, the eigenvalues of the $\modT$ operation here are $\pm e^{\pm 2\pi i/16}$.  
As seen in (\ref{stcft}), the only way of picking up such a factor is from $\chi_\sigma$ or $\bar\chi_\sigma$. 
This guarantees that the matrix $M_{\sigma}$ must obey
\begin{align}
\ZxGG \longrightarrow {\cal Z}_{M_\sigma}=
\bar{v}\begin{bmatrix}
&a & \\
a^*&&b\\
&b^*&
\end{bmatrix}v
\end{align}
for some complex numbers $a$ and $b$. The partition function for a single horizontal line also has a simple form: the corresponding matrix $M$ is diagonal. The reason is that a horizontal defect commutes with the transfer matrix locally, so it presumably commutes with the full energy-momentum tensor\footnote{incidentally, the fourth object in this paper conventionally labelled $T$} generating conformal transformations. Since the Verma module is built by acting on the highest weight state by these generators, the operator creating a horizontal defect acts identically on all states in a given module. Therefore the continuum partition function for a single horizontal duality defect is of the form
\begin{align}
\ZGxG \longrightarrow\ 
\bar{v}\begin{bmatrix}
\alpha& & \\
&\beta&\\
&&\gamma
\end{bmatrix}v
\end{align}
for real numbers $\alpha, \beta$ and $\gamma$.  Horizontal and vertical duality defects are related by the $\modS$ modular transformation, and the resulting consistency condition is
\begin{align}
\modS \Bigg(\ZGxG \Bigg) = \ZxGG\ \longrightarrow\  
s^{\dagger}\begin{bmatrix}
\alpha& & \\
&\beta&\\
&&\gamma
\end{bmatrix}s
 = 
 \begin{bmatrix}
&a & \\
a^*&&b\\
&b^*&
\end{bmatrix}\ .
\end{align}
Carrying through the algebra gives $\beta=0$, $\gamma = -\alpha$, and $a =b= -\gamma/\sqrt{2}$.
Thus both partition functions are determined up to an overall constant $\alpha_\sigma$:
\begin{align}
\ZxGG  \longrightarrow  \frac{\alpha_\sigma}{\sqrt{2}}\, \bar{v}
 \begin{bmatrix}
&1 & \\
1&&1\\
&1&
\end{bmatrix}v\ ,
\qquad\qquad
\ZGxG  \longrightarrow 
\alpha_\sigma\, 
\bar{v}\begin{bmatrix}
1& & \\
&&\\
&&-1
\end{bmatrix}v\ .
\label{vduality}
\end{align}
The partition function for a vertical spin-flip defect line is found in a similar fashion, giving 
\begin{align}
\ZxRRw  \longrightarrow   \alpha_\psi \,
\bar{v}\begin{bmatrix}
& &1 \\
&1&\\
1&&
\end{bmatrix}v
\ , \qquad\qquad
\ZRxRw  \longrightarrow   \alpha_\psi\,
\bar{v}\begin{bmatrix}
1& & \\
&-1&\\
&&1
\end{bmatrix}v
\label{vspinflip}
\end{align}
for some constant $\alpha_\psi$.

To determine the overall constants requires one additional input. The operators creating horizontal defects obey the fusion algebra, so the partition function ${\cal Z}_{a,b}$  in the presence of two horizontal defects of types $a$ and $b$ must obey
\begin{align}
{\cal Z}_{a,b}=\sum_c N_{ab}^c {\cal Z}_{M_c}\ .
\label{ZN}
\end{align}
This is illustrated for two duality defects in (\ref{id2}). Now consider the limit where torus is rectangular and much longer in the vertical direction, so $\tau$ is imaginary and large so that $q\to 0$.  When two horizontal defects are taken very far apart in this limit, the partition function must therefore factorise into
${\cal Z}_{a,b} \to {\cal Z}_{M_a} {\cal Z}_{M_b}\ .$
Comparing this with (\ref{ZN}) means that the overall constants in the partition functions must also obey the fusion algebra:
\begin{align}
\alpha_{a}\alpha_b = \sum_c  N_{ab}^c\alpha_c\ .
\end{align}
This yields, $\alpha_\mathds{1}=1$  and because $D_\sigma^2=1 + D_\psi$ and $D_\psi^2=1$, this yields $\alpha_\sigma=\pm \sqrt{2}$ and $\alpha_\psi=1$. 

This is enough to find the continuum partition functions for all nine independent defect configurations.
The three basic ones are
\begin{align}
\Z \longrightarrow 
\bar{v}\begin{bmatrix}
1&& \\
&1&\\
&&1
\end{bmatrix}v, \quad
\ZxGG \longrightarrow \pm 
\bar{v} \begin{bmatrix}
&1 & \\
1&&1\\
&1&
\end{bmatrix}v, \quad
\ZxRRw \longrightarrow  
\bar{v}\begin{bmatrix}
& &1 \\
&1&\\
1&&
\end{bmatrix}v.
\label{threebasic}
\end{align}
The partition functions with horizontal defects alone are given in (\ref{vduality},\ref{vspinflip}). 
The Dehn twist $\modT$  gives then
\begin{align}
	&\ZRRxw = \mathbin{\rotatebox[origin=c]{90}{$\ZRRxw$}} \longrightarrow \bar{v} \left[\begin{array}{@{}c@{\ }c@{\ }c@{}} &&-1 \\ &1& \\ -1&& \end{array}\right] v,\quad
	&\ZGGx \longrightarrow \bar{v} \left[\begin{array}{@{}c@{}c@{}c@{}} &\zeta_{16}& \\ \zeta_{16}^{-1}&&\zeta_{16}^7 \\ &\zeta_{16}^9& \end{array}\right] v
\end{align}
where we chose the $+$ sign in $\alpha_\sigma$. 
Lastly, we have three non-trivial partition functions which involve combinations of $S$ and $T$:
\begin{align}
&\ZRGGw \longrightarrow \bar{v} \begin{bmatrix} &i& \\ -i&&-i \\ &i& \end{bmatrix} v,\quad
&\ZGGRw &\longrightarrow \bar{v} \left[\begin{array}{@{}c@{}c@{}c@{}} &\zeta_{16}^{-3}& \\ \zeta_{16}^{3}&&\zeta_{16}^{-5} \\ &\zeta_{16}^{5}& \end{array}\right] v, \quad
&\ZGRGw \longrightarrow  \sqrt{2}\bar{v} \begin{bmatrix} &&i \\ && \\ -i&& \end{bmatrix} v.
\label{lastthree}
\end{align}

This gives all ten partition functions, and we have several nice consistency checks between CFT and lattice. The constraint \ref{zzdualagain} is immediate:
\begin{align}
	\Z - \ZRxRw  \;=\;  2 \bar\chi_\sigma \chi_\sigma  \;=\;  \ZxRRw + \ZRRxw \;.
\end{align}
By explicitly plugging in $q$ real into the first of (\ref{lastthree}) yields 
\begin{align}
\ZRGGw\ =\ 0\ \hbox{}.
\end{align}
just as derived from an $F$ move in (\ref{ZDpsi2}) and from the double degeneracy in (\ref{ZDpsi}). From the same derivations, every term in the partition function of a single vertical duality defect must appear with coefficient that is a multiple of two. This is indeed apparent for $q$ real in the middle relation in (\ref{threebasic}). 

%

\section{Duality on other surfaces}
\label{sec:highergenus}
In this section we briefly describe how to perform Kramers-Wannier duality in the Ising model on other surfaces. Our approach makes this calculation very staightforward. We begin with the disc and annulus, and finish off with a pair of pants. 
These three objects can be used to cover any 2-manifold, and we discuss how the corresponding duality works in this more general setting.

\subsection{Disc and annulus}
We begin with the Ising model on a disc. 
One way to construct a lattice that is topologically a disc is by starting with a square lattice, choosing a closed path ${\cal C}$ along the lattice and discarding all edges, and vertices on the exterior of this loop. 
The spins live on the sites of the lattice, and we define a vector space ${\cal B}$ whose basis states are the configurations of spins along the boundary, e.g. $\ket{B}  = \ket{l_{i_1} l_{i_2} \hdots l_{i_L}}$.

We first describe the partition function with a given fixed boundary condition $\ket{B}$.
The weight per configuration is given by
\begin{align}
e^{-\beta H(\{ h_i \})} = \prod_{p} \BWx{\alpha_p}{\beta_p}{\delta_p}{\gamma_p} \times \prod_{v \in D\backslash \partial D} d_v \times \prod_{v \in \partial D} \sqrt{d_v}
\label{discweight}
\end{align}
where $D$ is comprised of all sites in the disc, and $v \in \partial D$ all those on the boundary. 
The disc partition function is
\begin{align}
Z(D, \ket{B}) = \sum_{ \{ h_i\} \  : \  h_{i \in \partial D} =  l_i} e^{-\beta H( \{ h_i \}) }\ =\  \ZDisc \Bigg|_{\ket{B}}
\end{align}
where the schematic picture on the right represents the partition function defined on the left.
The partition function is complex linear in the boundary condition,
\begin{align}
\ZDisc \Bigg|_{\alpha \ket{B_1}+\beta \ket{B_2}}  =\ \alpha \times \ZDisc \Bigg|_{ \ket{B_1}}+\beta \times \ZDisc \Bigg|_{\ket{B_2}}.
\end{align} 
Hence understanding how the duality acts on a disc with one fixed boundary condition allows us to understand it for arbitrary boundary conditions. 

The technique is the same as what we used for the torus, but much simpler since there are no nontrivial cycles. 
We insert a closed loop in the bulk and push it out to the boundary using the defect commutation relations and eventually find the partition function on the dual lattice with dual boundary conditions found by acting with $\D_\sigma$ on the original boundary condition, 
\begin{align}
\ZDisc \Bigg|_{\ket{B}} =  \frac{1}{\sqrt{2}}\, \ZDiscGLoop \Bigg|_{\ket{B}} =\ \frac{1}{\sqrt{2}}\, \ZDiscGDual \Bigg|_{\ket{B}}  = \frac{1}{\sqrt{2}} \, \ZDiscG \Bigg|_{\D_\sigma \ket{B}}\ .
\label{Bdual}
\end{align}
If $|B\rangle$ yields a conformal boundary condition in the continuum limit (i.e.\ conformal invariance is preserved in the presence of the boundary), then that given by $\D_\sigma|B\rangle$ must also be conformal. The factor of $\sqrt{2}$ in (\ref{freefixed}) is the relative non-integer `ground-state degeneracy' familiar in conformal field theory \cite{AffleckLudwig}; its log is the difference in the ground-state entropies coming from the two boundary conditions. At the critical point it says that the partition function with free boundary conditions on a given lattice on the disc is $\sqrt{2}$ times that on the dual lattice with fixed boundary conditions. It arises very directly here as a consequence of the fact that the non-zero eigenvalues of $\D_\sigma$ have magnitude $\sqrt{2}$. 

When the boundary conditions are fixed-up $\ket{u}=\ket{\cdots 000\cdots}$ or fixed-down $|d\rangle=\ket{\cdots111\cdots}$, then acting with duality gives
\begin{align}
\D_\sigma \ket{u}=
\D_\sigma \ket{d} = \Ket{ \cdots \left(\tfrac{ \ket{0} +\ket{1}}{\sqrt{2}}\right)
 \left(\tfrac{ \ket{0} +\ket{1}}{\sqrt{2}}\right) \left(\tfrac{ \ket{0} +\ket{1}}{\sqrt{2}}\right)\cdots }
\equiv\ket{\rm free} \ .
\label{freefixed}
\end{align}
This free boundary condition is the equal-amplitude sum over all possible boundary spin configurations. Thus the partition function with fixed-up or fixed-down is equal to that with the dual couplings and free boundary conditions, up to the $\sqrt{2}$ in \eqref{Bdual}.
As we have taken pains to emphasise, the operator $\D_\sigma$ is not invertible. Thus the duality transform (\ref{Bdual}) with $\ket{B}=\ket{\rm free}$ does not return $\ket{+}$ or $\ket{-}$. Rather, it gives the sum of the two: 
\begin{align}
\mathcal{D}_\sigma \ket{\rm free} = \D_\sigma^2 \ket{u} = \ket{u}\, +\, \D_\psi \ket{u}\ =\ \ket{u}\, +\, \ket{d}\ .
\end{align}
The ground-state boundary entropy for $\ket{B}=\ket{u}+\ket{d}$ is indeed $\ln2$ greater than that for $\ket{u}$. 

The extension to an annulus is straightforward. We just choose two paths on the square lattice, one enclosing the other, and to the spins along each path we associate a boundary condition $\ket{B_1}$ and $\ket{B_2}$. The analog of the duality transformation (\ref{Bdual}) for the annulus was essentially already done in (\ref{nucleate1}, \ref{nucleate2}). We nucleate a duality defect loop inside the annulus and then do an $F$ move to fuse together opposite ends of the loop. The result is two duality defects wrapped around the cycle, with a linear combination of a spin flip defect and an identity defect connecting them. Fusing one duality defect with each boundary changes the two boundary conditions gives
\begin{align}
\ZAnnulus = \frac{1}{\sqrt2}\left(\  \ZAnnulusDual + \ZAnnulusDualPsi\ \right).
\end{align}
In the second term, a spin-flip defect then stretches from the inside to the outside.
If $\ket{B_1}$ and $\ket{B_2}$ are both fixed initially, then the resulting inside and outside boundary conditions are both free. The spin-flip defect stretching across does not affect the free boundary conditions, since $\sigma^x(\ket{0}+\ket{1})=(\ket{0}+\ket{1})$.

\subsection{Manifolds }
\begin{figure}[h] \centerline{%
 \xymatrix @!0 @M=2mm @R=22mm @C=38mm{
 (a)&(b)&(c)&(d) \\
 \Grapha &\Graphb & \Graphc & \Graphd
 }}
\caption{%
	In part (a) we show a graph and label some of its vertices, edges, and faces. In part (b) we show its dual graph and label some of its dual edges, dual vertices and dual faces. Part (c) depicts how one constructs the dual graph. Duality maps vertices to faces and vice verse, and edges to edges: $f \leftrightarrow \widehat{v}$, $e \leftrightarrow \widehat{e}$ and $v \leftrightarrow \widehat{f}$.	 Part (d) shows how one can parse the space into quadrilaterals that we use to depict the Boltzmann weights. 
	}
	\label{graph}
\end{figure}
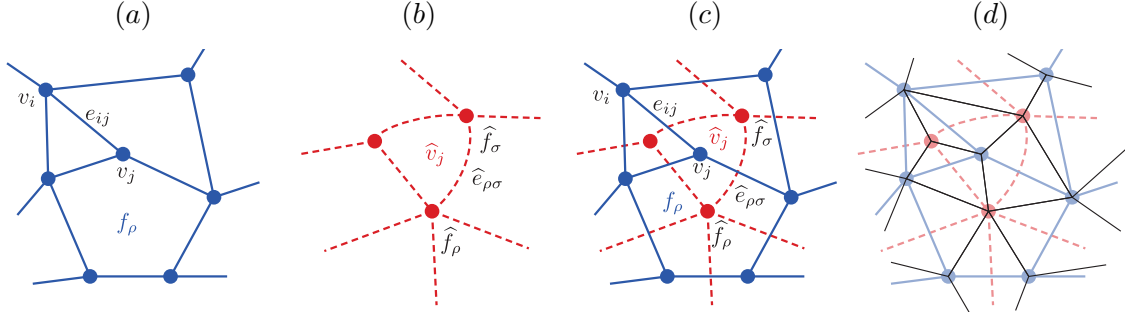

Here we use the duality defect commutation relations to study Kramers-Wannier duality of the Ising model on graphs embedded in orientable 2-manifolds. Procedures for defining integrable models on general manifolds have been developed in e.g.\ \cite{Mercat2001} and \cite{Bobenko2002}.
The basic idea for constructing the lattice is to take a graph embedded in the manifold, and overlay it with its geometric dual. The edges of the embedded graph and its geometric dual then naturally create a skeleton for the quadrilaterals; see Fig.~\ref{graph}d. 
As above, we then can define the Boltzmann weights for the spins on these quadrilaterals. Since the defect commutation relations are local, it is then simple to define defect lines as seams between the quadriaterals as well. Then we use the pants decomposition theorem to dissect the manifold into manageable pieces, apply the duality to each pair of pants, and then glue them back together. As with the torus, everything can be done by using the schematic pictures, since we have gone to great pains to show how to justify them locally.

We begin with a compact, connected, orientable 2-manifold $\Sigma$ with no boundary components.
Following \cite{Mercat2001} we choose a cellular decomposition of the manifold $\Sigma$. One can think of this as a locally planar embedding of a simple 2-connected graph $G$ in $\Sigma$; both \cite{mohar2001} and \cite{gross1987} give excellent expositions of the graph theory needed.
The decomposition is a set of vertices $V$, edges $E$, and faces $F$ which we collectively denote $\Gamma$. 
The sets $V$, $E$ and $F$ are disjoint, and chosen so that $V \cup E \cup F = \Sigma$.
The collection of edges and vertices define a graph embedded in the manifold $\Sigma$. 
Each face $F$ is homeomorphic to the disc; see Fig.~\ref{graph}a. 
The dual decomposition is then easy to find, as in  
Fig.~\ref{graph}abc. 
To each face $f_\rho \in F$ we define a dual vertex $\widehat{f}_\rho \in \widehat{V} \subset F$. 
To every pair of faces $f_\rho ,f_\sigma \in F$ that share an edge $e \in E$ we define a dual edge $\widehat{e} \in \widehat{E}$ that connects the vertices $\widehat{f}_\rho$ and $\widehat{f}_\sigma$ and intersects $e$ transversally. 
To each vertex $v_i \in V$ we define a dual face $\widehat{v}_i \in \widehat{F}$ whose boundary coincides with the $\widehat{e}$ that intersect the edges $e \in E$ connected to $v_i$.
This defines the dual decomposition $\widehat{\Gamma}$. 
Here we use unoriented edges so that $\widehat{\widehat{\Gamma}} = \Gamma$. 

The Ising model is defined by taking spins $h_i = 0,1$ to reside on the vertices $v_i \in V$, or for the dual model, on $\widehat{v}_\rho\in \widehat{V}$. Interactions occur either between every two spins that share an edge $e \in E$, or $\widehat{e}\in \widehat{E}$. 
These interactions are parametrized by a coupling $u_{e}$ and $u_{\widehat{e}}$ respectively for each edge. 
Each quadrilateral is comprised of the four vertices who lie on the endpoints of a given $e$ and its dual edge $\widehat{e}$; see Fig.~\ref{graph}d. 
The Boltzmann weight is defined on the quadrilateral as,
\begin{align}
	&\BWGraph  =\;
	\begin{cases}		
		\cos{u_{e}}	&	h_i=h_j , \\
		\sin{u_{e}}	&	h_i \ne h_j, 
		\end{cases}
		\quad \qquad
  \BWGraphDual  =\;
	\begin{cases}		
		\cos\left({\frac\pi4-{u}_{\widehat{e}}}\right)	&	\widehat{h}_\rho=\widehat{h}_\sigma , \\
		\sin\left({\frac\pi4- {u}_{\widehat{e}}}\right)	&	\widehat{h}_\rho\ne\widehat{h}_\sigma .
		\end{cases}
\end{align}
The partition function is given by the sum over all heights of the product over all Boltzmann weights multiplied by the weight per site $d_j$.

We now describe how to perform Kramers-Wannier duality on the Ising partition function defined by the spectral parameters $u_{e}$ and the graph defined through $\Gamma$.
The duality works in much the same way as it did for the torus, disc, and annulus. 
We insert a small duality defect loop around one Boltzmann weight, just as in  (\ref{eq:IsingClosedLoop}). We then stretch it across the entire system and perform various manipulations using $F$ moves, leaving us with linear relations between the partition function and its dual.
For this to work, as before we must have ${u}_{\widehat{e}}=u_e$.

\begin{figure}[h] \centerline{%
 \xymatrix @!0 @M=4mm  @C=58mm{
 \GraphCuta \ar[r]&\GraphCutb
 }}
\caption{%
	We cut the graph by choosing a closed path on the vertices in $V \cup \widehat{V}$ that does not intersect $E\cup \widehat{E}$, e.g.\ runs along the quadrilaterals. This forces every other vertex along the path to be on the dual lattice.
	}
	\label{GraphCut}
\end{figure}
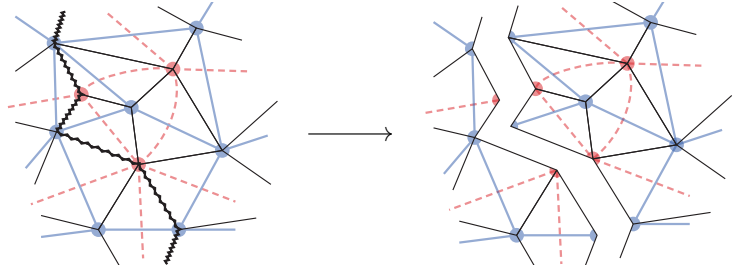

The process of dragging the duality defect across the entire manifold is most easily done by first dissecting the manifold into more manageable parts, performing the duality on each part and gluing it back together. 
Any orientable, compact, connected 2-manifold admits a pants decomposition~\cite{Hatcher1980}. 
A pants decomposition is a way of splitting the manifold into disjoint unions of pants -- two-manifolds which are homeomorphic to the three-punctured sphere.
We choose a pants decomposition ${\cal P}$ of $\Sigma$ 
such that for each $p \in {\cal P}$ the boundary $\partial p$ does not intersect any of the edges in $E$ or $\widehat{E}$. 
As illustrated in Fig.~\ref{GraphCut},  $\partial p$ then goes through the vertices, and we can unambiguously identify each edge in $E$ and $\widehat{E}$ as being on one side of the cut or the other. 
Any pants decomposition of $\Sigma$ is homotopic to one of this form.
We also choose to distribute the vertex weight per site equally across the boundary, i.e., we assign $\sqrt{d_v}$ for each boundary site and $d_v$ for all others as usual.

Each pair of pants has three boundary components and each boundary component has a fixed spin configuration. The partition function on it is then
\begin{align}
 Z(p, \ket{\alpha}, \ket{\beta},\ket{\gamma}) \equiv \ZPants.
\end{align}
Here we have denoted the fixed spin configurations as $\ket{\alpha}, \ket{\beta}$, and $\ket{\gamma}$.
The pants partition function is complex linear in its three arguments: $Z(p,\alpha \ket{\alpha} + \alpha' \ket{\alpha'},*,*) = \alpha Z(p,\ket{\alpha},*,*) +\alpha' Z(p, \ket{\alpha'},*,*)$.
 
Two pants partition functions are glued together by identifying spins on one boundary of each partition function and then summing over all spin configurations along the (former) boundary. 
For example, the partition function on the punctured torus with boundary condition $\ket{\gamma}$ at the puncture can be found from $Z(p,\ket{\alpha},\ket{\beta},\ket{\gamma})$ by identifying and summing over $\ket{\alpha}$ and $\ket{\beta}$:
\begin{align}
Z(T^2, \gamma) = \sum_{\alpha} Z(p, \ket{\alpha}, \ket{\beta},\ket{\gamma}) \delta_{\alpha \beta} = \sum_\alpha \ZPuncturedTorus
\end{align}
The toroidal partition function then can be found by gluing in the disc partition function, i.e.\ identifying its boundary spins with those of this punctured torus, and then summing over all spins. 
An obvious generalization allows defect lines to terminate at the punctures as well. 
These can be glued together in a similar fashion when the terminating defect lines are of the same type and meet at the same point. 
Fusion constraints dictate that the defect lines terminating at the three punctures must always be able to fuse to the identity.

The pants with the various defect lines terminating at the boundary span the entire configuration space. 
Starting with a pair of pants with no duality defect lines present, 
we nucleate a duality defect, and manipulate it like we did on the torus.
Precisely, we wrap it around the pair of pants and fuse it with itself and then move it behind the pants:
\begin{align}
\ZPants = \frac{1}{\sqrt{2}} \ZPantsa =  \frac{1}{\sqrt{2}} \ZPantsb = \frac{1}{2} \sum_{a = \mathds{1},\psi}\  \ZPantsc =\frac{1}{2} \sum_{a = \mathds{1},\psi}\  \ZPantsd
\end{align}
For clarity, we omit the shading, but all these manipulations apply both on and off the critical point. Next we fuse together the duality defect lines near the spin-flip defect, use the $F$-move (\ref{FSymbol}), and then remove a bubble using (\ref{threerules}):
\begin{align}
\frac{1}{2} \sum_{a = \mathds{1},\psi}\  \ZPantsd = \frac{1}{4} \sum_{a,b,c = \mathds{1},\psi}\  \ZPantse = 2^{-\frac32} \sum_{a,b,c = \mathds{1},\psi}N_{ab}^c\  \ZPantsf
\end{align}
We are left with a sum over configurations involving $\mathds{1}$ and $\psi$ defect lines terminating at duality defects on the boundary.
Explictly, the relation between the partition function and its dual is given by:
\begin{align}
Z(p,\ket{\alpha},\ket{\beta},\ket{\gamma}) = 2^{-\frac32}\sum_{a,b,c = \mathds{1}, \psi} N_{ab}^c Z^d(p,\D_\sigma(a)\ket{\alpha},\,\D_\sigma(b)\ket{\beta},\,\D_\sigma(c)\ket{\gamma})
\label{pantsduality}
\end{align}
where  $\D_{\sigma}(\mathds{1})=\D_\sigma$ is the standard duality defect creation operator, while $\D_\sigma(\psi)$ creates a duality defect with a spin-flip defect emanating from it.
The superscript in $Z^d$ is a reminder that it is the partition function for the dual lattice, while $Z$ is on the original. 

It is now easy to sew the pants together to obtain any manifold desired. 
When sewing two boundaries together, one can directly identify spins across the boundary, with the appropriate vertex weights inserted, then fuse the duality defects together. Equivalently, one can act on the boundary with the duality defects, then sew the boundaries together again by directly identifying each spin configuration in the sum.
As an example, we sew two pants together with the former prescription. 
Using (\ref{pantsduality}) and then doing our (we hope by now) standard manipulations gives
\begin{align}
\PantsGluea =\frac{1}{2^3} \PantsGlueb =\frac{\sqrt{2}}{2^3}\sum_{\begin{smallmatrix} a,b,c,\\a', b',c', \\ x,y \end{smallmatrix}} N^c_{xy} \delta_{c,c'}N^{c'}_{xy} \PantsGluec\ .
\end{align}

With these building blocks we can construct the duality on any oriented, compact, 2-manifold. 
Allowing for arbitrary initial defect line configurations is straightforward, as is allowing for a boundary.
To insert a boundary, we just take one pair of pants to terminate along the boundary of interest and discard the vertices, edges, and faces on one side of the boundary. The vertex weight per site associated to the boundary is given by $\sqrt{d_{h_i}}$ for each $ i \in \partial p$, as on the disc (\ref{discweight}). 
We suspect that similar methods can be used to describe dualities on non-orientable manifolds as well.


\section{Conclusion}

We have examined in considerable detail topological defects in the Ising lattice model, showing how their branching and fusing can be characterized by a set of very precise rules. A key ingredient in our construction was the connection of the microscopic definition of the model to the macroscopic behaviour of the defects. In particular, we showed that the rules used to define the microscopic degrees of freedom and those governing branching and fusing of defects were one and the same. These rules go by the name of a fusion category, and much is known about them both from the physics point of view and the mathematical one. Indeed, integrable lattice models such as the Ising model played a fundamental role in understanding them at the beginning of the story, when knot and link invariants were first being derived \cite{Wadati1989}. We believe the time is right for revisiting this connection. There are many other fusion categories known, and we have taken pains here to exhibit much of the general structure so that the generalisations given in part~II appear natural.

Our study of defects make it straightforward to find twisted boundary conditions, where a modified form of translation invariance still holds. At the critical point, these result in conformal boundary conditions, again of fundamental importance in the continuum \cite{Cardy1989}. The calculation in a two-dimensional classical model gives a simple and direct way to find the analogous boundary conditions in quantum spin chains, much easier that the traditional brute force methods. Off the critical point, it makes it easy to find an unpaired Majorana zero mode at the domain wall present with duality-twisted boundary conditions on the spin chain.  

It is quite clear from work on conformal field theory that defects play a fundamental role there \cite{Frohlich2004,Frohlich:2006}. Our results indicate that they play a fundamental role in understanding the connection 
between lattice and continuum as well. The calculation of the conformal spin in the continuum via a Dehn twist in the lattice model provides a way of matching a given lattice model with its corresponding conformal field theory. This is quite a strong constraint, since in rational conformal field theories, the dimensions of operators can take on only certain rational values. In the Ising case, this of course is not necessary; which conformal field theory describes the continuum limit has long been known \cite{BPZ:CFT:84} and indeed been rigorously proven \cite{Smirnov2012}. However, in other models typically much less is known, and so our technique may prove quite useful.

Even still in Ising, we have shown how understanding the branching and fusing of the duality defect allows a derivation of the modular transformation matrices valid both in the lattice and continuum. It also gives a simple method for understanding Kramers-Wannier duality on the torus and higher-genus surfaces on and off the critical point. All these results can be greatly generalised, and in part~II we plan to start exploring them in the context of height models. A few of the many generalisations conceivable are allowing the defect loops to fluctuate, exploring boundary-condition changing operators from defects terminating at boundaries, and building a 2d quantum theory from these topological defects.

\bigskip
{\it Acknowledments}\quad We thank Jason Alicea for illuminating discussions. DA gratefully acknowledges the support of the NSERC PGSD program.
RM is grateful for support from the Sherman Fairchild Foundation and the Institute for Quantum Information and Matter.

\appendix
\section{The zero mode}
\label{zero-mode}
Here we provide a more in-depth analysis of the zero mode appearing due to the presence of the duality defect, as discussed in section \ref{sec:Majorana}.
It is most easily studied by rewriting the Hamiltonian in terms of free-fermionic variables.
Unlike conventional zero modes (e.g. \cite{Kitaev2001}), it does not provide a pairing between states of equal energy in different sectors; that here follows from Kramers pairing. However, we explain below that when the fermions are the physical degrees of freedom, the corresponding boundary conditions for the Hamiltonian change. In this case, the zero mode here does provide a natural pairing between equal energy states, i.e.\ is a strong zero mode in the sense of \cite{Fendley2015}. 

For convenience here we undo the unitary transformation done to get the Hamiltonian (\ref{Hduality}), and study the quantum Hamiltonian arising directly from taking the Hamiltonian limit of the transfer matrix:
\begin{align}
H^{d\pm} =& - J\sum_{j=1}^{r-1}\sigma^z_j \sigma^z_{j+1} -\sum_{j =2}^{r-1}  \sigma^x_j
 -\sum_{j=r}^{L-1} \sigma^z_j \sigma^z_{j+1} - J\sum_{j=r}^{L} \sigma^x_j \mp \sigma_{L}^z \sigma_{1}^x\ .
 \label{Hduality1}
\end{align}
The ${\mathbb Z}_2$ symmetry generator commuting with $H^{d\pm}$ is modified to $\Omega=i\sigma^z_1\D_\psi$.
It turns out to be quite useful to consider both $\pm$ signs on the term arising from the duality defect. Since the unitary transformation $U=\sigma^z_1$ toggles this sign, $H^{d+}$ and $H^{d-}$ have the same spectra.

\begin{figure}[htbp]
	\begin{center}
	\includegraphics[width=5.5in]{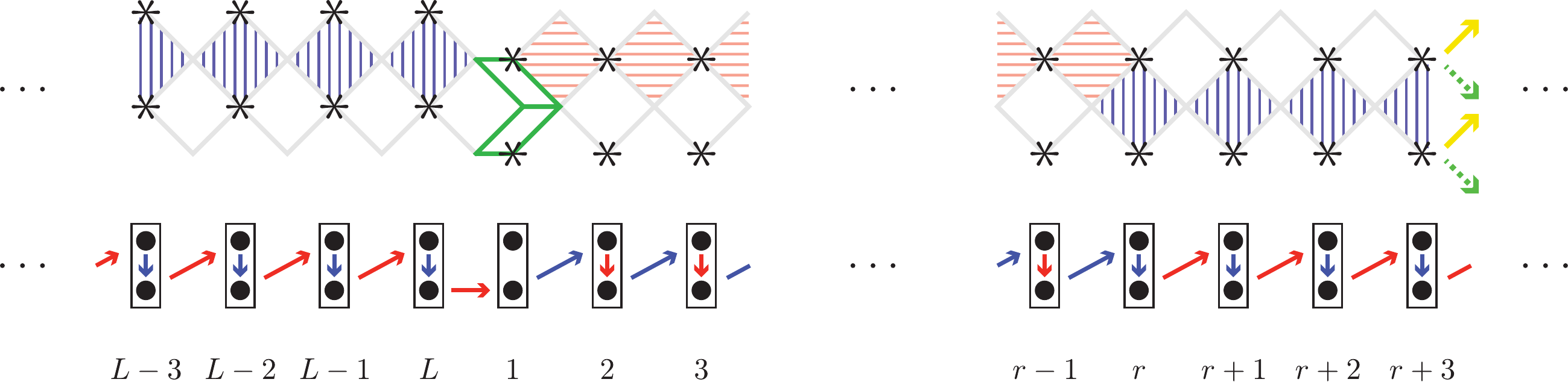}
	\end{center}
	\caption{Anisotripic limit shown in pictorial form when translated over to Majorana variables. The relation between the staggering of the spectral parameters and the hatchings is given in Fig.~\ref{Gluing}.}
		\label{fig:Ising-to-Majorana}
\end{figure}

The free fermionic operators are defined using the Jordan-Wigner transformation:
\begin{align}
  a_j  = \Big( \prod_{k=1}^{j-1} \sigma_k^x \Big)\sigma_j^z ,  \qquad\quad b_j = i \Big( \prod_{k=1}^{j-1} \sigma_k^x \Big)\sigma_j^z \sigma_j^x,
\end{align}
so that $a_j$ and $b_j$ satisfy $\{ a_j, a_{j'} \} =\{ b_j, b_{j'} \} = 2\delta_{j,j'}$ and $\{ a_j, b_{j'} \}=0$. The set $a_j$ and $b_j$ form a set of real ``Majorana'' fermions. We have defined the strings so that they terminate at the duality defect, just as explained in section (\ref{sec:Majorana}).
These operators satisfy $\sigma_j^z \sigma_{j+1}^z = - i b_j a_{j+1}$ and $\sigma_j^x = -i a_j b_j$. 
The boundary term $\sigma_L^z \sigma_1^x$ becomes $\Omega (-ib_L b_1)$. The 
Hamiltonian (\ref{Hduality1}) in terms of fermions is thus
\begin{align}
H^{d\pm} =&  J\sum_{j=1}^{r-1} i  b_j a_{j+1} + \sum_{j=r}^{L-1} i b_j a_{j+1}+\sum_{j =2}^{r-1} i  a_j b_j + J \sum_{j=r}^{L} i a_j b_j \pm i  \Omega b_L b_1
\label{Hf}
\end{align}
where the domain wall remains at site $r$. 
The mapping to Majorana operators is shown schematically in Fig.~\ref{fig:Ising-to-Majorana}. One very interesting fact is that $a_1$ is left entirely out of the Hamiltonian.
At the critical point $J=1$, the Hamiltonian therefore obeys a fermionic version of translation invariance relating the other $2L-1$ fermions. This gives a simple explanation of why the effective length is $L_{\rm eff}=L-1/2$, as derived above.

This Hamiltonian is non-local when written in terms of the fermions. The reason is that the operator $\Omega$ appears in the Hamiltonian, and in terms of fermions, $\Omega = - b_1 \prod_{j = 2}^L (-ia_j b_j)$. One can of course work in sectors of fixed $\Omega$, but for the situation where the fermions are the physical degrees of freedom, it is more appropriate to modify the Hamiltonian. We discuss this below.
Note also that the operator $\Omega$ is a product of an odd number of Majorana operators and does not involve $a_1$. As a result it obeys the following algebra with respect to the Majorana operators:
\begin{align}
\Omega a_1 = -a_1 \Omega,\quad \Omega a_{j\neq1} =  a_{j\neq1} \Omega, \quad \text{and} \quad \Omega b_j = b_j \Omega\ .
\end{align}
This means that even though $a_1=\sigma^z_1$ does not appear in the Hamiltonian $H^{d\pm}$, it does not commute with it. It thus is not a zero mode, but rather is a boundary-condition-changing operator: $a_1H^{d\pm}a_1= H^{d\mp}$.
\comment{
An eigenstate $\ket{\psi_{\pm}, \omega}$ of $H^{d\pm}$ and $\Omega$, with eigenvalues $E^{d\pm}$ and $\omega$ respectively, therefore obeys
\begin{align}
H^{d \mp} a_1 \ket{\psi_{\pm}, \omega} &=a_1 H^{d \pm}a_1 a_1 \ket{\psi_{\pm}, \omega}  = E^{d \pm} a_1 \ket{\psi_{\pm}, \omega},  \\
\Omega a_1 \ket{\psi_{\pm}, \omega} &= - \omega a_1\ket{\psi_{\pm}, \omega} .
\end{align}
Therefore $a_1=\sigma^z_1$ identifies states of $H^{d\pm}$ with charge $\omega$ and states of $H^{d\mp}$ of charge $-\omega$.
}


The spectrum of the Hamiltonian in a fixed charge sector can be found by solving for eigen-operators $\Lambda_n$ such that $[ H^{d\pm}, \Lambda_n] = E_n \Lambda_n$ where the $\Lambda_n$ are linear in the Majorana variables. A particularly interesting mode is the one at zero energy: $\Lambda$ such that $[H^{d\pm},\Lambda] = 0$. The calculation of $\Lambda$ is straightforward via the iterative construction. We define the operators
\begin{equation}
\begin{aligned}
A & =  {J^{r-2}}{a_2} + \cdots +  {J}{a_{r-1}}+ a_r + a_{r+1}  + J{a_{r+2}}+ \cdots   + {J^{L-r-1}}{a_L}\\
B &= J^{2-r} b_1 + \cdots + J^{-1} b_{r-2} +b_{r-1} + b_r + J^{-1} b_{r+1} + \cdots +J^{r-L} b_L\ , \ .
\end{aligned}
\end{equation}
With a little work one finds that these satisfy
\begin{equation}
\begin{aligned}
[H ^{d\pm}, \pm{\Omega} B] &= 2 i  ( J^{2-r} b_L -  J^{r-L} b_1),  \\
[H^{d\pm}, A ] & = 2i(J ^{r-2}b_1 - J^{L-r}b_L).
\end{aligned}
\end{equation}
There are two linear combinations of $A$ and $B$ that commute with $H^{d\pm}$, but only one zero mode for a given value of $J$. The reason is that in the $L\to\infty$ limit, when acting on a normalizable state, the zero mode must give another normalizable one. The normalizable zero modes are
\begin{align}
&\Lambda  =\;
	\begin{cases}		
		   A \pm  {\Omega}J^{L-2}B	&	|J|<1, \\
		 J^{2-L}A \pm {\Omega}B   	&	|J|>1. 
		\end{cases}
		\label{ZeroMode}
\end{align}
The normalization is easy to check. Because the fermionic operators all $a_j$ and $b_j$ anticommute and square to $1$, $\{A,B\}=0$ and
\begin{align}
A^2=\frac{2-J^{2r-2}-J^{2L-2r}}{1-J^2} \ ,\quad B^2=\frac{2-J^{2-2r}-J^{2r-2L-2}}{1-J^{-2}}
\end{align}
so that $\lim_{L\to\infty} \Lambda^2$ is finite except at the critical point $|J|=1$.
The magnitude of each term in $\Lambda$ exponentially decreases as the index is moved away from $r$, but because of the explicit $\Omega$ the zero mode is non-local even in fermionic variables. 

As noted above, the presence of $\Omega$ inside the Hamiltonian (\ref{Hduality1}) makes it local in terms of the spins, but non-local in terms of the fermions. 
It is precisely this factor of $\Omega$ on the boundary link of the fermions that spoils the conventional wisdom attributed to the zero mode found here. 
A conventional Majorana zero mode commutes with the Hamiltonian up to exponentially small corrections, but does not commute with the global charge (proposals for obtaining them in the lab are reviewed in \cite{Aliceareview}). 
The spectrum reflects this by a near-perfect degeneracy between states in the different global charge sectors,  a degeneracy that becomes exact as the exponentially small corrections go to zero.
The zero mode (\ref{ZeroMode}) found here \emph{exactly} commutes with the Hamiltonian.
This is implied by the fact that the duality defect is free to move. 
If the zero mode were not exact, then changing the location of the duality defect would presumably change the energy of the zero mode, contradicting the freedom of movement. 
Note also that $[\Omega,\Lambda]=0$, so that the zero mode does not map between $\Omega=\pm 1$ sectors.

A local and physical boundary condition for fermions yielding conventional zero modes comes from omitting the $\Omega$ in (\ref{Hduality1})
to obtain
\begin{align}
\tilde{H}^\pm &=  \sum_{j=1}^{r-1} i J b_j a_{j+1} + \sum_{j=r}^{L-1} i b_j a_{j+1}+\sum_{j =2}^{r-1} i  a_j b_j + \sum_{j=r}^{L} iJ a_j b_j \pm i b_L b_1\ .
\end{align}
These can be rewritten in terms of the previous Hamiltonian via a ``mixed'' Hamiltonian:
\begin{align}
\tilde{H}^{\pm} &= H^{d+}P_\pm + H^{d-} P_\mp,\quad \text{where} \quad P_\pm = \frac{1\pm \Omega}{2}\ .
\end{align}
It is then straightforward to construct the eigenstates of $H^{d\pm}$ from those of $\tilde{H}^{\pm}$.

\comment{The subspace spanned by $\ket{\tilde{E}^\pm,f}$ and $a_1\ket{\tilde{E}^\pm,f} $ can equivalently be spanned by the states $e^{-i \pi s a_1/4} \ket{\tilde{E}^\pm,+}$ for $s = \pm1$. With the feature that,
\Omega e^{-i \pi s a_1/4} \ket{\tilde{E}^\pm,+} = s e^{-i \pi s a_1/4} \ket{\tilde{E}^\pm,+}.
\end{align}
and similarly with $e^{-i \pi s a_1/4} \ket{\tilde{E}^-,+}$. 
So that the eigenvectors of 
\begin{align}
H^{d\pm} = \tilde{H}^+ P_\pm + \tilde{H}^- P_\mp
\end{align}
are given by  $e^{\mp i \pi a_1/4} \ket{\tilde{E}^+,+}$ and $e^{\pm i \pi a_1/4} \ket{\tilde{E}^-,+}$. Hence given the eigenvectors of the mixed sector Hamiltonian, one can find the eigenvectors of the duality twisted Hamiltonian. 
}

A zero mode  $\widetilde{\Lambda}$ for $\tilde{H}^{\pm}$ can be obtained as above, yielding simply (\ref{ZeroMode}) with $\Omega$ omitted. 
Hence the zero mode in the physical fermionic Hamiltonian is truly localized around the domain wall. Another difference is that there are now two zero modes: $[a_1,\tilde{H}^{\pm} ]=0$ as well. Moreover, now the zero modes do map between sectors.
Namely, there is a second ${\mathbb Z}_2$ symmetry generator  $(-1)^F = i a_1 \Omega$ obeying
\begin{equation}
[(-1)^F, \tilde{H}^{\pm} ]=0\ ,\qquad \{a_n,(-1)^F\}=\{b_n,(-1)^F\}=\{\widetilde{\Lambda},(-1)^F\}=0\ .
\end{equation}
Because the zero modes do not commute with $(-1)^F$ and square to 1, they toggle between the sectors with $(-1)^F=\pm 1$. Because they commute with $\tilde{H}^{\pm}$, the states related by $a_1$ or $\widetilde{\Lambda}$ must have the same energy. Thus $a_1$ and $\widetilde{\Lambda}$ are both strong zero modes, localised at the duality defect and by the domain wall respectively.

\phantomsection
\addcontentsline{toc}{section}{References}

\bibliography{references}

\begin{thebibliography}{52}%
\makeatletter
\providecommand \@ifxundefined [1]{%
 \@ifx{#1\undefined}
}%
\providecommand \@ifnum [1]{%
 \ifnum #1\expandafter \@firstoftwo
 \else \expandafter \@secondoftwo
 \fi
}%
\providecommand \@ifx [1]{%
 \ifx #1\expandafter \@firstoftwo
 \else \expandafter \@secondoftwo
 \fi
}%
\providecommand \natexlab [1]{#1}%
\providecommand \enquote  [1]{``#1''}%
\providecommand \bibnamefont  [1]{#1}%
\providecommand \bibfnamefont [1]{#1}%
\providecommand \citenamefont [1]{#1}%
\providecommand \href@noop [0]{\@secondoftwo}%
\providecommand \href [0]{\begingroup \@sanitize@url \@href}%
\providecommand \@href[1]{\@@startlink{#1}\@@href}%
\providecommand \@@href[1]{\endgroup#1\@@endlink}%
\providecommand \@sanitize@url [0]{\catcode `\\12\catcode `\$12\catcode
  `\&12\catcode `\#12\catcode `\^12\catcode `\_12\catcode `\%12\relax}%
\providecommand \@@startlink[1]{}%
\providecommand \@@endlink[0]{}%
\providecommand \url  [0]{\begingroup\@sanitize@url \@url }%
\providecommand \@url [1]{\endgroup\@href {#1}{\urlprefix }}%
\providecommand \urlprefix  [0]{URL }%
\providecommand \Eprint [0]{\href }%
\providecommand \doibase [0]{http://dx.doi.org/}%
\providecommand \selectlanguage [0]{\@gobble}%
\providecommand \bibinfo  [0]{\@secondoftwo}%
\providecommand \bibfield  [0]{\@secondoftwo}%
\providecommand \translation [1]{[#1]}%
\providecommand \BibitemOpen [0]{}%
\providecommand \bibitemStop [0]{}%
\providecommand \bibitemNoStop [0]{.\EOS\space}%
\providecommand \EOS [0]{\spacefactor3000\relax}%
\providecommand \BibitemShut  [1]{\csname bibitem#1\endcsname}%
\let\auto@bib@innerbib\@empty
\bibitem [{\citenamefont {{Perk}}\ and\ \citenamefont
  {{Au-Yang}}(2006)}]{Perk2006}%
  \BibitemOpen
  \bibfield  {author} {\bibinfo {author} {\bibfnamefont {J.~H.~H.}\
  \bibnamefont {{Perk}}}\ and\ \bibinfo {author} {\bibfnamefont
  {H.}~\bibnamefont {{Au-Yang}}},\ }\href@noop {} {\bibfield  {journal}
  {\bibinfo  {journal} {contribution to Encyclopedia of Mathematical Physics}\
  } (\bibinfo {year} {2006})},\ \Eprint {http://arxiv.org/abs/math-ph/0606053}
  {math-ph/0606053} \BibitemShut {NoStop}%
\bibitem [{\citenamefont {Baxter}(1982)}]{Baxter1982}%
  \BibitemOpen
  \bibfield  {author} {\bibinfo {author} {\bibfnamefont {R.~J.}\ \bibnamefont
  {Baxter}},\ }\href {http://physics.anu.edu.au/theophys/baxter_book.php}
  {\emph {\bibinfo {title} {{Exactly solved models in statistical
  mechanics}}}}\ (\bibinfo  {publisher} {Academic},\ \bibinfo {year}
  {1982})\BibitemShut {NoStop}%
\bibitem [{\citenamefont {Wadati}\ \emph {et~al.}(1989)\citenamefont {Wadati},
  \citenamefont {Deguchi},\ and\ \citenamefont {Akutsu}}]{Wadati1989}%
  \BibitemOpen
  \bibfield  {author} {\bibinfo {author} {\bibfnamefont {M.}~\bibnamefont
  {Wadati}}, \bibinfo {author} {\bibfnamefont {T.}~\bibnamefont {Deguchi}}, \
  and\ \bibinfo {author} {\bibfnamefont {Y.}~\bibnamefont {Akutsu}},\ }\href
  {\doibase 10.1016/0370-1573(89)90123-3} {\bibfield  {journal} {\bibinfo
  {journal} {Phys. Rept.}\ }\textbf {\bibinfo {volume} {180}},\ \bibinfo
  {pages} {247} (\bibinfo {year} {1989})}\BibitemShut {NoStop}%
\bibitem [{\citenamefont {Chui}\ \emph {et~al.}(2001)\citenamefont {Chui},
  \citenamefont {Mercat}, \citenamefont {Orrick},\ and\ \citenamefont
  {Pearce}}]{Pearce2001}%
  \BibitemOpen
  \bibfield  {author} {\bibinfo {author} {\bibfnamefont {C.}~\bibnamefont
  {Chui}}, \bibinfo {author} {\bibfnamefont {C.}~\bibnamefont {Mercat}},
  \bibinfo {author} {\bibfnamefont {W.~P.}\ \bibnamefont {Orrick}}, \ and\
  \bibinfo {author} {\bibfnamefont {P.~A.}\ \bibnamefont {Pearce}},\ }\href
  {\doibase http://dx.doi.org/10.1016/S0370-2693(01)00982-0} {\bibfield
  {journal} {\bibinfo  {journal} {Phys.\ Lett.\ B}\ }\textbf {\bibinfo {volume}
  {517}},\ \bibinfo {pages} {429 } (\bibinfo {year} {2001})}\BibitemShut
  {NoStop}%
\bibitem [{\citenamefont {Chui}\ \emph {et~al.}(2003)\citenamefont {Chui},
  \citenamefont {Mercat},\ and\ \citenamefont {Pearce}}]{Pearce2003}%
  \BibitemOpen
  \bibfield  {author} {\bibinfo {author} {\bibfnamefont {C.~O.}\ \bibnamefont
  {Chui}}, \bibinfo {author} {\bibfnamefont {C.}~\bibnamefont {Mercat}}, \ and\
  \bibinfo {author} {\bibfnamefont {P.~A.}\ \bibnamefont {Pearce}},\ }\href
  {\doibase 10.1088/0305-4470/36/11/301} {\bibfield  {journal} {\bibinfo
  {journal} {J. Phys.}\ }\textbf {\bibinfo {volume} {A36}},\ \bibinfo {pages}
  {2623} (\bibinfo {year} {2003})},\ \Eprint
  {http://arxiv.org/abs/hep-th/0210301} {arXiv:hep-th/0210301} \BibitemShut
  {NoStop}%
\bibitem [{\citenamefont {Kramers}\ and\ \citenamefont
  {Wannier}(1941)}]{Kramers1941}%
  \BibitemOpen
  \bibfield  {author} {\bibinfo {author} {\bibfnamefont {H.~A.}\ \bibnamefont
  {Kramers}}\ and\ \bibinfo {author} {\bibfnamefont {G.~H.}\ \bibnamefont
  {Wannier}},\ }\href {\doibase 10.1103/PhysRev.60.252} {\bibfield  {journal}
  {\bibinfo  {journal} {Phys. Rev.}\ }\textbf {\bibinfo {volume} {60}},\
  \bibinfo {pages} {252} (\bibinfo {year} {1941})}\BibitemShut {NoStop}%
\bibitem [{\citenamefont {Feiguin}\ \emph {et~al.}(2007)\citenamefont
  {Feiguin}, \citenamefont {Trebst}, \citenamefont {Ludwig}, \citenamefont
  {Troyer}, \citenamefont {Kitaev}, \citenamefont {Wang},\ and\ \citenamefont
  {Freedman}}]{Feiguin2007}%
  \BibitemOpen
  \bibfield  {author} {\bibinfo {author} {\bibfnamefont {A.}~\bibnamefont
  {Feiguin}}, \bibinfo {author} {\bibfnamefont {S.}~\bibnamefont {Trebst}},
  \bibinfo {author} {\bibfnamefont {A.~W.~W.}\ \bibnamefont {Ludwig}}, \bibinfo
  {author} {\bibfnamefont {M.}~\bibnamefont {Troyer}}, \bibinfo {author}
  {\bibfnamefont {A.}~\bibnamefont {Kitaev}}, \bibinfo {author} {\bibfnamefont
  {Z.}~\bibnamefont {Wang}}, \ and\ \bibinfo {author} {\bibfnamefont {M.~H.}\
  \bibnamefont {Freedman}},\ }\href {\doibase 10.1103/PhysRevLett.98.160409}
  {\bibfield  {journal} {\bibinfo  {journal} {Phys. Rev. Lett.}\ }\textbf
  {\bibinfo {volume} {98}},\ \bibinfo {pages} {160409} (\bibinfo {year}
  {2007})}\BibitemShut {NoStop}%
\bibitem [{\citenamefont {Gaiotto}\ \emph {et~al.}(2015)\citenamefont
  {Gaiotto}, \citenamefont {Kapustin}, \citenamefont {Seiberg},\ and\
  \citenamefont {Willett}}]{Gaiotto}%
  \BibitemOpen
  \bibfield  {author} {\bibinfo {author} {\bibfnamefont {D.}~\bibnamefont
  {Gaiotto}}, \bibinfo {author} {\bibfnamefont {A.}~\bibnamefont {Kapustin}},
  \bibinfo {author} {\bibfnamefont {N.}~\bibnamefont {Seiberg}}, \ and\
  \bibinfo {author} {\bibfnamefont {B.}~\bibnamefont {Willett}},\ }\href
  {\doibase 10.1007/JHEP02(2015)172} {\bibfield  {journal} {\bibinfo  {journal}
  {JHEP}\ }\textbf {\bibinfo {volume} {02}},\ \bibinfo {pages} {172} (\bibinfo
  {year} {2015})},\ \Eprint {http://arxiv.org/abs/1412.5148} {arXiv:1412.5148}
  \BibitemShut {NoStop}%
\bibitem [{\citenamefont {Schutz}(1993)}]{Schutz1993}%
  \BibitemOpen
  \bibfield  {author} {\bibinfo {author} {\bibfnamefont {G.}~\bibnamefont
  {Schutz}},\ }\href@noop {} {\bibfield  {journal} {\bibinfo  {journal} {J.
  Phys. A}\ }\textbf {\bibinfo {volume} {26}},\ \bibinfo {pages} {4555}
  (\bibinfo {year} {1993})}\BibitemShut {NoStop}%
\bibitem [{\citenamefont {Oshikawa}\ and\ \citenamefont
  {Affleck}(1997)}]{Affleck1997}%
  \BibitemOpen
  \bibfield  {author} {\bibinfo {author} {\bibfnamefont {M.}~\bibnamefont
  {Oshikawa}}\ and\ \bibinfo {author} {\bibfnamefont {I.}~\bibnamefont
  {Affleck}},\ }\href {\doibase
  http://dx.doi.org/10.1016/S0550-3213(97)00219-8} {\bibfield  {journal}
  {\bibinfo  {journal} {Nucl.\ Phys.\ B}\ }\textbf {\bibinfo {volume} {495}},\
  \bibinfo {pages} {533 } (\bibinfo {year} {1997})}\BibitemShut {NoStop}%
\bibitem [{\citenamefont {Grimm}(2002{\natexlab{a}})}]{Grimm2003}%
  \BibitemOpen
  \bibfield  {author} {\bibinfo {author} {\bibfnamefont {U.}~\bibnamefont
  {Grimm}},\ }in\ \href {http://alice.cern.ch/format/showfull?sysnb=2338204}
  {\emph {\bibinfo {booktitle} {{Proceedings, 24th International Colloquium on
  Group Theoretical Methods in Physics (GROUP 24)}}}}\ (\bibinfo {year}
  {2002})\ pp.\ \bibinfo {pages} {395--398},\ \Eprint
  {http://arxiv.org/abs/hep-th/0209048} {arXiv:hep-th/0209048} \BibitemShut
  {NoStop}%
\bibitem [{\citenamefont {Fr\"{o}hlich}\ \emph {et~al.}(2004)\citenamefont
  {Fr\"{o}hlich}, \citenamefont {Fuchs}, \citenamefont {Runkel},\ and\
  \citenamefont {Schweigert}}]{Frohlich2004}%
  \BibitemOpen
  \bibfield  {author} {\bibinfo {author} {\bibfnamefont {J.}~\bibnamefont
  {Fr\"{o}hlich}}, \bibinfo {author} {\bibfnamefont {J.}~\bibnamefont {Fuchs}},
  \bibinfo {author} {\bibfnamefont {I.}~\bibnamefont {Runkel}}, \ and\ \bibinfo
  {author} {\bibfnamefont {C.}~\bibnamefont {Schweigert}},\ }\href {\doibase
  10.1103/PhysRevLett.93.070601} {\bibfield  {journal} {\bibinfo  {journal}
  {Phys. Rev. Lett.}\ }\textbf {\bibinfo {volume} {93}},\ \bibinfo {pages}
  {070601} (\bibinfo {year} {2004})}\BibitemShut {NoStop}%
\bibitem [{\citenamefont {Moore}\ and\ \citenamefont
  {Seiberg}(1989{\natexlab{a}})}]{MSReview89}%
  \BibitemOpen
  \bibfield  {author} {\bibinfo {author} {\bibfnamefont {G.~W.}\ \bibnamefont
  {Moore}}\ and\ \bibinfo {author} {\bibfnamefont {N.}~\bibnamefont
  {Seiberg}},\ }in\ \href {http://alice.cern.ch/format/showfull?sysnb=0113749}
  {\emph {\bibinfo {booktitle} {{1989 Banff NATO ASI: Physics, Geometry and
  Topology Banff, Canada, August 14-25, 1989}}}}\ (\bibinfo {year}
  {1989})\BibitemShut {NoStop}%
\bibitem [{\citenamefont {Cardy}(1986)}]{Cardy1986}%
  \BibitemOpen
  \bibfield  {author} {\bibinfo {author} {\bibfnamefont {J.~L.}\ \bibnamefont
  {Cardy}},\ }\href {\doibase 10.1016/0550-3213(86)90552-3} {\bibfield
  {journal} {\bibinfo  {journal} {Nucl. Phys. B}\ }\textbf {\bibinfo {volume}
  {270}},\ \bibinfo {pages} {186} (\bibinfo {year} {1986})}\BibitemShut
  {NoStop}%
\bibitem [{\citenamefont {Cardy}(1989)}]{Cardy1989}%
  \BibitemOpen
  \bibfield  {author} {\bibinfo {author} {\bibfnamefont {J.~L.}\ \bibnamefont
  {Cardy}},\ }\href {\doibase 10.1016/0550-3213(89)90521-X} {\bibfield
  {journal} {\bibinfo  {journal} {Nucl. Phys.}\ }\textbf {\bibinfo {volume}
  {B324}},\ \bibinfo {pages} {581} (\bibinfo {year} {1989})}\BibitemShut
  {NoStop}%
\bibitem [{\citenamefont {Petkova}\ and\ \citenamefont
  {Zuber}(2001{\natexlab{a}})}]{Petkova2001}%
  \BibitemOpen
  \bibfield  {author} {\bibinfo {author} {\bibfnamefont {V.}~\bibnamefont
  {Petkova}}\ and\ \bibinfo {author} {\bibfnamefont {J.-B.}\ \bibnamefont
  {Zuber}},\ }\href {\doibase http://dx.doi.org/10.1016/S0370-2693(01)00276-3}
  {\bibfield  {journal} {\bibinfo  {journal} {Phys. Lett.\ B}\ }\textbf
  {\bibinfo {volume} {504}},\ \bibinfo {pages} {157 } (\bibinfo {year}
  {2001}{\natexlab{a}})}\BibitemShut {NoStop}%
\bibitem [{\citenamefont {Petkova}\ and\ \citenamefont
  {Zuber}(2001{\natexlab{b}})}]{Petkova2001b}%
  \BibitemOpen
  \bibfield  {author} {\bibinfo {author} {\bibfnamefont {V.}~\bibnamefont
  {Petkova}}\ and\ \bibinfo {author} {\bibfnamefont {J.-B.}\ \bibnamefont
  {Zuber}},\ }\href {\doibase http://dx.doi.org/10.1016/S0550-3213(01)00096-7}
  {\bibfield  {journal} {\bibinfo  {journal} {Nuclear Physics B}\ }\textbf
  {\bibinfo {volume} {603}},\ \bibinfo {pages} {449 } (\bibinfo {year}
  {2001}{\natexlab{b}})}\BibitemShut {NoStop}%
\bibitem [{\citenamefont {Fuchs}\ \emph {et~al.}(2002)\citenamefont {Fuchs},
  \citenamefont {Runkel},\ and\ \citenamefont {Schweigert}}]{Runkel2002}%
  \BibitemOpen
  \bibfield  {author} {\bibinfo {author} {\bibfnamefont {J.}~\bibnamefont
  {Fuchs}}, \bibinfo {author} {\bibfnamefont {I.}~\bibnamefont {Runkel}}, \
  and\ \bibinfo {author} {\bibfnamefont {C.}~\bibnamefont {Schweigert}},\
  }\href {\doibase http://dx.doi.org/10.1016/S0550-3213(02)00744-7} {\bibfield
  {journal} {\bibinfo  {journal} {Nuclear Physics B}\ }\textbf {\bibinfo
  {volume} {646}},\ \bibinfo {pages} {353 } (\bibinfo {year}
  {2002})}\BibitemShut {NoStop}%
\bibitem [{\citenamefont {Fuchs}\ \emph
  {et~al.}(2004{\natexlab{a}})\citenamefont {Fuchs}, \citenamefont {Runkel},\
  and\ \citenamefont {Schweigert}}]{Runkel2004}%
  \BibitemOpen
  \bibfield  {author} {\bibinfo {author} {\bibfnamefont {J.}~\bibnamefont
  {Fuchs}}, \bibinfo {author} {\bibfnamefont {I.}~\bibnamefont {Runkel}}, \
  and\ \bibinfo {author} {\bibfnamefont {C.}~\bibnamefont {Schweigert}},\
  }\href {\doibase http://dx.doi.org/10.1016/j.nuclphysb.2003.11.026}
  {\bibfield  {journal} {\bibinfo  {journal} {Nuclear Physics B}\ }\textbf
  {\bibinfo {volume} {678}},\ \bibinfo {pages} {511 } (\bibinfo {year}
  {2004}{\natexlab{a}})}\BibitemShut {NoStop}%
\bibitem [{\citenamefont {Fuchs}\ \emph
  {et~al.}(2004{\natexlab{b}})\citenamefont {Fuchs}, \citenamefont {Runkel},\
  and\ \citenamefont {Schweigert}}]{Runkel2004b}%
  \BibitemOpen
  \bibfield  {author} {\bibinfo {author} {\bibfnamefont {J.}~\bibnamefont
  {Fuchs}}, \bibinfo {author} {\bibfnamefont {I.}~\bibnamefont {Runkel}}, \
  and\ \bibinfo {author} {\bibfnamefont {C.}~\bibnamefont {Schweigert}},\
  }\href {\doibase http://dx.doi.org/10.1016/j.nuclphysb.2004.05.014}
  {\bibfield  {journal} {\bibinfo  {journal} {Nuclear Physics B}\ }\textbf
  {\bibinfo {volume} {694}},\ \bibinfo {pages} {277 } (\bibinfo {year}
  {2004}{\natexlab{b}})}\BibitemShut {NoStop}%
\bibitem [{\citenamefont {Fuchs}\ \emph {et~al.}(2005)\citenamefont {Fuchs},
  \citenamefont {Runkel},\ and\ \citenamefont {Schweigert}}]{Runkel2005}%
  \BibitemOpen
  \bibfield  {author} {\bibinfo {author} {\bibfnamefont {J.}~\bibnamefont
  {Fuchs}}, \bibinfo {author} {\bibfnamefont {I.}~\bibnamefont {Runkel}}, \
  and\ \bibinfo {author} {\bibfnamefont {C.}~\bibnamefont {Schweigert}},\
  }\href {\doibase http://dx.doi.org/10.1016/j.nuclphysb.2005.03.018}
  {\bibfield  {journal} {\bibinfo  {journal} {Nuclear Physics B}\ }\textbf
  {\bibinfo {volume} {715}},\ \bibinfo {pages} {539 } (\bibinfo {year}
  {2005})}\BibitemShut {NoStop}%
\bibitem [{\citenamefont {Fjelstad}\ \emph {et~al.}(2006)\citenamefont
  {Fjelstad}, \citenamefont {Fuchs}, \citenamefont {Runkel},\ and\
  \citenamefont {Schweigert}}]{Runkel2006}%
  \BibitemOpen
  \bibfield  {author} {\bibinfo {author} {\bibfnamefont {J.}~\bibnamefont
  {Fjelstad}}, \bibinfo {author} {\bibfnamefont {J.}~\bibnamefont {Fuchs}},
  \bibinfo {author} {\bibfnamefont {I.}~\bibnamefont {Runkel}}, \ and\ \bibinfo
  {author} {\bibfnamefont {C.}~\bibnamefont {Schweigert}},\ }\href@noop {}
  {\bibfield  {journal} {\bibinfo  {journal} {Theory and Applications of
  Categories}\ }\textbf {\bibinfo {volume} {16}},\ \bibinfo {pages} {342}
  (\bibinfo {year} {2006})}\BibitemShut {NoStop}%
\bibitem [{\citenamefont {Fr\"{o}hlich}\ \emph {et~al.}(2007)\citenamefont
  {Fr\"{o}hlich}, \citenamefont {Fuchs}, \citenamefont {Runkel},\ and\
  \citenamefont {Schweigert}}]{Frohlich:2006}%
  \BibitemOpen
  \bibfield  {author} {\bibinfo {author} {\bibfnamefont {J.}~\bibnamefont
  {Fr\"{o}hlich}}, \bibinfo {author} {\bibfnamefont {J.}~\bibnamefont {Fuchs}},
  \bibinfo {author} {\bibfnamefont {I.}~\bibnamefont {Runkel}}, \ and\ \bibinfo
  {author} {\bibfnamefont {C.}~\bibnamefont {Schweigert}},\ }\href {\doibase
  10.1016/j.nuclphysb.2006.11.017} {\bibfield  {journal} {\bibinfo  {journal}
  {Nucl. Phys.}\ }\textbf {\bibinfo {volume} {B763}},\ \bibinfo {pages} {354}
  (\bibinfo {year} {2007})},\ \Eprint {http://arxiv.org/abs/hep-th/0607247}
  {arXiv:hep-th/0607247} \BibitemShut {NoStop}%
\bibitem [{\citenamefont {Affleck}\ and\ \citenamefont
  {Ludwig}(1991)}]{AffleckLudwig}%
  \BibitemOpen
  \bibfield  {author} {\bibinfo {author} {\bibfnamefont {I.}~\bibnamefont
  {Affleck}}\ and\ \bibinfo {author} {\bibfnamefont {A.~W.~W.}\ \bibnamefont
  {Ludwig}},\ }\href {\doibase 10.1103/PhysRevLett.67.161} {\bibfield
  {journal} {\bibinfo  {journal} {Phys. Rev. Lett.}\ }\textbf {\bibinfo
  {volume} {67}},\ \bibinfo {pages} {161} (\bibinfo {year} {1991})}\BibitemShut
  {NoStop}%
\bibitem [{\citenamefont {McCoy}\ and\ \citenamefont {Wu}(2014)}]{mccoywubook}%
  \BibitemOpen
  \bibfield  {author} {\bibinfo {author} {\bibfnamefont {B.}~\bibnamefont
  {McCoy}}\ and\ \bibinfo {author} {\bibfnamefont {T.}~\bibnamefont {Wu}},\
  }\href@noop {} {\emph {\bibinfo {title} {The Two-Dimensional Ising Model:
  Second Edition}}},\ Dover books on physics\ (\bibinfo  {publisher} {Dover
  Publications},\ \bibinfo {year} {2014})\BibitemShut {NoStop}%
\bibitem [{\citenamefont {Choi}\ \emph {et~al.}(1989)\citenamefont {Choi},
  \citenamefont {Kim},\ and\ \citenamefont {Pearce}}]{Pearce1989}%
  \BibitemOpen
  \bibfield  {author} {\bibinfo {author} {\bibfnamefont {J.-Y.}\ \bibnamefont
  {Choi}}, \bibinfo {author} {\bibfnamefont {D.}~\bibnamefont {Kim}}, \ and\
  \bibinfo {author} {\bibfnamefont {P.~A.}\ \bibnamefont {Pearce}},\ }\href
  {http://stacks.iop.org/0305-4470/22/i=10/a=020} {\bibfield  {journal}
  {\bibinfo  {journal} {Journal of Physics A: Mathematical and General}\
  }\textbf {\bibinfo {volume} {22}},\ \bibinfo {pages} {1661} (\bibinfo {year}
  {1989})}\BibitemShut {NoStop}%
\bibitem [{\citenamefont {Kadanoff}\ and\ \citenamefont
  {Ceva}(1971)}]{KadanoffCeva}%
  \BibitemOpen
  \bibfield  {author} {\bibinfo {author} {\bibfnamefont {L.~P.}\ \bibnamefont
  {Kadanoff}}\ and\ \bibinfo {author} {\bibfnamefont {H.}~\bibnamefont
  {Ceva}},\ }\href {\doibase 10.1103/PhysRevB.3.3918} {\bibfield  {journal}
  {\bibinfo  {journal} {Phys. Rev.}\ }\textbf {\bibinfo {volume} {B3}},\
  \bibinfo {pages} {3918} (\bibinfo {year} {1971})}\BibitemShut {NoStop}%
\bibitem [{\citenamefont {Pasquier}\ and\ \citenamefont
  {Saleur}(1990)}]{Pasquier1990}%
  \BibitemOpen
  \bibfield  {author} {\bibinfo {author} {\bibfnamefont {V.}~\bibnamefont
  {Pasquier}}\ and\ \bibinfo {author} {\bibfnamefont {H.}~\bibnamefont
  {Saleur}},\ }\href {\doibase http://dx.doi.org/10.1016/0550-3213(90)90122-T}
  {\bibfield  {journal} {\bibinfo  {journal} {Nuclear Physics B}\ }\textbf
  {\bibinfo {volume} {330}},\ \bibinfo {pages} {523 } (\bibinfo {year}
  {1990})}\BibitemShut {NoStop}%
\bibitem [{\citenamefont {Levy}(1991)}]{Levy1991}%
  \BibitemOpen
  \bibfield  {author} {\bibinfo {author} {\bibfnamefont {D.}~\bibnamefont
  {Levy}},\ }\href {\doibase 10.1103/PhysRevLett.67.1971} {\bibfield  {journal}
  {\bibinfo  {journal} {Phys. Rev. Lett.}\ }\textbf {\bibinfo {volume} {67}},\
  \bibinfo {pages} {1971} (\bibinfo {year} {1991})}\BibitemShut {NoStop}%
\bibitem [{\citenamefont {Grimm}\ and\ \citenamefont
  {Sch\"{u}tz}(1993)}]{GrimmSchutz1993}%
  \BibitemOpen
  \bibfield  {author} {\bibinfo {author} {\bibfnamefont {U.}~\bibnamefont
  {Grimm}}\ and\ \bibinfo {author} {\bibfnamefont {G.}~\bibnamefont
  {Sch\"{u}tz}},\ }\href {\doibase 10.1007/BF01049955} {\bibfield  {journal}
  {\bibinfo  {journal} {J.\ Stat.\ Phys.}\ }\textbf {\bibinfo {volume} {71}},\
  \bibinfo {pages} {923} (\bibinfo {year} {1993})}\BibitemShut {NoStop}%
\bibitem [{\citenamefont {Grimm}(2002{\natexlab{b}})}]{Grimm2002}%
  \BibitemOpen
  \bibfield  {author} {\bibinfo {author} {\bibfnamefont {U.}~\bibnamefont
  {Grimm}},\ }\href@noop {} {\bibfield  {journal} {\bibinfo  {journal} {J.
  Phys. A}\ }\textbf {\bibinfo {volume} {35}},\ \bibinfo {pages} {L25}
  (\bibinfo {year} {2002}{\natexlab{b}})}\BibitemShut {NoStop}%
\bibitem [{\citenamefont {Poghosyan}\ \emph {et~al.}(2015)\citenamefont
  {Poghosyan}, \citenamefont {Kenna},\ and\ \citenamefont
  {Izmailian}}]{Armen2015}%
  \BibitemOpen
  \bibfield  {author} {\bibinfo {author} {\bibfnamefont {A.}~\bibnamefont
  {Poghosyan}}, \bibinfo {author} {\bibfnamefont {R.}~\bibnamefont {Kenna}}, \
  and\ \bibinfo {author} {\bibfnamefont {N.}~\bibnamefont {Izmailian}},\ }\href
  {http://stacks.iop.org/0295-5075/111/i=6/a=60010} {\bibfield  {journal}
  {\bibinfo  {journal} {Europhys. Lett.}\ }\textbf {\bibinfo {volume} {111}},\
  \bibinfo {pages} {60010} (\bibinfo {year} {2015})}\BibitemShut {NoStop}%
\bibitem [{\citenamefont {Alicea}(2012)}]{Aliceareview}%
  \BibitemOpen
  \bibfield  {author} {\bibinfo {author} {\bibfnamefont {J.}~\bibnamefont
  {Alicea}},\ }\href {\doibase 10.1088/0034-4885/75/7/076501} {\bibfield
  {journal} {\bibinfo  {journal} {Rept. Prog. Phys.}\ }\textbf {\bibinfo
  {volume} {75}},\ \bibinfo {pages} {076501} (\bibinfo {year} {2012})},\
  \Eprint {http://arxiv.org/abs/1202.1293} {arXiv:1202.1293
  [cond-mat.supr-con]} \BibitemShut {NoStop}%
\bibitem [{\citenamefont {O'Brien}\ \emph {et~al.}(1996)\citenamefont
  {O'Brien}, \citenamefont {Pearce},\ and\ \citenamefont
  {Warnaar}}]{OBrien1996}%
  \BibitemOpen
  \bibfield  {author} {\bibinfo {author} {\bibfnamefont {D.~L.}\ \bibnamefont
  {O'Brien}}, \bibinfo {author} {\bibfnamefont {P.~A.}\ \bibnamefont {Pearce}},
  \ and\ \bibinfo {author} {\bibfnamefont {S.~O.}\ \bibnamefont {Warnaar}},\
  }\href {\doibase http://dx.doi.org/10.1016/S0378-4371(96)00055-6} {\bibfield
  {journal} {\bibinfo  {journal} {Physica A}\ }\textbf {\bibinfo {volume}
  {228}},\ \bibinfo {pages} {63 } (\bibinfo {year} {1996})}\BibitemShut
  {NoStop}%
\bibitem [{\citenamefont {Chui}\ and\ \citenamefont {Pearce}(2002)}]{Chui2002}%
  \BibitemOpen
  \bibfield  {author} {\bibinfo {author} {\bibfnamefont {C.~H.~O.}\
  \bibnamefont {Chui}}\ and\ \bibinfo {author} {\bibfnamefont {P.~A.}\
  \bibnamefont {Pearce}},\ }\href {\doibase 10.1023/A:1015113909363} {\bibfield
   {journal} {\bibinfo  {journal} {J. Stat. Phys.}\ }\textbf {\bibinfo {volume}
  {107}},\ \bibinfo {pages} {1167} (\bibinfo {year} {2002})}\BibitemShut
  {NoStop}%
\bibitem [{\citenamefont {Kitaev}(2001)}]{Kitaev2001}%
  \BibitemOpen
  \bibfield  {author} {\bibinfo {author} {\bibfnamefont {A.~Y.}\ \bibnamefont
  {Kitaev}},\ }\href@noop {} {\bibfield  {journal} {\bibinfo  {journal}
  {Physics-Uspekhi}\ }\textbf {\bibinfo {volume} {44}},\ \bibinfo {pages} {131}
  (\bibinfo {year} {2001})}\BibitemShut {NoStop}%
\bibitem [{\citenamefont {{Fendley}}(2015)}]{Fendley2015}%
  \BibitemOpen
  \bibfield  {author} {\bibinfo {author} {\bibfnamefont {P.}~\bibnamefont
  {{Fendley}}},\ }\href@noop {} {\bibfield  {journal} {\bibinfo  {journal}
  {e-print}\ } (\bibinfo {year} {2015})},\ \Eprint
  {http://arxiv.org/abs/1512.03441} {arXiv:1512.03441 [cond-mat.stat-mech]}
  \BibitemShut {NoStop}%
\bibitem [{\citenamefont {Moore}\ and\ \citenamefont
  {Seiberg}(1989{\natexlab{b}})}]{MooreSeiberg:89}%
  \BibitemOpen
  \bibfield  {author} {\bibinfo {author} {\bibfnamefont {G.}~\bibnamefont
  {Moore}}\ and\ \bibinfo {author} {\bibfnamefont {N.}~\bibnamefont
  {Seiberg}},\ }\href {\doibase 10.1007/BF01238857} {\bibfield  {journal}
  {\bibinfo  {journal} {Commun. Math. Phys.}\ }\textbf {\bibinfo {volume}
  {123}},\ \bibinfo {pages} {177} (\bibinfo {year}
  {1989}{\natexlab{b}})}\BibitemShut {NoStop}%
\bibitem [{\citenamefont {Bonderson}(2007)}]{Bondersonthesis}%
  \BibitemOpen
  \bibfield  {author} {\bibinfo {author} {\bibfnamefont {P.~H.}\ \bibnamefont
  {Bonderson}},\ }\emph {\bibinfo {title} {{Non-abelian anyons and
  interferometry}}},\ \href
  {http://etd.caltech.edu/etd/available/etd-06042007-101617/unrestricted/thesis.pdf}
  {Ph.D. thesis},\ \bibinfo  {school} {Caltech} (\bibinfo {year}
  {2007})\BibitemShut {NoStop}%
\bibitem [{\citenamefont {Kitaev}(2006)}]{Kitaev2006}%
  \BibitemOpen
  \bibfield  {author} {\bibinfo {author} {\bibfnamefont {A.}~\bibnamefont
  {Kitaev}},\ }\href {\doibase 10.1016/j.aop.2005.10.005} {\bibfield  {journal}
  {\bibinfo  {journal} {Annals Phys.}\ }\textbf {\bibinfo {volume} {321}},\
  \bibinfo {pages} {2} (\bibinfo {year} {2006})}\BibitemShut {NoStop}%
\bibitem [{\citenamefont {Andrews}\ \emph {et~al.}(1984)\citenamefont
  {Andrews}, \citenamefont {Baxter},\ and\ \citenamefont
  {Forrester}}]{Andrews1984}%
  \BibitemOpen
  \bibfield  {author} {\bibinfo {author} {\bibfnamefont {G.}~\bibnamefont
  {Andrews}}, \bibinfo {author} {\bibfnamefont {R.}~\bibnamefont {Baxter}}, \
  and\ \bibinfo {author} {\bibfnamefont {P.}~\bibnamefont {Forrester}},\ }\href
  {\doibase 10.1007/BF01014383} {\bibfield  {journal} {\bibinfo  {journal} {J.\
  Stat.\ Phys.}\ }\textbf {\bibinfo {volume} {35}},\ \bibinfo {pages} {193}
  (\bibinfo {year} {1984})}\BibitemShut {NoStop}%
\bibitem [{\citenamefont {Tesler}(2000)}]{Tesler}%
  \BibitemOpen
  \bibfield  {author} {\bibinfo {author} {\bibfnamefont {G.}~\bibnamefont
  {Tesler}},\ }\href {\doibase http://dx.doi.org/10.1006/jctb.1999.1941}
  {\bibfield  {journal} {\bibinfo  {journal} {Journal of Combinatorial Theory,
  Series B}\ }\textbf {\bibinfo {volume} {78}},\ \bibinfo {pages} {198 }
  (\bibinfo {year} {2000})}\BibitemShut {NoStop}%
\bibitem [{\citenamefont {Ginsparg}(1988)}]{Ginsparg}%
  \BibitemOpen
  \bibfield  {author} {\bibinfo {author} {\bibfnamefont {P.~H.}\ \bibnamefont
  {Ginsparg}},\ }in\ \href
  {http://inspirehep.net/record/265020/files/arXiv:hep-th_9108028.pdf} {\emph
  {\bibinfo {booktitle} {{Les Houches Summer School in Theoretical Physics:
  Fields, Strings, Critical Phenomena Les Houches, France, June 28-August 5,
  1988}}}}\ (\bibinfo {year} {1988})\ \Eprint
  {http://arxiv.org/abs/hep-th/9108028} {arXiv:hep-th/9108028} \BibitemShut
  {NoStop}%
\bibitem [{\citenamefont {Bl\"ote}\ \emph {et~al.}(1986)\citenamefont
  {Bl\"ote}, \citenamefont {Cardy},\ and\ \citenamefont
  {Nightingale}}]{Blote1986}%
  \BibitemOpen
  \bibfield  {author} {\bibinfo {author} {\bibfnamefont {H.~W.~J.}\
  \bibnamefont {Bl\"ote}}, \bibinfo {author} {\bibfnamefont {J.~L.}\
  \bibnamefont {Cardy}}, \ and\ \bibinfo {author} {\bibfnamefont {M.~P.}\
  \bibnamefont {Nightingale}},\ }\href {\doibase 10.1103/PhysRevLett.56.742}
  {\bibfield  {journal} {\bibinfo  {journal} {Phys. Rev. Lett.}\ }\textbf
  {\bibinfo {volume} {56}},\ \bibinfo {pages} {742} (\bibinfo {year}
  {1986})}\BibitemShut {NoStop}%
\bibitem [{\citenamefont {Affleck}(1986)}]{Affleck1986}%
  \BibitemOpen
  \bibfield  {author} {\bibinfo {author} {\bibfnamefont {I.}~\bibnamefont
  {Affleck}},\ }\href {\doibase 10.1103/PhysRevLett.56.746} {\bibfield
  {journal} {\bibinfo  {journal} {Phys. Rev. Lett.}\ }\textbf {\bibinfo
  {volume} {56}},\ \bibinfo {pages} {746} (\bibinfo {year} {1986})}\BibitemShut
  {NoStop}%
\bibitem [{\citenamefont {Belavin}\ \emph {et~al.}(1984)\citenamefont
  {Belavin}, \citenamefont {Polyakov},\ and\ \citenamefont
  {Zamolodchikov}}]{BPZ:CFT:84}%
  \BibitemOpen
  \bibfield  {author} {\bibinfo {author} {\bibfnamefont {A.~A.}\ \bibnamefont
  {Belavin}}, \bibinfo {author} {\bibfnamefont {A.~M.}\ \bibnamefont
  {Polyakov}}, \ and\ \bibinfo {author} {\bibfnamefont {A.~B.}\ \bibnamefont
  {Zamolodchikov}},\ }\href {\doibase 10.1016/0550-3213(84)90052-X} {\bibfield
  {journal} {\bibinfo  {journal} {Nucl. Phys. B}\ }\textbf {\bibinfo {volume}
  {241}},\ \bibinfo {pages} {333} (\bibinfo {year} {1984})}\BibitemShut
  {NoStop}%
\bibitem [{\citenamefont {Duminil-Copin}\ and\ \citenamefont
  {Smirnov}(2012)}]{Smirnov2012}%
  \BibitemOpen
  \bibfield  {author} {\bibinfo {author} {\bibfnamefont {H.}~\bibnamefont
  {Duminil-Copin}}\ and\ \bibinfo {author} {\bibfnamefont {S.}~\bibnamefont
  {Smirnov}},\ }in\ \href@noop {} {\emph {\bibinfo {booktitle} {Probability and
  Statistical Physics in Two and More Dimensions, Clay Mathematics
  Proceedings}}},\ Vol.~\bibinfo {volume} {15}\ (\bibinfo {year} {2012})\ pp.\
  \bibinfo {pages} {213--276},\ \Eprint {http://arxiv.org/abs/1109.1549}
  {arXiv:1109.1549} \BibitemShut {NoStop}%
\bibitem [{\citenamefont {Mercat}(2001)}]{Mercat2001}%
  \BibitemOpen
  \bibfield  {author} {\bibinfo {author} {\bibfnamefont {C.}~\bibnamefont
  {Mercat}},\ }\href {\doibase 10.1007/s002200000348} {\bibfield  {journal}
  {\bibinfo  {journal} {Comm. Math. Phys.}\ }\textbf {\bibinfo {volume}
  {218}},\ \bibinfo {pages} {177} (\bibinfo {year} {2001})}\BibitemShut
  {NoStop}%
\bibitem [{\citenamefont {Bobenko}\ and\ \citenamefont
  {Suris}(2002)}]{Bobenko2002}%
  \BibitemOpen
  \bibfield  {author} {\bibinfo {author} {\bibfnamefont {A.~I.}\ \bibnamefont
  {Bobenko}}\ and\ \bibinfo {author} {\bibfnamefont {Y.~B.}\ \bibnamefont
  {Suris}},\ }\href {\doibase 10.1155/S1073792802110075} {\bibfield  {journal}
  {\bibinfo  {journal} {International Mathematics Research Notices}\ }\textbf
  {\bibinfo {volume} {2002}},\ \bibinfo {pages} {573} (\bibinfo {year}
  {2002})},\ \Eprint
  {http://arxiv.org/abs/http://imrn.oxfordjournals.org/content/2002/11/573.full.pdf+html}
  {http://imrn.oxfordjournals.org/content/2002/11/573.full.pdf+html}
  \BibitemShut {NoStop}%
\bibitem [{\citenamefont {Mohar}\ and\ \citenamefont
  {Thomassen}(2001)}]{mohar2001}%
  \BibitemOpen
  \bibfield  {author} {\bibinfo {author} {\bibfnamefont {B.}~\bibnamefont
  {Mohar}}\ and\ \bibinfo {author} {\bibfnamefont {C.}~\bibnamefont
  {Thomassen}},\ }\href@noop {} {\emph {\bibinfo {title} {Graphs on
  surfaces}}},\ Vol.~\bibinfo {volume} {10}\ (\bibinfo  {publisher} {JHU
  Press},\ \bibinfo {year} {2001})\BibitemShut {NoStop}%
\bibitem [{\citenamefont {Gross}\ and\ \citenamefont
  {Tucker}(1987)}]{gross1987}%
  \BibitemOpen
  \bibfield  {author} {\bibinfo {author} {\bibfnamefont {J.~L.}\ \bibnamefont
  {Gross}}\ and\ \bibinfo {author} {\bibfnamefont {T.~W.}\ \bibnamefont
  {Tucker}},\ }\href@noop {} {\emph {\bibinfo {title} {Topological graph
  theory}}}\ (\bibinfo  {publisher} {Courier Corporation},\ \bibinfo {year}
  {1987})\BibitemShut {NoStop}%
\bibitem [{\citenamefont {Hatcher}\ and\ \citenamefont
  {Thurston}(1980)}]{Hatcher1980}%
  \BibitemOpen
  \bibfield  {author} {\bibinfo {author} {\bibfnamefont {A.}~\bibnamefont
  {Hatcher}}\ and\ \bibinfo {author} {\bibfnamefont {W.}~\bibnamefont
  {Thurston}},\ }\href {\doibase
  http://dx.doi.org/10.1016/0040-9383(80)90009-9} {\bibfield  {journal}
  {\bibinfo  {journal} {Topology}\ }\textbf {\bibinfo {volume} {19}},\ \bibinfo
  {pages} {221 } (\bibinfo {year} {1980})}\BibitemShut {NoStop}%
\end{thebibliography}%
\bibliographystyle{apsrev4-1}

\clearpage

\end{document}